\documentclass[11pt,a4paper]{article}
\usepackage{jheppub}
\usepackage[sort&compress,numbers]{natbib}
\usepackage[utf8]{inputenc}
\usepackage{todonotes}
\usepackage{amsmath}
\usepackage{float}
\usepackage{amssymb}
\usepackage{bm}
\usepackage{tikz}
\tikzset{>=latex}
\usepackage{graphicx}
\usepackage{changepage}
\usepackage{siunitx}
\usepackage[section]{placeins}
\usepackage[caption=false]{subfig}
\usepackage{siunitx}
\usepackage{pgfplots}
\usepackage{hyperref}
\usetikzlibrary{pgfplots.fillbetween}
\usetikzlibrary{math}

\numberwithin{equation}{section}

\setcitestyle{square}

\begin{document}

\title{A detailed study of the stability of vortons}

\author[1]{Richard A. Battye}
\author[2]{Steven J. Cotterill}
\author{Jonathan A. Pearson}

\affiliation[1,2]{Jodrell Bank Centre for Astrophysics, Department of Physics and Astronomy, University of Manchester, Manchester, M13 9PL, UK}

\emailAdd{richard.battye@manchester.ac.uk}
\emailAdd{steven.cotterill@manchester.ac.uk}
\emailAdd{jpoffline@gmail.com}

\abstract{We construct and simulate the dynamics of gauged vortons - circular loops of cosmic string supported by the angular momentum of trapped charge and current and provide additional details on the fully stable vorton that we have previously presented. We find that their existence and dynamical properties can be accurately predicted by an analysis based on infinite, straight superconducting strings if an additional constraint on their phase frequency is satisfied. We show a good quantitative agreement with the thin string approximation (TSA) and provide evidence that curvature corrections are inversely proportional to the vorton radius. This is verified with an energy minimisation algorithm that produces vorton solutions and subsequent axial and full three dimensional evolution codes. We find that we can predict the frequencies of each mode of oscillation, determine which modes are unstable and calculate the growth rate of the unstable modes to a high degree of accuracy.}

\date{\today}

\maketitle

\flushbottom

\newpage

\section{Introduction}

Spontaneously broken symmetries may have produced a variety of topological solitons in the early universe \cite{V&Sbook}.
Cosmic strings are one dimensional topological defects that cannot have free ends and therefore must either be infinite or form into loops. In 1985, Witten \cite{Witten1985} showed that strings can behave as superconducting wires with either fermionic or bosonic charge carriers. In the fermionic case, there are trapped zero modes along the string due to the Yukawa coupling between the fermion and the vortex (string) forming scalar field. In this paper, we will only consider the bosonic case in which there are two, coupled scalar fields and the onset of superconductivity is due to the second scalar field developing an expectation value inside the core of the string.

Strings will form if the first homotopy group of the vacuum manifold is non-trivial \cite{M&Sbook,V&Sbook}. The particular symmetry group of the theory does not matter as long as this condition is satisfied and as such there is a large class of symmetry breaking processes which have the potential to produce superconducting strings. The simplest of these is a $U(1)\times U(1)$ model (one breaks to form strings while the other remains unbroken in the vacuum and is usually assumed to be associated with electromagnetism) which is typically used in the literature with the assumption that other models with more complicated group structures will behave in a similar way. We will also adopt this approach but it is important to remember that the phenomenon is more general.

It was pointed out in \cite{Ostriker1986} that stable loops of superconducting string may be possible within this theory and that - due to the very large mass per unit length of GUT scale strings - they could easily overclose the universe. Early studies \cite{Copeland1987,Hill1988,Amsterdamski1988,Babul1988,Davis1988a,Haws1988} were focused on the currents completely cancelling the string tension and producing a ``cosmic spring'', however the current never becomes large enough to significantly reduce the tension \cite{Peter1992} due to the effect of current quenching \cite{Davis1988a}. Including the effects of the gauge fields allows for a static state to form in which the tension is balanced by the repulsive magnetic fields, although this only occurs in a narrow region of the parameter space \cite{Hill1988}. It has been conjectured in \cite{Peter1993} that generating the required current is sufficiently difficult that springs will be irrelevant in a cosmological context. Spinning, current carrying loops are an extension, proposed in \cite{Davis1988b,Davis1988c}, that carry charge as well as current. Stable loops of this type (known as vortons) are supported by conservation of angular momentum and require significantly lower currents than the non-spinning alternative. They are expected to be stable to radial perturbations from energetic considerations, but their stability to non-axial effects remains an open problem.

We will study a model with a vortex forming scalar field, $\phi$, with a local $U(1)_\phi$ symmetry and a condensate scalar field, $\sigma$, with a global $U(1)_\sigma$ symmetry. We will work in the neutral current limit that was examined in \cite{Davis1988c,Peter1992} as it simplifies the semi-analytic method and numerical simulations, but we do not expect this simplification to significantly alter our conclusions. Additionally, it was shown in \cite{Peter1992b} that this limit is a very good approximation for realistic values of the electromagnetic coupling constant. The gauge field associated with $\phi$ plays a crucial role because it is responsible for cancelling the contribution to the energy from the winding number and therefore preventing the associated logarithmic divergence. 

The Lagrangian density for this model is
\begin{equation}
    \mathcal{L} = (\mathcal{D}_\mu\phi)(\mathcal{D}^\mu\phi)^* + \partial_\mu\sigma\partial^\mu\sigma^* - \frac{1}{4}F_{\mu\nu}F^{\mu\nu} - \frac{\lambda_\phi}{4}(|\phi|^2 - \eta_\phi^2)^2 - \frac{\lambda_\sigma}{4}(|\sigma|^2 - \eta_\sigma^2)^2 - \beta|\phi|^2|\sigma|^2 + \frac{\lambda_\sigma}{4}\eta_\sigma^4,
\end{equation}
where as usual, $\mathcal{D}_\mu = \partial_\mu - igA_\mu$, $F_{\mu\nu} = \partial_\mu A_\nu - \partial_\nu A_\mu$ and $g$ is the gauge coupling. The constant term at the end is a convenient addition that sets the energy of the vacuum state to zero but otherwise has no impact on the dynamics. The parameters $\eta_\phi$, $\eta_\sigma$, $\lambda_\phi$, $\lambda_\sigma$, $\beta$ and $g$ are all real positive constants.

The model consists of a Higgs potential for each field with an additional coupling term. The parameters will be chosen such that the $U(1)_\phi$ symmetry is broken in the vacuum and the $U(1)_\sigma$ symmetry is unbroken everywhere, except along the string core. This occurs because the coupling term prevents the $U(1)_\sigma$ symmetry from breaking but this term vanishes along the string core. The broken $U(1)_\phi$ symmetry sets $|\phi| = \eta_\phi$ in the vacuum, but the phase is undetermined. Along any closed path the topology of the vacuum manifold ensures that the phase must change by $2\pi n$ where $n \in \mathbb{Z}$ is the winding number. Any non-zero value of $n$ signals the existence of a string enclosed by the path. Strings with $n>1$ are typically unstable, so we always use $n=1$ in this work. If the string forms a closed loop then, as the $U(1)_\sigma$ symmetry is broken along the string, the condensate will also have an associated winding number, $N \in \mathbb{Z}$.

There is a conserved current associated with the global $U(1)_\sigma$ symmetry,
\begin{equation}
    \mathcal{J}_\mu = \frac{1}{2i}(\sigma^*\partial_\mu\sigma - \sigma\partial_\mu\sigma^*),
\end{equation}
due to Noether's theorem, that satisfies $\partial_\mu \mathcal{J}^\mu = 0$. Typically, this 4-current is split into spatial and temporal components and the word current sometimes refers only to the spatial part. The time component can be used to define the Noether charge,
\begin{equation}
    Q = \int d^3 x \mathcal{J}_0 ,
\end{equation}
which is conserved, $\dot{Q} = 0$. The winding number of the condensate generates the current and the phase frequency generates the charge. Both must be non-zero for the loop to have non-zero angular momentum.

In order to gain some intuition on the physics of superconducting strings, we will briefly discuss some ideas from non-superconducting cosmic strings. The simplest gauge theory that can produce cosmic strings is the Abelian-Higgs model (one complex scalar field with a local $U(1)$ symmetry). The (vacuum) masses of the Higgs and gauge fields are $m_\phi^2 = \frac{1}{2}\lambda_\phi\eta_\phi^2$ and $m_A^2 = g^2\eta_\phi^2$ respectively. The length scale of each field - defined by the width of the region in which the field does not take on its asymptotic value - is related to the mass by $r\approx m^{-1}$. Through a series of rescalings it can be shown that the only significant parameter in the model is the ratio of the two masses. At critical coupling (also known as the BPS limit) the masses are equal such that $g^2_{\text{BPS}} = \frac{1}{2}\lambda_\phi$, which corresponds to the associated length scales being equal, and the vector and scalar forces cancel. Bogomol'nyi was able to reduce the second order field equations to two coupled, first order equations at critical coupling \cite{Bogomol'nyi1976}. Working in this limit led to significant advancements in the study of topological defects, such as a closed form solution to the static field equations of a monopole by Prasad \& Sommerfield \cite{Prasad1975}. We choose to parameterise our gauge coupling with $G = g/g_\text{BPS}$ in order to preserve this intuition. String solutions with $G>1$ are categorised as type I, while those with $G<1$ are called type II. There is a clear distinction between these cases in the Abelian-Higgs model because the force between two parallel strings is attractive in the type I regime and repulsive in the type II regime (there is no interaction at critical coupling). This distinction is blurred for superconducting strings, but we will use the same categorisation nonetheless because it has implications for the length scales of the fields. The (non-vacuum value) gauge fields of type I strings are confined to the string, while they extend outside for type II strings. 

Much of the literature has been focused on the solutions for a straight, superconducting string and infering the existence of static loops by attempting to produce sufficient currents to eliminate the tension or balance it with other effects. Studies that attempt to construct static loop solutions and investigate the dynamics are less common. Global vorton solutions have been constructed in \cite{Lemperiere2003b} by modifying the interaction term to steepen the trapping potential and using the straight string ansatz to predict the vorton radius. Global vortons have also been constructed without this modification in \cite{Radu2008} and \cite{Battye2009a} without the use of the straight string ansatz. The former found solutions in the sigma model limit and then relaxed the conditions to construct vortons that are close to this limit. This produces vortons with a similar width and radius. The latter produced larger, thin vortons (much closer to those relevant to cosmology) via a gradient flow algorithm and tested their stability in full 3D simulations. They found that their solutions were radially stable, but unstable to non-axial perturbations created either by the boundary conditions or the discretisation of the simulation grid. A fully gauged spring solution was explored in \cite{Doudoulakis2007}, but no energy minimising solutions were found. The extension to spinning loops was considered in \cite{Garaud2013} who successfully constructed gauged vortons with small $N$ that are very thick, as in \cite{Radu2008}. Vorton dynamics were also discussed and it was claimed that the smallest vortons with large charge were fully stable, although this analysis was performed in the global limit. We will make comparisons with \cite{Lemperiere2003b,Battye2009a} throughout this paper, but will not directly compare to \cite{Radu2008,Garaud2013} since their vortons, being much smaller, are in a different regime to ours.

There has been more substantial progress in the study of kinky vortons - a $(2+1)$ dimensional analogue of vortons. The lower dimensionality allowed for an exact analytic solution to be found in \cite{Battye2008} and an improved numerical treatment. Kinky vortons can naturally form in a system with random initial conditions \cite{Battye2009c} and the thin string approximation (TSA) has been successfully applied to predict their radii and stability properties \cite{Battye2009b}.

It is our aim to extend previous work by constructing and simulating the dynamics of gauged vortons and investigate the parameter space in which solutions can be formed. We make an ansatz so that the winding number, N, and charge, Q, of the condensate can be fixed and used as additional parameters. We also use techniques from lattice gauge theory to preserve the local $U(1)$ symmetry on the lattice and make sure that the additional unenforced gauge condition is satisfied during dynamical simulations \cite{Moriarty1988}. In particular, we will predict the existence and stability of vortons with a semi-analytic method \cite{Lemperiere2003a,Carter1993} that utilises the solutions for a straight superconducting string and makes a thin string approximation. This approach was successful when applied to kinky vortons, but, prior to this work, has yet to be confirmed for vortons.

 Vortons can have applications in both condensed matter physics and cosmology. For the former, numerical studies have shown that stable Skyrmions can exist in two-component Bose-Einstein condensates \cite{Battye2002} that closely resemble vortons, although there has been no experimental evidence of their existence. For the latter, vortons can potentially be produced over a large range of energy scales, and the precise details can have a significant impact on their cosmological consequences. Stable vortons formed at a high energy phase transition would come to dominate the universe too early, interfering with the successes of standard cosmology, while those formed at lower energies may be beneficial by contributing to dark matter. In \cite{Brandenberger1996} the mechanics of vorton formation is discussed and estimates are given for the vorton abundance. They claimed that models with stable vortons cannot allow superconductivity to become active above $\sim 10^9 \text{GeV}$ or vortons will disrupt nucleosynthesis, and that only a small fraction of initial loops survive to become vortons when superconductivity becomes active at low energies. However, there is some disagreement about the fate of high energy vortons. In \cite{Martins1998a,Martins1998b} the authors examine chiral vortons and suggest that there are, in fact, three regimes - high energy strings which produce no vortons (in contrast with \cite{Brandenberger1996}), intermediate strings which produce vortons, but also start matter domination too early, and low energy strings which could contribute to dark matter - with electroweak scale vortons able to make approximately $6\%$ of the critical density.
 
 Both analyses assume that vortons are absolutely stable objects, but at the time there was no conclusive numerical evidence that stable vortons existed. Previous numerical studies have either found them to be unstable, or they have not sufficiently tested their stability to be sure. In order to rule out GUT theories which produce vortons, they will need to last long enough to be problematic (a few minutes if they are to survive until nucleosynthesis) and they will need to be stable over much longer time scales if they are to be a component of dark matter. An understanding of vorton stability is clearly crucial for considering their role in cosmology. We have observed that there are two main types of instabilities to consider - growing distortions to the shape of the vorton, which we call extrinsic instabilities, and growing oscillations in the width of the string, which we call pinching instabilities. In this paper we will thoroughly test the predictions of the TSA with respect to extrinsic instabilities, but leave an in-depth discussion of the pinching instabilities to a follow-up paper. We recently presented in \cite{PhysRevLett.127.241601} our discovery of a fully stable vorton by making use of the thin string approximation. In this paper we expand upon the theory and methodology behind this, and discuss the comparisons between TSA predictions and our field theory simulations in much greater detail.

\section{Analytic Approaches}

Through a series of rescalings the dimensionality of the parameter space can be reduced. Let $\Tilde{x}=\eta_\phi x$, $\Tilde{\phi} = \eta_\phi^{-1}\phi$, $\Tilde{\sigma} = \eta_\phi^{-1}\sigma$ and $\Tilde{A_\mu} = \eta_\phi^{-1}A_\mu$ such that the Lagrangian can be rewritten as
\begin{equation}
    \mathcal{L} = \eta_\phi^4\bigg[ (\Tilde{\mathcal{D}}_\mu\Tilde{\phi})(\Tilde{\mathcal{D}}^\mu\Tilde{\phi})^* + \Tilde{\partial}_\mu\Tilde{\sigma}\Tilde{\partial}^\mu\Tilde{\sigma}^* - \frac{1}{4}\Tilde{F}_{\mu\nu}\Tilde{F}^{\mu\nu} -\frac{1}{4}\lambda_\phi(|\Tilde{\phi}|^2 - 1)^2 - \frac{1}{4}\lambda_\sigma\bigg(|\Tilde{\sigma}|^2 - \frac{\eta_\sigma^2}{\eta_\phi^2}\bigg)^2 - \beta|\Tilde{\phi}|^2|\Tilde{\sigma}|^2 \bigg],
\end{equation}
where we have removed the constant term since it has no effect on the string dynamics. Making the addition rescaling $\Bar{x} = g\Tilde{x}$,
\begin{equation}
    \mathcal{L} = g^2\eta_\phi^4\bigg[ (\Bar{\mathcal{D}}_\mu\Tilde{\phi})(\Bar{\mathcal{D}}^\mu\Tilde{\phi})^* + \Bar{\partial}_\mu\Tilde{\sigma}\Bar{\partial}^\mu\Tilde{\sigma}^* - \frac{1}{4}\Bar{F}_{\mu\nu}\Bar{F}^{\mu\nu} -\frac{\lambda_\phi}{4g^2}(|\Tilde{\phi}|^2 - 1)^2 - \frac{\lambda_\sigma}{4g^2}\bigg(|\Tilde{\sigma}|^2 - \frac{\eta_\sigma^2}{\eta_\phi^2}\bigg)^2 - \frac{\beta}{g^2}|\Tilde{\phi}|^2|\Tilde{\sigma}|^2 \bigg],
\end{equation}
where we have defined $\Bar{\mathcal{D}}_\mu = \Bar{\partial}_\mu - i\Tilde{A}_\mu$ and $\Bar{F}_{\mu\nu} = \Bar{\partial}_\mu\Tilde{A}_\nu - \Bar{\partial}_\nu\Tilde{A}_\mu$. This reveals that the only significant parameters in the model are
\begin{equation}
    \zeta_\phi:=\frac{\lambda_\phi}{2g^2}, \quad \zeta_\sigma:=\frac{\lambda_\sigma}{2g^2}, \quad \xi:= \frac{\beta}{g^2}, \quad \alpha:=\frac{\eta_\phi}{\eta_\sigma}.
\end{equation}
This is a continuation of the rescaling arguments for the Abelian-Higgs model in which $\zeta_\phi$ is the only significant parameter. A more intuitive, equivalent set of reduced parameters is 
\begin{equation}
    G = \frac{g}{g_\text{BPS}}, \quad \frac{\lambda_\sigma}{\lambda_\phi}, \quad \frac{\beta}{\lambda_\phi}, \quad \frac{\eta_\sigma}{\eta_\phi}.
\end{equation}
where $g_\text{BPS}^2 = \lambda_\phi/2$. It is now clear that we can fix both $\eta_\phi$ and $\lambda_\phi$ with no loss of generality. This removes a length scale from the problem by fixing the mass of the vortex field ($m^2_\phi = \frac{1}{2}\lambda_\phi\eta_\phi^2$). For the rest of this paper we choose to set $\eta_\phi = \lambda_\phi = 1$, but comparisons with any other choice can be easily made by rescaling length scales and field magnitudes in the appropriate way.

Straight string solutions can be used to approximate a loop of string by identifying the ends and neglecting the effects of curvature on the field profiles. We can, therefore, gain some insight into the physics of vortons by examining infinite, straight, superconducting strings since the equations of motion are much easier to solve because they can be reduced to a one dimensional problem. For a cylindrically symmetric, infinite, straight string directed along the z-axis, the vortex field will have the form, $\phi = e^{in\theta}|\phi|$, where $n$ is the winding number of the vortex and the magnitude of the field only depends on the radial coordinate, $\rho$. We make the ansatz $\sigma = e^{i(\omega t + kz)}|\sigma|$ and look for solutions where the only non-zero component of the gauge field is $A_\theta$. The $t$ and $z$ derivatives of the condensate field can be absorbed into the potential, $V \to V - \chi|\sigma|^2 $, where $\chi = \omega^2 - k^2$. Solutions with $\chi<0$ are referred to as magnetic, whereas those with $\chi>0$ are electric and $\chi=0$ are chiral. Now the field equations for a static string reduce to a set of coupled, one dimensional, differential equations

\begin{equation} \label{eq: 1D EoMs start}
    \frac{d^2|\phi|}{d\rho^2} + \frac{1}{\rho}\frac{d|\phi|}{d\rho} - \bigg[\frac{1}{2}\lambda_\phi(|\phi|^2 - \eta_\phi^2) + \beta|\sigma|^2 + \bigg(\frac{n-gA_\theta}{\rho}\bigg)^2\bigg]|\phi| = 0,
\end{equation}

\begin{equation} \label{eq: sigma straight static EoM}
    \frac{d^2|\sigma|}{d\rho^2} + \frac{1}{\rho}\frac{d|\sigma|}{d\rho} - \bigg[\frac{1}{2}\lambda_\sigma(|\sigma|^2 - \eta_\sigma^2) + \beta|\phi|^2 - \chi \bigg]|\sigma| = 0,
\end{equation}

\begin{equation} \label{eq: 1D EoMs end}
    \frac{d^2A_\theta}{d\rho^2} - \frac{1}{\rho}\frac{dA_\theta}{d\rho} + 2g|\phi|^2(n - gA_\theta) = 0.
\end{equation}

\begin{figure}[!t]
    \centering
    \subfloat[$\eta_\sigma=0.35$, $\lambda_\sigma=36$, $\beta=6.6$ and $G=0.2$ (parameter set A) with $\chi=1.074$.\protect\footnotemark This is an example of an electric string.]{
        \centering
        \includegraphics[trim={0.5cm 0 1.5cm 0},clip,width=0.46\linewidth]{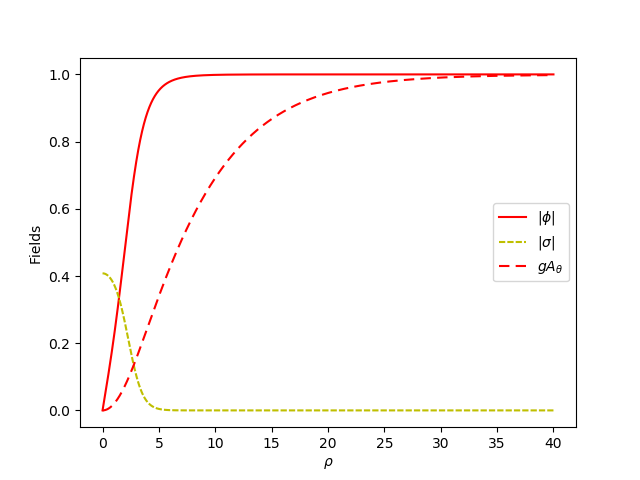}
        \label{fig: LS straight string profile}
    }\hspace{1em}
    \subfloat[$\eta_\sigma=0.61$, $\lambda_\sigma=10$, $\beta=3$ and $G=0.5$ (parameter set B) with $\chi=-0.01$. This string is mildly magnetic, but close to the chiral limit.]{
        \centering
        \includegraphics[trim={0.5cm 0 1.5cm 0},clip,width=0.46\linewidth]{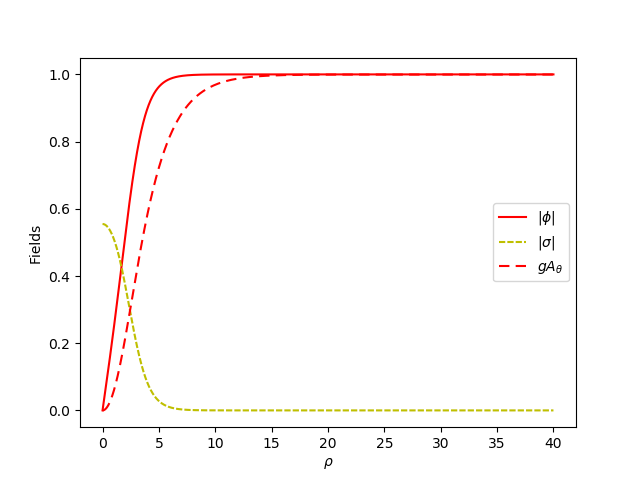}
        \label{fig: CB straight string profile}
    }
    \caption{Straight string profiles for two different parameter sets. Notice that the length scales associated with the vortex and condensate fields are roughly the same while the range of the gauge field is different and can be reduced by increasing $G$.}
    \label{fig:straight string profiles}
\end{figure}

\noindent From these equations it is clear that the field profiles depend only on the combination $\chi$, not the individual values of $\omega$ and $k$. Along the string, the $U(1)_\phi$ symmetry is unbroken so $|\phi(\rho=0)|= A_\theta(\rho=0) = 0$. Far away from the string, the fields take their vacuum values so $|\phi(\rho=\infty)| = \eta_\phi$ and $|\sigma(\rho=\infty)| = 0$. The two remaining boundary conditions are $|\sigma'(\rho=0)| = 0$ and $A_\theta(\rho = \infty) = n/g$ which come from symmetry arguments and the requirement that the total energy is finite, respectively. These equations can be easily solved numerically for a given parameter set and choice of $\chi$. We use a successive over-relaxation (SOR) routine, on a grid with $\Delta\rho = 0.01$ and $0\leq\rho\leq 100$, to compute the radial profile functions for each of the fields. Figure \ref{fig:straight string profiles} shows two examples of superconducting string solutions, for different sets of parameters. Figure \ref{fig: LS straight string profile} shows an electric string while Figure \ref{fig: CB straight string profile} shows a mildly magnetic one.

\footnotetext{This is equivalent (under a length rescaling) to a parameter set used in \cite{Lemperiere2003b} except that we do not modify the potential here ($\beta$ governs the strength of the $\beta|\phi|^2|\sigma|^2$ interaction, rather than the $\beta'|\phi|^6|\sigma|^2$ interaction) and we have set the gauge coupling to a non-zero value.}

However, one should be careful with this method in the electric regime as there is a maximum value of $\chi$ that can be achieved for realistic strings with a fixed charge per unit length, $q$, rather than fixed $\chi$ - which is merely a convenient construction, not a conserved quantity. The maximum $\chi$ is achieved at a finite value of $q$ and then $\chi$ decreases as more charge is added. The method described above will only find one of the two solutions below this limit (the one with lower charge, since it has lower energy) and will produce unrealistic solutions above the limit. To access the full, realistic range of solutions, we can use a Lorentz boost to transform into the frame with $k=0$, $\chi = \omega^2$ (see section \ref{sec: EoS}) and replace $\chi$ in equation (\ref{eq: sigma straight static EoM}) with $(q_p/\Sigma_2)^2$, where the subscript $p$ indicates that it is the charge per unit length in the purely electric frame ($k=0$) which is the smallest possible value that $q$ can take. Other frames are also valid of course, but require $k^2$ to also be specified, which would be an arbitrary choice. Here, we have used the fact that, under our ansatz for $\sigma$,
\begin{equation}
    q = \int\rho\mathcal{J}_0 d\rho d\theta = 2\pi\omega\int\rho|\sigma|^2 d\rho , 
\end{equation}
and we have defined,
\begin{equation}\label{eq: sigma def}
    \Sigma_n = 2\pi\int\rho|\sigma|^n d\rho .
\end{equation}
It is the contribution of $\Sigma_2$ that allows for two solutions at each choice of $\chi$ in the electric regime (which we will often refer to as the higher and lower charge branches) when the increase in $q_p$ is compensated for by the increase in $\Sigma_2$ at $q_p>\Sigma_2(q_p)/\Sigma_2'(q_p)$. Comparing the derivatives of both sides of this inequality with respect to $q_p$ suggests that it is reasonable to assume that this condition will always apply above some critical charge. In the magnetic regime, where $\chi = -k^2$ in the appropriate frame, there is only a single solution.

\subsection{Parameter space} \label{sec: parameter space}

The existence of superconducting strings is not a generic feature of the parameter space. There are several conditions that must be satisfied which set constraints on the range of $\chi$ for which solutions exist. In much of the literature, one of the conditions that is enforced is that the global minimum of the potential must be $|\phi| = \eta_\phi$, $\sigma = 0$ to guarantee the stability of the vacuum. This condition is satisfied if
\begin{equation} \label{eq: minima cond}
  \lambda_\phi \eta_\phi^4 > \lambda_\sigma \eta_\sigma^4.
\end{equation} 
However, if $\eta_\phi>\eta_\sigma$ the Universe may settle into the vacuum state with a broken $U(1)_\phi$ symmetry whether condition (\ref{eq: minima cond}) is satisfied or not, simply because the phase transition occurs at a higher energy scale (earlier time) and the coupling between the fields will then prevent the second phase transition. This will still be stable in the classical theory due to the potential energy barrier between the local and global minima and superconducting strings are still able to form in such a system. Realistic models may require that condition (\ref{eq: minima cond}) is satisfied to ensure the stability of the vacuum against quantum effects, but it is important to make the distinction that it is not a classical requirement, and we will not be considering quantum effects in this paper. It may also be possible to create models in which the decay of the false vacuum by quantum tunneling is sufficiently unlikely that it could last until the present day, or at least long enough to be cosmologically relevant. Parameter set B (shown in Table \ref{tab: parameters}) is the only one that we will be using that does not satisfy this condition.

In the vicinity of a superconducting string, the effective potential is modified by $\chi$ \cite{Lemperiere2003a} such that
\begin{equation}
    \frac{1}{4}\lambda_\sigma(|\sigma|^2 - \eta_\sigma^2)^2 \to \frac{1}{4}\lambda_\sigma\bigg(|\sigma|^2 - \eta_\sigma^2 -\frac{2\chi}{\lambda_\sigma}\bigg)^2,
\end{equation}
where we have left out the constant terms. This changes the position and depth of one of the minima. Condition (\ref{eq: minima cond}) can be extended to incorporate this change,
\begin{equation} \label{eq: minima cond chi}
    \lambda_\phi \eta_\phi^4 > \lambda_\sigma\bigg(\eta_\sigma^2 + \frac{2\chi}{\lambda_\sigma}\bigg)^2.
\end{equation}
It must be remembered that for any given model, the parameters are fixed while, in general, $\chi$ will be different for each string. As such, this condition is not universal and only serves to set limits on $\chi$, while equation (\ref{eq: minima cond}) sets limits on the model parameters. Again, it is easy to find superconducting string solutions that do not satisfy the new condition, but these may be unstable to quantum effects. As we are only focusing on classical physics in this paper we will ignore both of these inequalities, however, whether the vacuum state is the global minima of the potential, or just a local one, will still have important consequences. In fact, most of the vortons that we have constructed do not satisfy condition (\ref{eq: minima cond chi}) as it much easier to find vortons that satisfy the phase condition (discussed at the end of section \ref{sec: semi-analytic}) while remaining numerically feasible in this regime.

Localisation of the condensate to the string and stability of the vacuum requires that the mass term for the condensate is positive far from the string core,
\begin{equation} \label{eq: mass max cond}
    m_\sigma^2(|\phi| = \eta_\phi) = \beta\eta_\phi^2 - \frac{1}{2}\lambda_\sigma\eta_\sigma^2 - \chi > 0,
\end{equation}
while the formation of a condensate at the centre of the string requires the mass there to be negative,
\begin{equation} \label{eq: mass min cond}
    m_\sigma^2(|\phi| = 0) = -\frac{1}{2}\lambda_\sigma\eta_\sigma^2 - \chi < 0.
\end{equation}
This sets another upper and lower bound on the allowed values of $\chi$. In practice, the dependence of solutions on $\chi$ has a much richer structure than these simple bounds, and we will denote the more complicated upper and lower limits by $\chi_\text{max}$ and $\chi_\text{min}$ respectively. It is well known that condition (\ref{eq: mass min cond}) is insufficient because the gradient energy cost of condensate formation must also be considered. In order to determine when a condensate will form we can consider small fluctuations of the form $\delta\sigma e^{i\nu t}$ around a non-superconducting string solution - one with $\sigma = 0$ everywhere. This results in a Schr\"{o}dinger-like equation,
\begin{equation} \label{eq: schrodinger}
    -\frac{d^2\delta\sigma}{d\rho^2} - \frac{1}{\rho}\frac{d\delta\sigma}{d\rho} + \beta|\phi|^2\delta\sigma = \Big(\nu^2 + \frac{1}{2}\lambda_\sigma\eta_\sigma^2 + \chi\Big)\delta\sigma,
\end{equation}
and the perturbation will be unstable if $\nu^2<0$ and a condensate will form. If the dimensionless strength of the potential is large, defined as the depth times the square of the width ($\sim \beta/\lambda_\phi$), then $\chi_\text{min}$ can be predicted by comparison with a harmonic oscillator \cite{Haws1988,Lemperiere2003a}. More generally, we can solve the equations of motion in the absence of the condensate field to find $\phi(\rho)$ and use this to numerically calculate the smallest eigenvalue, $\gamma$, of the left hand side of the Schr\"{o}dinger-like equation. Since the term in the brackets must be equal to one of the eigenvalues, $\chi_\text{min} = \gamma - \frac{1}{2}\lambda_\sigma\eta_\sigma^2$ ($\nu^2 = 0$) is the critical value.

The maximum value of $\chi$ is more complicated. For $\beta<\frac{1}{2}\sqrt{\lambda_\phi\lambda_\sigma}$, the condensate becomes delocalised from the string and $U(1)_\sigma$ is broken in the vacuum above $\chi_\text{max} = \beta\eta_\phi^2 - \frac{1}{2}\lambda_\sigma\eta_\sigma^2$. However, for $\beta>\frac{1}{2}\sqrt{\lambda_\phi\lambda_\sigma}$ the vacuum state will become a local, rather than global minima before reaching this limit (this inequality has previously appeared in other works on topological defects, such as in \cite{Battye2002,Battye2010} where it was called the phase separation condition). This occurs at $\chi_\text{eq}^{+} = \frac{1}{2}\eta_\phi^2\sqrt{\lambda_\phi\lambda_\sigma} - \frac{1}{2}\lambda_\sigma\eta_\sigma^2$. Although superconducting string solutions do exist above this critical value, there is another $\chi_\text{max}$ above which they do not, and attempting to numerically find solutions results in the flipping of the vacuum so that $U(1)_\sigma$ is broken and $U(1)_\phi$ is restored. We do not currently have a satisfactory way to predict this additional limit and it is unclear under what circumstances it will be the lower of the two limits - it may always be the lower limit when $\beta>\frac{1}{2}\sqrt{\lambda_\phi\lambda_\sigma}$. We suspect that it is caused by the vacuum state having sufficient energy to overcome an energy barrier (perhaps set by the saddle point of the potential) and relax to the true vacuum.

\begin{figure}[!t]
    \centering
    \subfloat[$\eta_\sigma=0.15$, $\lambda_\sigma=500$, $\beta=10$ and $G=0.5$ (parameter set H) with $q_p=10$. The condensate extends (albeit with only a small magnitude) far from the core of the string.]{
        \centering
        \includegraphics[trim={0.5cm 0 1.5cm 0},clip,width=0.46\linewidth]{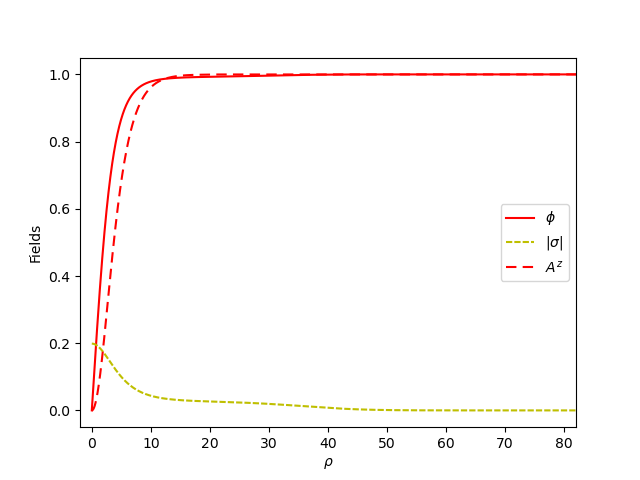}
        \label{fig: pH higher charge branch profile}
    }\hspace{1em}
    \subfloat[$\eta_\sigma=1$, $\lambda_\sigma=2/3$, $\beta=2/3$ and $G=0.1$ (parameter set E) with $q_p=400$. The condensate remains localised to the core of the string but this has widened significantly.\protect\footnotemark]{
        \centering
        \includegraphics[trim={0.5cm 0 1.5cm 0},clip,width=0.46\linewidth]{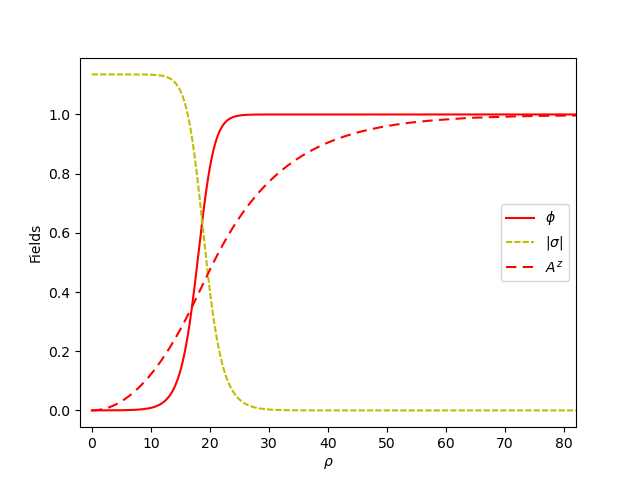}
        \label{fig: pE higher charge branch profile}
    }
    \caption{Straight string profiles on the higher charge branch. These plots make the qualitative differences clear between parameter sets with $\beta<\frac{1}{2}\sqrt{\lambda_\phi\lambda_\sigma}$ (left) and those with $\beta>\frac{1}{2}\sqrt{\lambda_\phi\lambda_\sigma}$ (right).}
    \label{fig: higher charge branch profiles}
\end{figure}

\footnotetext{These field profiles are very similar to those of the large charge Q-Monopole-Ball recently presented in \cite{Bai2021}, except for a string rather than a monopole.}

However, as previously mentioned, $\chi>\chi_\text{max}$ is not realistic because $\chi$ decreases as more charge is added to the string, beyond this limit. The phase separation condition still separates qualitatively different behaviours of strings with large charge per unit length, which we demonstrate in Figure \ref{fig: higher charge branch profiles}. For $\beta<\frac{1}{2}\sqrt{\lambda_\phi\lambda_\sigma}$, $U(1)_\sigma$ is not broken everywhere at large $q$, (as this would require $q\to\infty$) but extends further and further from the core of the string as more charge is added. The size of the condensate at the string core, and the width of the string, are only marginally affected. If $\beta>\frac{1}{2}\sqrt{\lambda_\phi\lambda_\sigma}$, the condensate always remains localised to the string and, at large $q$, the vacuum only flips in the vicinity of the string (not everywhere because, again, this would require infinite charge). In essence, this widens the core of the string, with the width increasing with charge. Additionally, the effective mass of the condensate in the vacuum always remains larger than zero, suppressing tunneling processes that would lead to the emission of charge from the string. Instead, we expect that the maximum charge that can be supported by the string will be limited by the onset of longitudinal instabilities, that we will later suggest should always occur in the regime where $\chi'(q_p)<0$, under reasonable assumptions. As a result, we often focus purely on the lower charge solution at a given $\chi$, and ignore the higher charge solution since we expect that it will not be able to produce a stable vorton.

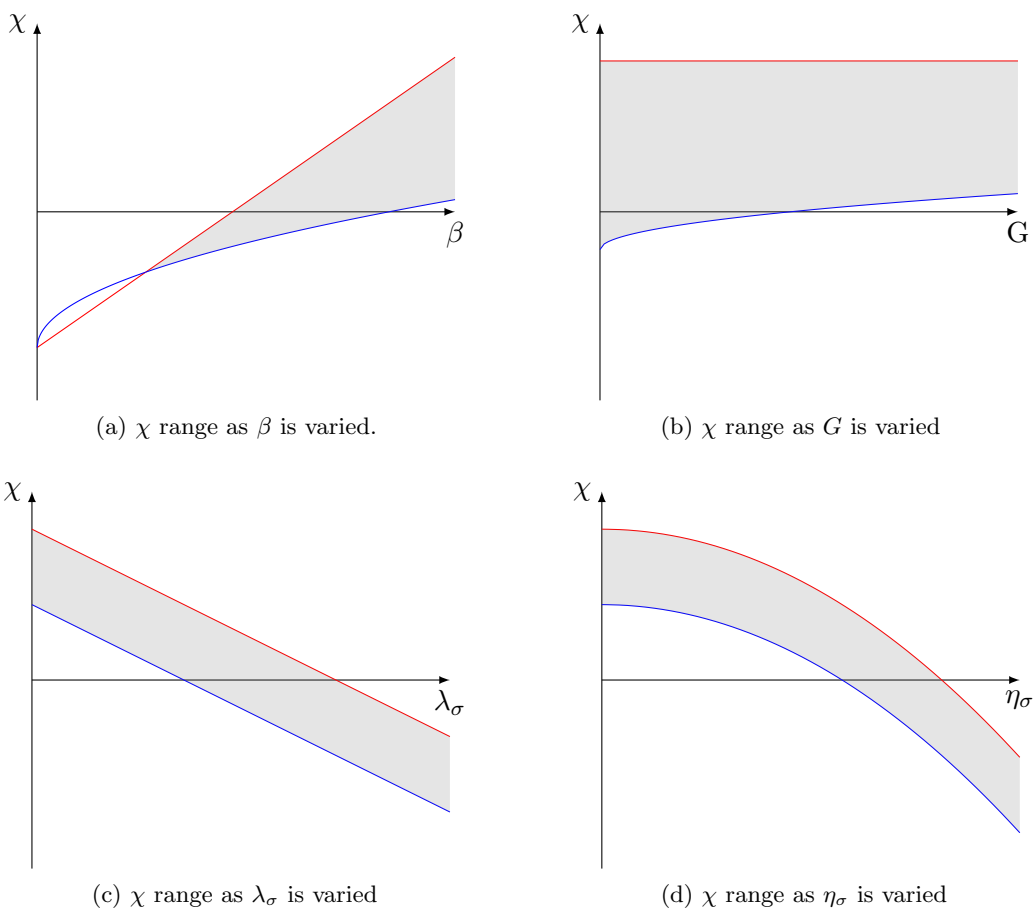
\begin{figure}[t]
    \centering
    \subfloat[$\chi$ range as $\beta$ is varied.]{
        \centering
        \begin{tikzpicture}
            % Set up axis
            \draw [->] (0,-2.5) -- (0,2.5);
            \draw [->] (0,0) -- (5.5,0);
            \node [left] at (0,2.5) {$\chi$};
            \node [below] at (5.5,0) {$\beta$};
            
            % Plot on axis
            \draw[domain=0:1.428, samples=100, variable=\x, red] plot ({\x}, {0.7*\x - 1.8});
            \draw[domain=0:1.428, samples=100, variable=\x, blue] plot ({\x}, {sqrt(0.7*\x) - 1.8});
            \draw[domain=1.428:5.5, samples=100, variable=\x, red, name path=A] plot ({\x}, {0.7*\x - 1.8});
            \draw[domain=1.428:5.5, samples=100, variable=\x, blue, name path=B] plot ({\x}, {sqrt(0.7*\x) - 1.8});
            \tikzfillbetween[of=A and B]{gray, opacity=0.2};
            
        \end{tikzpicture}
    } \hspace{2em}
    \subfloat[$\chi$ range as $G$ is varied]{
        \centering
        \begin{tikzpicture}
            \draw [->] (0,-2.5) -- (0,2.5);
            \draw [->] (0,0) -- (5.5,0);
            \node [left] at (0,2.5) {$\chi$};
            \node [below] at (5.5,0) {G};
            
            \draw[domain=0:5.5, smooth, variable=\x, red, name path=A] plot({\x}, {2});
            \draw[domain=0:5.5, samples=100, variable=\x, blue, name path=B] plot({\x}, {sqrt(0.1*\x)-0.5});
            \tikzfillbetween[of=A and B]{gray, opacity=0.2};
        \end{tikzpicture}
    }
    \\ 
    \subfloat[$\chi$ range as $\lambda_\sigma$ is varied]{
        \centering
        \begin{tikzpicture}
            \draw [->] (0,-2.5) -- (0,2.5);
            \draw [->] (0,0) -- (5.5,0);
            \node [left] at (0,2.5) {$\chi$};
            \node [below] at (5.5,0) {$\lambda_\sigma$};
            
            \draw[domain=0:5.5, smooth, variable=\x, red, name path=A] plot({\x}, {2 - 0.5*\x});
            \draw[domain=0:5.5, smooth, variable=\x, blue, name path=B] plot({\x}, {1 - 0.5*\x});
            \tikzfillbetween[of=A and B]{gray, opacity=0.2};
        \end{tikzpicture}    
    } \hspace{2em}
    \subfloat[$\chi$ range as $\eta_\sigma$ is varied]{
        \centering
        \begin{tikzpicture}
            \draw [->] (0,-2.5) -- (0,2.5);
            \draw [->] (0,0) -- (5.5,0);
            \node [left] at (0,2.5) {$\chi$};
            \node [below] at (5.5,0) {$\eta_\sigma$};
            
            \draw[domain=0:5.5, smooth, variable=\x, red, name path=A] plot({\x}, {2 - 0.1*\x*\x});
            \draw[domain=0:5.5, smooth, variable=\x, blue, name path=B] plot({\x}, {1 - 0.1*\x*\x});
            \tikzfillbetween[of=A and B]{gray, opacity=0.2};
        \end{tikzpicture}    
    }
    \caption{Plots showing (qualitatively) how the range of $\chi$ for which superconducting strings can form changes with each parameter while the others are kept fixed. The red and blue lines respectively show the dependency of $\chi_\text{max}$ and $\chi_\text{min}$ on each parameter and the shaded area shows the region of the parameter space in which superconducting strings can form. Note that we have not attempted to include the behaviour of $\chi_\text{max}$ when the vacuum is a local minima of the potential as we do not know the exact dependence of the new upper limit on each parameter. The upper plots demonstrate that the width of the $\chi$ range changes with $\beta$ and (to a lesser extent) $G$, while it is clear from the lower two plots that $\lambda_\sigma$ and $\eta_\sigma$ only shift the range - linearly for the former and quadratically for the latter.}
    \label{fig: chi range}
\end{figure}

In Figure \ref{fig: chi range} we demonstrate how the limits on $\chi$ vary with each parameter (and all others kept fixed), ignoring the new upper limit when $\beta>\frac{1}{2}\sqrt{\lambda_\phi\lambda_\sigma}$ as we have been unable to predict how this behaves. Of particular interest is the effect of $\eta_\sigma$, which is useful for modifying the stability properties of vortons. Both $\lambda_\sigma$ and $\eta_\sigma$ have the effect of shifting the accessible range of $\chi$, but $\eta_\sigma$ has the additional benefit of not changing the shape of the integrated quantities (as a function of $\chi$) that are introduced in section \ref{sec: semi-analytic} for use in the semi-analytic method. This is because only the combination $\chi + \frac{1}{2}\lambda_\sigma\eta_\sigma^2$ enters the equations of motion which can be kept constant by adjusting both $\chi$ and $\eta_\sigma$ simultaneously (for fixed $\lambda_\sigma$). Since the sound speeds, which are introduced in section \ref{sec: EoS}, are the only relevant quantity that depend on $\chi$ separately from $\eta_\sigma$, this provides a useful technique for scanning the parameter space for potentially stable vortons.

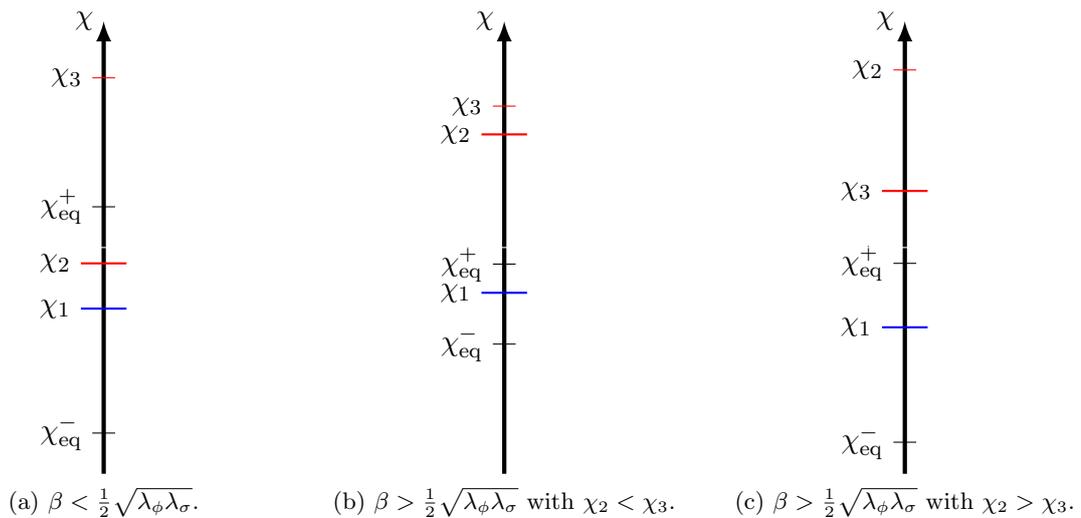
\begin{figure}[t]
    \centering
    \subfloat[$\beta < \frac{1}{2}\sqrt{\lambda_\phi\lambda_\sigma}$.]{
        \centering
        \begin{tikzpicture}
            \tikzmath{\lam = 16; \b = 1; \e = 0.4; \a = 0.2; \ws = 0.15; \wb = 0.3; \ls = 3; \c1 = \ls*(\a - 0.5*\lam*\e*\e)/4; \c2 = \ls*(\b - 0.5*\lam*\e*\e)/4; \c3 = \ls*3/4; \cmeq = \ls*(-0.5*sqrt(\lam) - 0.5*\lam*\e*\e)/4; \cpeq = \ls*(0.5*sqrt(\lam) - 0.5*\lam*\e*\e)/4;}
        
            \draw [->, ultra thick] (0,-\ls) -- (0,\ls);
            \draw [blue, thick] (-\wb,\c1) -- (\wb,\c1);
            \draw [red, thick] (-\wb,\c2) -- (\wb,\c2);
            \draw [red] (-\ws,\c3) -- (\ws,\c3);
            \draw (-\ws,\cmeq) -- (\ws,\cmeq);
            \draw (-\ws,\cpeq) -- (\ws,\cpeq);
            
            \node [left] at (0,\ls) {$\chi$};
            \node [left] at (-\wb,\c1) {$\chi_1$};
            \node [left] at (-\wb,\c2) {$\chi_2$};
            \node [left] at (-\ws,\c3) {$\chi_3$};
            \node [left] at (-\ws,\cmeq) {$\chi_\text{eq}^{-}$};
            \node [left] at (-\ws,\cpeq) {$\chi_\text{eq}^{+}$};
            
            \draw [white] (-2.5,0) -- (2.5,0);
        \end{tikzpicture}
    }
    \subfloat[$\beta > \frac{1}{2}\sqrt{\lambda_\phi\lambda_\sigma}$ with $\chi_2 < \chi_3$.]{
        \centering
        \begin{tikzpicture}
            \tikzmath{\lam = 2; \b = 3; \e = 1; \a = 0.2; \ws = 0.15; \wb = 0.3; \ls = 3; \c1 = \ls*(\a - 0.5*\lam*\e*\e)/4; \c2 = \ls*(\b - 0.5*\lam*\e*\e)/4; \c3 = \ls*2.5/4; \cmeq = \ls*(-0.5*sqrt(\lam) - 0.5*\lam*\e*\e)/4; \cpeq = \ls*(0.5*sqrt(\lam) - 0.5*\lam*\e*\e)/4;}
        
            \draw [->, ultra thick] (0,-\ls) -- (0,\ls);
            \draw [blue, thick] (-\wb,\c1) -- (\wb,\c1);
            \draw [red, thick] (-\wb,\c2) -- (\wb,\c2);
            \draw [red] (-\ws,\c3) -- (\ws,\c3);
            \draw (-\ws,\cmeq) -- (\ws,\cmeq);
            \draw (-\ws,\cpeq) -- (\ws,\cpeq);
            
            \node [left] at (0,\ls) {$\chi$};
            \node [left] at (-\wb,\c1) {$\chi_1$};
            \node [left] at (-\wb,\c2) {$\chi_2$};
            \node [left] at (-\ws,\c3) {$\chi_3$};
            \node [left] at (-\ws,\cmeq) {$\chi_\text{eq}^{-}$};
            \node [left] at (-\ws,\cpeq) {$\chi_\text{eq}^{+}$};
            
            \draw [white] (-2.5,0) -- (2.5,0);
        \end{tikzpicture}
    }
    \subfloat[$\beta > \frac{1}{2}\sqrt{\lambda_\phi\lambda_\sigma}$ with $\chi_2 > \chi_3$.]{
        \centering
        \begin{tikzpicture}
            \tikzmath{\lam = 10; \b = 5; \e = 0.61; \a = 0.45; \ws = 0.15; \wb = 0.3; \ls = 3; \c1 = \ls*(\a - 0.5*\lam*\e*\e)/4; \c2 = \ls*(\b - 0.5*\lam*\e*\e)/4; \c3 = \ls*1/4; \cmeq = \ls*(-0.5*sqrt(\lam) - 0.5*\lam*\e*\e)/4; \cpeq = \ls*(0.5*sqrt(\lam) - 0.5*\lam*\e*\e)/4;}
        
            \draw [->, ultra thick] (0,-\ls) -- (0,\ls);
            \draw [blue, thick] (-\wb,\c1) -- (\wb,\c1);
            \draw [red] (-\ws,\c2) -- (\ws,\c2);
            \draw [red, thick] (-\wb,\c3) -- (\wb,\c3);
            \draw (-\ws,\cmeq) -- (\ws,\cmeq);
            \draw (-\ws,\cpeq) -- (\ws,\cpeq);
            
            \node [left] at (0,\ls) {$\chi$};
            \node [left] at (-\wb,\c1) {$\chi_1$};
            \node [left] at (-\ws,\c2) {$\chi_2$};
            \node [left] at (-\wb,\c3) {$\chi_3$};
            \node [left] at (-\ws,\cmeq) {$\chi_\text{eq}^{-}$};
            \node [left] at (-\ws,\cpeq) {$\chi_\text{eq}^{+}$};
            
            \draw [white] (-2.5,0) -- (2.5,0);
        \end{tikzpicture}
    }
    \caption{The constraints on $\chi$ for three different illustrative cases. Here, we have defined $\chi_1 = \gamma - \frac{1}{2}\lambda_\sigma\eta_\sigma^2$, $\chi_2 = \beta\eta_\phi^2 - \frac{1}{2}\lambda_\sigma\eta_\sigma^2$ and $\chi_3$ is the additional constraint that is not yet understood. Lower limits are coloured blue, upper limits are coloured red and the range of $\chi$ for which superconducting strings exist is marked with wider and thicker lines. Additionally, we have marked the points $\chi_\text{eq}^{\pm} = \pm\frac{1}{2}\eta_\phi^2\sqrt{\lambda_\phi\lambda_\sigma} - \frac{1}{2}\lambda_\sigma\eta_\sigma^2$ inside which the vacuum lies at the global minima of the effective potential. For $\beta < \frac{1}{2}\sqrt{\lambda_\phi\lambda_\sigma}$ the vacuum is always at the global minima within the $\chi$ range. For $\beta > \frac{1}{2}\sqrt{\lambda_\phi\lambda_\sigma}$, strings in part of the $\chi$ range will have a vacuum that is only a local minima of the effective potential. The $\chi$ range may either remain bounded from above by $\chi_2$ or by a new constraint, $\chi_3$. The former corresponds to requiring that $U(1)_\sigma$ remains unbroken in the vacuum so that the condensate remains localised to the string and we believe the latter constraint prevents the local minima vacuum state from climbing over the potential barrier and reaching the global minima vacuum state (which has $U(1)_\phi$ unbroken and $U(1)_\sigma$ broken).}
\end{figure}

\begin{table}[t]
    \centering
    \begin{tabular}{l c c c c c c c c c c c}
        \hline
        Set & $\lambda_\phi$ & $\lambda_\sigma$ & $\eta_\phi$ & $\eta_\sigma$ & $\beta$ & $\beta'$ & $G$ & $\chi_\text{min}$ & $\chi_\text{max}$ & $\chi_\text{eq}^-$ & $\chi_\text{eq}^+$ \\
        \hline
        A & $1$ & $36$ & $1$ & $0.35$ & $6.6$ & $0$ & $0.2$ & $-0.182$ & $1.24$ & $-5.21$ & $0.795$ \\
        B & $1$ & $10$ & $1$ & $0.61$ & $3$ & $0$ & $0.5$ & $-0.423$ & $0.01$ & $-3.44$ & $-0.279$ \\
        C & $1$ & $36$ & $1$ & $0.35$ & $6.6$ & $0$ & $1$ & $0.421$ & $1.54$ & $-5.21$ & $0.795$ \\
        D & $1$ & $\frac{2}{3}$ & $1$ & $1$ & $\frac{2}{3}$ & $0$ & $0$ & $0.124$ & $0.134$ & $-0.742$ & $0.075$ \\
        E & $1$ & $\frac{2}{3}$ & $1$ & $1$ & $\frac{2}{3}$ & $0$ & $0.1$ & $0.137$ & $0.144$ & $-0.742$ & $0.075$ \\
        F & $1$ & $36$ & $1$ & $0.35$ & $0$ & $6.6$ & $0.2$ & $-1.37$ & $0.12$ & $-5.21$ & $0.795$ \\
        G & $1$ & $900$ & $1$ & $0.1825$ & $20$ & $0$ & $0.2$ & $-11.3$ & $0.79$ & $-30$ & $0.012$ \\
        H & $1$ & $500$ & $1$ & $0.15$ & $10$ & $0$ & $0.5$ & $-2.74$ & $4.38$ & $-16.8$ & $5.56$ \\
        \hline
    \end{tabular}
    \caption{The sets of parameters used in this paper. The range of $\chi$ for which localised superconducting strings exist and the range for which the vacuum state is the global minima of the effective potential are also listed. Set D has been studied in \cite{Battye2009a} (and set E is the gauged extension of this), while set F is the gauged extension of a parameter set that was studied in \cite{Lemperiere2003b} (with A and C being related parameter sets).}
    \label{tab: parameters}
\end{table}

In Table \ref{tab: parameters} we list the parameter sets that we will be using throughout this paper. We also give the $\chi_\text{eq}$ points and the approximate values of $\chi_\text{min}$ (determined by solving the eigenvalue equation of (\ref{eq: schrodinger}) numerically) and $\chi_\text{max}$ (often determined by trial and error as most of the the parameter sets we use are limited by the constraint described above, that is not fully understood). The strength of the modified interaction term $\beta'|\phi|^6|\sigma|^2$ is also included here.
Sets A, C and F are similar to one of the parameter sets used in \cite{Lemperiere2003b} - the differences are that sets A and C do not have the modified interaction term and all sets have been rescaled and given a non-zero gauge coupling. Set B is an interesting choice as it does not respect condition (\ref{eq: minima cond}). Consequently, there will be a lower energy ground state even in a Universe with no superconducting strings (or far from a string so that the effect of $\chi$ is cut off). Since we are neglecting quantum effects we will not consider this a problem as it is a convenient set of parameters for studying the classical physics. Set D is equivalent to the set used in \cite{Battye2009a} after rescaling, and set E is just the gauged extension of this. Set G was chosen so that chiral vortons with a global minima vacuum state are possible while satisfying the phase frequency condition (discussed at the end of section \ref{sec: semi-analytic}). We will not discuss this parameter set much because it is inconvenient to run dynamical simulations, due to the large winding numbers required, but we mention it as a proof of principle that potentially stable vortons are possible to construct in parameter sets that are more realistic than set B. Finally, set H is the only parameter set that we have used which has $\beta<\frac{1}{2}\sqrt{\lambda_\phi\lambda_\sigma}$.

\subsection{Semi-analytic method} \label{sec: semi-analytic}

By assuming that the fields of a vorton solution are well approximated by a piece of straight, superconducting string that is wrapped into a loop, we can gain valuable insights into vorton dynamics. Using the energy of the string, the radii of vortons can be predicted \cite{Lemperiere2003a} and from the equation of state we can predict the intervals of stability to perturbations of different Fourier modes \cite{Carter1993}.

From the Lagrangian, the energy density of a static string is easily calculated. By inserting our ansatz for the fields, the total energy can be expressed as 
\begin{equation}
    E = \mu L + 2\pi L\int \rho d\rho\bigg\{ \bigg|\frac{d\sigma}{d\rho}\bigg|^2 + (\omega^2 + k^2)|\sigma|^2 + \frac{1}{4}\lambda_\sigma(|\sigma|^2 - \eta_\sigma^2)^2 + \beta|\phi|^2|\sigma|^2 - \frac{1}{4}\lambda_\sigma\eta_\sigma^4 \bigg\},
\end{equation}
where $L$ is the length of string and $\mu$ is the mass per unit length of the string which is defined by
\begin{equation}
    \mu = 2\pi\int \rho d\rho\bigg\{ \bigg|\frac{\partial\phi}{\partial\rho}\bigg|^2 + \bigg(\frac{n-gA_\theta}{\rho}\bigg)^2|\phi|^2 + \frac{1}{2\rho^2}\bigg(\frac{dA_\theta}{d\rho}\bigg)^2 + \frac{1}{4}\lambda_\phi(|\phi|^2 - \eta_\phi^2)^2 \bigg\}.
\end{equation}
The energy can be greatly simplified by using the static equation of motion (\ref{eq: sigma straight static EoM}). Multiplying this equation by $|\sigma|$, integrating over the entire volume and simplifying the derivatives with integration by parts gives,
\begin{equation} \label{eq: integrated sigma EoM}
      2\pi L\int \rho d\rho\bigg\{ \bigg|\frac{d\sigma}{d\rho}\bigg|^2 + \bigg[\frac{1}{2}\lambda_\sigma(|\sigma|^2 - \eta_\sigma^2) + \beta|\phi|^2 - \chi\bigg]|\sigma|^2\bigg\} = 0.
\end{equation}
Now this can be substituted into the energy so that
\begin{equation} \label{eq: energy EoM subbed}
    E = \mu L + 2\pi L \int\rho d\rho\bigg\{2\omega^2|\sigma|^2 - \frac{1}{4}\lambda_\sigma|\sigma|^4\bigg\}.
\end{equation}

At this point, we have managed to split the energy per unit length into the contribution due to $\phi$ and its associated gauge field - which is all contained within $\mu$ - and the contribution made by the condensate field - which is the rest of the expression. The energy can be written in a more convenient form by recognising that, under our ansatz, the Noether charge can be written as $Q =\omega\Sigma_2 L$ and therefore,
\begin{equation} \label{eq: semi-analytic energy}
    E = \bigg(\mu-\frac{1}{4}\lambda_\sigma\Sigma_4\bigg)L + \frac{2Q^2}{\Sigma_2L}.
\end{equation}
For a given parameter set, $\mu$ and $\Sigma_n$ are functions of $\chi$ only. The wavenumber $k$ is related to the winding number of a vorton by $kL = 2\pi N$, so that
\begin{equation} \label{eq: chi}
    \chi = \bigg(\frac{Q}{\Sigma_2 L}\bigg)^2 - \bigg(\frac{2\pi N}{L}\bigg)^2.
\end{equation}
Now a radial profile solution can be calculated and $\mu$, $\Sigma_2$ and $\Sigma_4$ can be computed for each value of $\chi$. It is easy to show that the minima of the energy only depends on the ratio $\mathcal{R}=\frac{N}{Q}$ and $\chi$ by dividing equation (\ref{eq: semi-analytic energy}) by Q
\begin{equation} \label{eq: energy div Q}
    \frac{E}{Q} = \bigg(\mu - \frac{1}{4}\lambda_\sigma\Sigma_4\bigg)\frac{L}{Q} + \frac{2}{\Sigma_2\frac{L}{Q}},
\end{equation}
and rearranging equation (\ref{eq: chi}) we find
\begin{equation}
    \frac{L}{Q} = \sqrt{\frac{ \Sigma_2^{-2} - (2\pi\mathcal{R})^2 }{\chi}}.
\end{equation}
Note that magnetic, chiral and electric vortons must have $\Sigma_2^{-1}<2\pi\mathcal{R}$, $\Sigma_2^{-1} = 2\pi\mathcal{R}$ and $\Sigma_2^{-1}>2\pi\mathcal{R}$ respectively. Since $\mathcal{R}$ is a conserved quantity, vortons form when the partial derivative of equation (\ref{eq: energy div Q}) with respect to $\chi$ is zero. There are two approaches for finding the minima of this function. The first is to specify $\mathcal{R}$ and then use an algorithm to iteratively approach the minima by changing $\chi$ and calculating the energy. Once the minima is found, the radii of vortons can be predicted and this scales linearly with $N$ (and $Q$) if the ratio is kept constant. The energy minimisation algorithm will need to be performed again if a different value of $\mathcal{R}$ is specified.

The second approach uses an analytic formula and approaches the problem slightly differently. Instead of specifying $\mathcal{R}$ and finding the energy minimising value of $\chi$, this method finds the value of $\mathcal{R}$ that will make a given value of $\chi$ be the energy minimising solution. As pointed out in \cite{Peter1992} the Lagrangian can be written as
\begin{equation} \label{eq: simplified Lagrangian}
    2\pi \int \rho\mathcal{L} \, d\rho = -\bigg(\mu - \frac{1}{4}\lambda_\sigma \Sigma_4\bigg),
\end{equation}
which can be derived using the same technique (substituting the static equations of motion) that was used to simplify the energy in equation (\ref{eq: energy EoM subbed}). The only explicit $\chi$ dependence in the Lagrangian density comes from the $t$ and $z$ derivatives that were absorbed into the potential. Therefore, $\frac{\partial\mathcal{L}}{\partial\chi} = |\sigma|^2$ and we can differentiate with respect to $\chi$ on both sides of equation (\ref{eq: simplified Lagrangian}) to get
\begin{equation} \label{eq: mu and sigma4 deriv}
    -\Sigma_2 = \mu' - \frac{1}{4}\lambda_\sigma\Sigma_4',
\end{equation}
where $'$ denotes a derivative with respect to $\chi$, unless otherwise implied. This is a very useful piece of information because it means that $\mu'$ and $\Sigma_4'$ are not required to calculate the derivative of $\frac{E}{Q}$ with respect to $\chi$. Setting this derivative to zero yields a quadratic equation for $(\frac{L}{Q}\big)^2$,
\begin{equation} \label{eq: L over Q quadratic}
    \bigg(\frac{L}{Q}\bigg)^4 + c_1\bigg(\frac{L}{Q}\bigg)^2 + c_2 = 0,
\end{equation}
where
\begin{equation}
    c_1 = \frac{2}{\Sigma_2}\bigg[\frac{\Sigma_2'}{\Sigma_2^2} - \frac{1}{2\chi\Sigma_2 + \mu - \frac{1}{4}\lambda_\sigma\Sigma_4}\bigg],
\end{equation}
and
\begin{equation}
    c_2 = -\frac{4\Sigma_2'}{\Sigma_2^4(2\chi\Sigma_2 + \mu - \frac{1}{4}\lambda_\sigma\Sigma_4)}.
\end{equation}
The two solutions to this quadratic are
\begin{equation} \label{eq: quadratic sols}
    \bigg(\frac{L}{Q}\bigg)^2 = \frac{2}{\Sigma_2(2\chi\Sigma_2 + \mu - \frac{1}{4}\lambda_\sigma\Sigma_4)} \quad \text{and} \quad \bigg(\frac{L}{Q}\bigg)^2 = -2\frac{\Sigma_2'}{\Sigma_2^3},
\end{equation}
which picks a single frame in which the string can be wrapped into an energy minimising vorton state by specifying its charge per unit length. The second solution initially appears as though it may be a physical solution along the higher charge branch, where $\Sigma_2'<0$. However, rewriting $\Sigma_2'(\chi)$ in terms of $\Sigma_2'(q_p)$ reveals that this solution requires a frame in which $q<q_p$, which is not possible. We will show in the next section that it is reasonable to assume that the other solution is positive because the quantity in the brackets is the energy per unit length in the electric regime, and the tension in the magnetic regime. Using this result we can see that wrapping each string solution into a loop will produce a vorton as long as the ratio between the winding number and charge satisfies
\begin{equation}
    \mathcal{R} = \frac{1}{2\pi\Sigma_2}\sqrt{1 - \frac{2\chi\Sigma_2}{2\chi\Sigma_2 + \mu - \frac{1}{4}\lambda_\sigma\Sigma_4}}.
\end{equation}
There are a few caveats to this method. It assumes that the vorton core is small compared to the radius and curvature effects on the field profiles and interactions between opposite ends of the loops can be neglected. For a straight string, $\chi$ is constant across all space, but for a vorton $\chi(\rho) = \omega^2 - (N/\rho)^2$, therefore the string must be thin enough that $\chi$ does not significantly vary across its cross section. Perhaps most importantly, this also changes the condition for the vacuum state to be stable. As $\rho \to \infty$, the $\sigma$ mass term becomes
\begin{equation} \label{eq: vorton formation constraint}
    m_\sigma^2(\rho=\infty) = \beta\eta_\phi^2 - \frac{1}{2}\lambda_\sigma\eta_\sigma^2 - \omega^2,
\end{equation}
which must be greater than zero. This condition is more strict than equation (\ref{eq: mass max cond}) because $\omega^2\geq\chi$. If the energy minimising stright string solution predicts a value for $\omega$ that is greater than this limit, then we should not expect to find a vorton solution at the corresponding predicted radii. It is possible to create artificial solutions that violate this condition if the boundaries are placed too close to the vorton, but a larger simulation will show that these are unphysical. Using equations (\ref{eq: quadratic sols}) and $Q = \omega\Sigma_2 L$ we can show that the vorton will have
\begin{equation} \label{eq: phase freq cond}
    \omega^2 = \chi + \frac{\mu - \frac{1}{4}\lambda_\sigma\Sigma_4}{2\Sigma_2},
\end{equation}
and then from the definition of $\chi$ is it clear that $k^2$ is equal to the second term.

\subsection{Stability to extrinsic oscillations} \label{sec: EoS}

Predicting the stability of vortons to different vibrational modes can be achieved by determining the equation of state of the superconducting string solutions. This stability analysis gave accurate results for kinky vortons and we will closely follow the approach taken in \cite{Battye2009b}, albeit without the benefits of an analytic solution. The energy momentum tensor is given by
\begin{equation}
    \mathcal{T}^\mu_\nu = 2g^{\mu\alpha}\frac{\partial\mathcal{L}}{\partial g^{\alpha\nu}} - \delta^\mu_\nu \mathcal{L},
\end{equation}
and substituting in the Lagrangian gives
\begin{equation}
    \mathcal{T}^{\mu\nu} = 2(\mathcal{D}^\mu\phi)(\mathcal{D}^\nu\phi)^* + 2\partial^\mu\sigma\partial^\nu\sigma^* - F^\mu_\alpha F^{\nu\alpha} - g^{\mu\nu}\mathcal{L}.
\end{equation}
The macroscopic energy-momentum tensor, $T^{ab}$ with $a,b \in {t,z}$ is calculated by integrating $\mathcal{T}^{ab}$ over the string cross-section. Using our ansatz for a static string gives the four components as 
\begin{equation}
    T^{tt} = 2\omega^2\Sigma_2 + \mu - \frac{1}{4}\lambda_\sigma\Sigma_4,
\end{equation}
\begin{equation}\label{eq: Ttz}
    T^{tz} = T^{zt} = 2k\omega\Sigma_2,
\end{equation}
\begin{equation}
    T^{zz} = 2k^2\Sigma_2 - \mu + \frac{1}{4}\lambda_\sigma\Sigma_4.
\end{equation}
The tension and energy per unit length are the eigenvalues of $T^{ab}$. In the frame in which the macroscopic tensor is diagonal, the tension per unit length is $T = -T^{zz}$ and the energy per unit length is $U = T^{tt}$. From (\ref{eq: Ttz}), this is achieved if either $\omega$ or $k$ is zero. 

Under a Lorentz boost of velocity $v$ in the $z$ direction, $\omega \to \gamma(\omega - vk)$ and $k \to \gamma(k - v\omega)$, and hence for $\chi<0$, $v=\omega/k$ will set $\omega \to 0$ ($\chi = -k^2$), while for $\chi>0$, $v=k/\omega$ sets $k \to 0$ ($\chi=\omega^2$), allowing the diagonalisation of $T^{ab}$. Therefore, the tension and energy per unit length are given by
\begin{equation}
    T = 
    \begin{cases}
        \mu - \frac{1}{4}\lambda_\sigma\Sigma_4 &\mbox{if } \chi>0, \\
        2\chi\Sigma_2 + \mu - \frac{1}{4}\lambda_\sigma\Sigma_4 &\mbox{if } \chi < 0,
    \end{cases}
\end{equation}
\begin{equation}
    U =
    \begin{cases}
        2\chi\Sigma_2 + \mu - \frac{1}{4}\lambda_\sigma\Sigma_4 &\mbox{if } \chi>0, \\
        \mu - \frac{1}{4}\lambda_\sigma\Sigma_4 &\mbox{if } \chi<0,
    \end{cases}
\end{equation}
with the two clearly equal in the chiral limit, $\chi=0$, where $T=U$. 

\begin{figure}[t]
    \centering
    \subfloat[$\eta_\sigma=0.35$, $\lambda_\sigma=36$, $\beta=6.6$ and $G=0.2$ (parameter set A).]{
        \centering
        \includegraphics[trim={0.5cm 0 1.5cm 0},clip,width=0.46\linewidth]{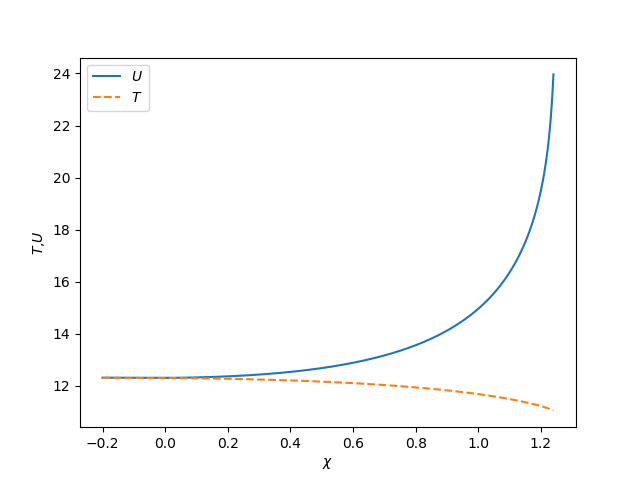}
    }\hspace{1em}
    \subfloat[$\eta_\sigma=0.61$, $\lambda_\sigma=10$, $\beta=3$ and $G=0.5$ (parameter set B).]{
        \centering
        \includegraphics[trim={0.5cm 0 1.5cm 0},clip,width=0.46\linewidth]{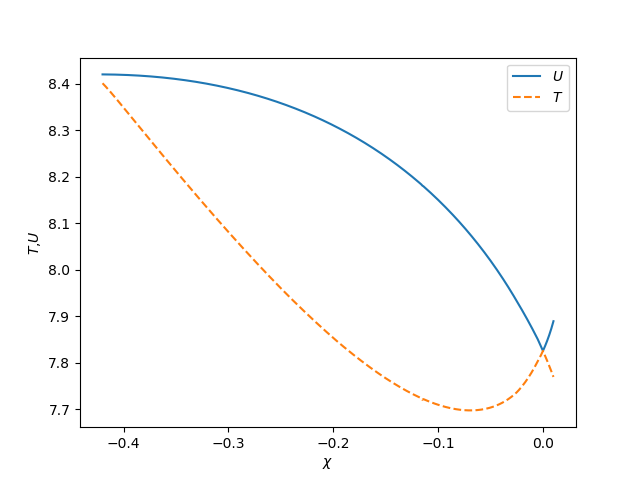} 
    }
    \caption{The tension and energy per unit length of string as a function of $\chi$. The functions converge at small $\chi$ because there is no condensation onto the string, while at $\chi = 0$ they are equal and discontinous.}
    \label{fig:tension and energy}
\end{figure}

We see that the equation of state is $U - T = 2|\chi|\Sigma_2$. We can use this result to further investigate our claim that it is reasonable to assume the first solution in equation (\ref{eq: quadratic sols}) will be positive. The solution can now be rewritten as
\begin{equation}
    \bigg(\frac{L}{Q}\bigg)^2 =
    \begin{cases}
        \frac{2}{U\Sigma_2} &\mbox{if } \chi>0, \\
        \frac{2}{T\Sigma_2} &\mbox{if } \chi<0.
    \end{cases}
\end{equation}
Therefore, it is guaranteed to be positive in the electric regime and will be positive in the magnetic regime so long as the tension is positive. This is a reasonable thing to assume and any negative tension strings will be unstable anyway \cite{Carter1989}. In figure \ref{fig:tension and energy} the energy per unit length and tension are plotted as a function of $\chi$ for parameter sets A and B. The range of $\chi$ is set by the approximate values of $\chi_\text{min}$ and $\chi_\text{max}$ as discussed in section \ref{sec: parameter space}. Note that both of these parameter sets have $\beta>\frac{1}{2}\sqrt{\lambda_\phi\lambda_\sigma}$ and, therefore, the upper limit on $\chi$ is determined by trial and error.

Having an expression for the equation of state allows both the transverse speed, $c_T$, and the longitudinal speed, $c_L$, to be calculated from
\begin{equation}
    c_T^2 = \frac{T}{U} \qquad c_L^2 = -\frac{dT}{dU}.
\end{equation}
These determine the propagation speeds of perturbations through the string and are, therefore, clearly  important quantities for determining the stability of vorton solutions. If either $c_T^2$ or $c_L^2$ are negative the string will be unstable, and they must also be less than $1$ or causality will be violated. Substituting in the expressions for tension and energy per unit length allows the sound speeds to be written as
\begin{equation}
    c_T^2 = \bigg(1 + \frac{2\chi\Sigma_2}{\mu - \frac{1}{4}\lambda_\sigma\Sigma_4}\bigg)^{-\text{sgn}(\chi)},
\end{equation}
\begin{equation}
    c_L^2 = \bigg(1+ \frac{2\chi\Sigma_2'}{\Sigma_2}\bigg)^{-\text{sgn}(\chi)},
\end{equation}
where we have used equation (\ref{eq: mu and sigma4 deriv}) to simplify the longitudinal sound speed. At $\chi_\text{max}$, $\Sigma_2'$ is undefined due to the turning point and $\Sigma_2'(\chi)<0$ along the higher charge branch of solutions. Additionally, we can write
\begin{equation}
    \frac{2\chi\Sigma_2'(\chi)}{\Sigma_2} = \frac{q_p\Sigma_2'(q_p)}{\Sigma_2 - q_p\Sigma'(q_p)} ,
\end{equation}
which is always less than $-1$ when $q_p\Sigma_2'(q_p)>\Sigma_2$, which is also the condition to be on the higher charge branch. Therefore, all solutions on the higher charge branch should be unstable to longitudinal perturbations, since $c_L^2<0$, and we can focus solely on the lower charge branch when searching for stable vortons. The accuracy of this prediction is another test of the TSA which we will not specifically address in this work, but will be discussed in our follow-up paper on pinching instabilities. The quantity, $\Sigma_2'(\chi)$, may be calculated by varying $\chi$, solving the static equations of motion and using a finite difference method. Alternatively, taking the derivative of equation (\ref{eq: sigma def}) with respect to $\chi$ gives
\begin{equation} \label{eq: sigma_2 derivative}
    \Sigma_2' = 4\pi \int \rho |\sigma| \frac{\partial|\sigma|}{\partial\chi} d\rho.
\end{equation}
Now if we perturb the equations of motion by taking $\chi \to \chi + \delta \chi$, we discover another set of coupled differential equations,

\begin{equation}
    \begin{gathered}
        \frac{\partial^2}{\partial\rho^2}\bigg(\frac{\partial|\phi|}{\partial\chi}\bigg) + \frac{1}{\rho}\frac{\partial}{\partial\rho}\bigg(\frac{\partial|\phi|}{\partial\rho}\bigg)  - \bigg[\frac{1}{2}\lambda_\phi(3|\phi|^2 - \eta_\phi^2) + \beta|\sigma|^2 + \bigg(\frac{n-gA_\theta}{\rho}\bigg)^2\bigg]\frac{\partial|\phi|}{\partial\chi} \\ - 2\beta|\phi||\sigma|\frac{\partial|\sigma|}{\partial\chi} + \frac{2g}{\rho^2}(n-gA_\theta)\frac{\partial A_\theta}{\partial\chi} = 0,
    \end{gathered}
\end{equation}

\begin{figure}[t]
    \centering
    \subfloat[$\eta_\sigma=0.35$, $\lambda_\sigma=36$, $\beta=6.6$ and $G=0.2$ (parameter set A).]{
        \centering
        \includegraphics[trim={0.5cm 0 1.5cm 0},clip,width=0.46\linewidth]{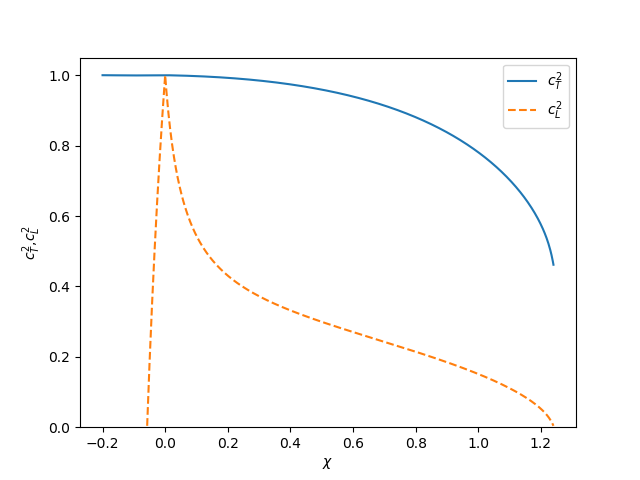}
    }\hspace{1em}
    \subfloat[$\eta_\sigma=0.61$, $\lambda_\sigma=10$, $\beta=3$ and $G=0.5$ (parameter set B).]{
        \centering
        \includegraphics[trim={0.5cm 0 1.5cm 0},clip,width=0.46\linewidth]{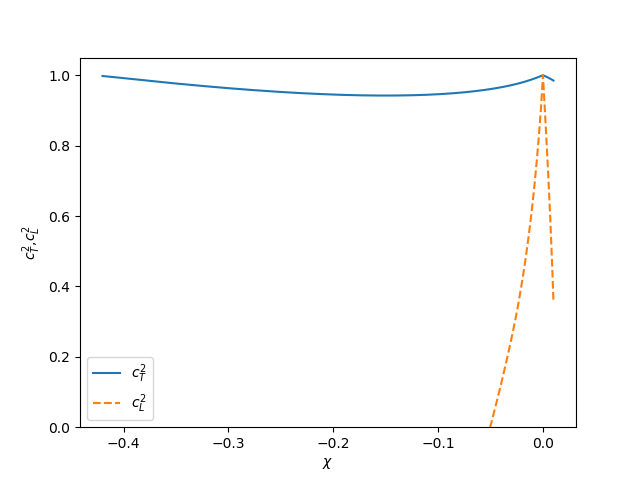}
    }
    \caption{The sound speeds as a function of $\chi$. Again, the functions are equal and discontinous at $\chi = 0$. Notice that in both cases $c_L^2 \leq c_T^2$, $\forall \chi$. This appears to always be true for superconducting strings, although we have not found an analytic argument to prove it.}
    \label{fig:sound_speeds}
\end{figure}

\vspace{-6pt}
\begin{equation}
    \frac{\partial^2}{\partial\rho^2}\bigg(\frac{\partial|\sigma|}{\partial\chi}\bigg) + \frac{1}{\rho}\frac{\partial}{\partial\rho}\bigg(\frac{\partial|\sigma|}{\partial\chi}\bigg) - \bigg[\frac{1}{2}\lambda_\sigma(3|\sigma|^2 - \eta_\sigma^2) + \beta|\phi|^2 - \chi\bigg]\frac{\partial|\sigma|}{\partial\chi} - 2\beta|\phi||\sigma|\frac{\partial|\phi|}{\partial\chi} + |\sigma| = 0,
\end{equation}

\begin{equation}
    \frac{\partial^2}{\partial\rho^2}\bigg(\frac{\partial A_\theta}{\partial\chi}\bigg) - \frac{1}{\rho}\frac{\partial}{\partial\rho}\bigg(\frac{\partial A_\theta}{\partial\chi}\bigg) - 2g^2|\phi|^2\frac{\partial A_\theta}{\partial\chi} + 4g|\phi|(n-gA_\theta)\frac{\partial|\phi|}{\partial\chi} = 0.
\end{equation}
For each string solution we can also solve this perturbed equation of motion to find the derivatives of each field with respect to $\chi$ as a function of the radial coordinate and subsequently calculate $\Sigma_2'$. Figure \ref{fig:sound_speeds} shows the sound speeds as a function of $\chi$ for parameter sets A and B.

\begin{figure}[t]
    \centering
    \includegraphics[scale=0.7]{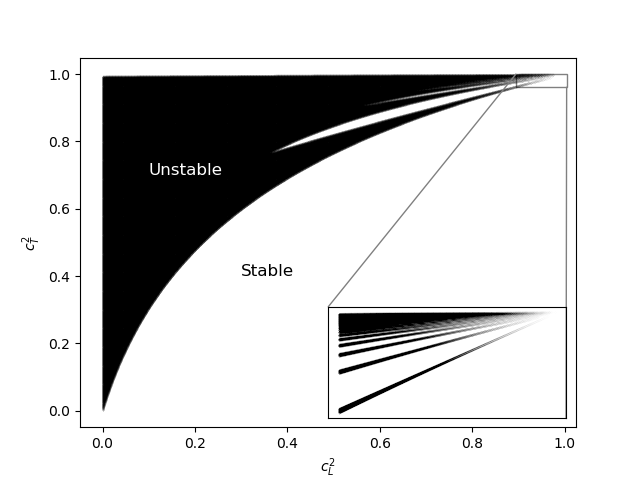}
    \caption{The regions of instability (black) for modes between $m=2$ and $m=100$, determined by finding the values of the sound speeds which give a negative discriminant. The large stable region corresponds to the region where $\Delta(m)=0$ has no real roots while the thin, stable slices are caused by $m$ being an integer. There are an infinite number of these regions as the limit $c_L^2 = c_T^2 = 1$ is approached and we show a few more stable slices by zooming into the region with $0.96\leq c_T^2\leq1$ and $0.9\leq c_L^2\leq1$. This makes it very likely that any parameter sets which admit chiral, superconducting strings will pass through stable regions.}
    \label{fig:cl_ct instability}
\end{figure}

The sound speeds in these parameter sets agree with the observation made in \cite{Peter1992} that the longitudinal speed appears to be, in general, less than the transverse speed. String loops with $c_L^2>c_T^2$ are stable to all perturbative modes (as will be shown later), but this is not true for the converse. The consequence of this is that stable vorton solutions are less likely to be produced, although it is by no means impossible as there remain pairs of sound speeds that are completely stable - see figure \ref{fig:cl_ct instability}.

\begin{figure}[t]
    \centering
    \subfloat[$\eta_\sigma=0.35$, $\lambda_\sigma=36$, $\beta=6.6$ and $G=0.2$ (parameter set A).]{
        \centering
        \includegraphics[trim={0.5cm 0 1.5cm 0},clip,width=0.46\linewidth]{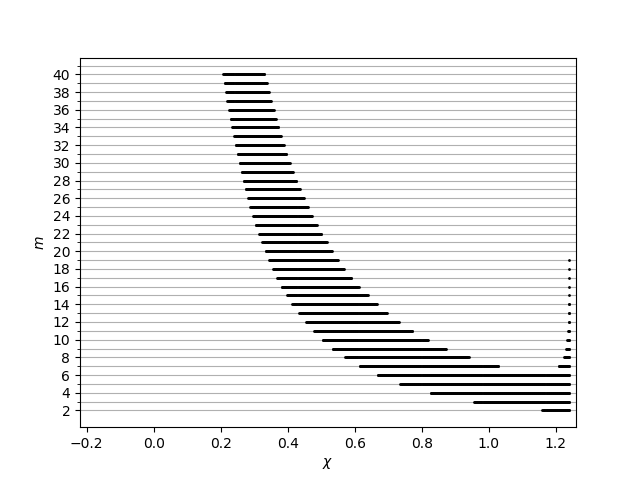}
        \label{fig: LS instability intervals}
    }\hspace{1em}
    \subfloat[$\eta_\sigma=0.61$, $\lambda_\sigma=10$, $\beta=3$ and $G=0.5$ (parameter set B).]{
        \centering
        \includegraphics[trim = {0.5cm 0 1.5cm 0},clip,width=0.46\linewidth]{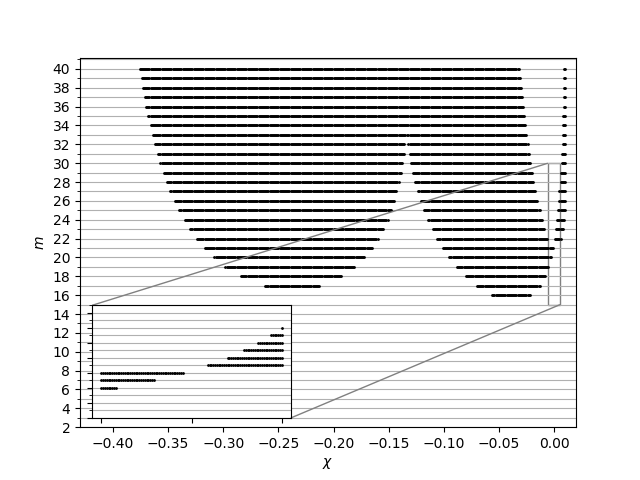}
        \label{fig: CB full instability intervals}
    }
    \caption{The black regions show the range of $\chi$ for which vortons are expected to be unstable for modes between $m=2$ and $m=40$. Only the regions $\chi>1.02$ (left) and $\chi>-0.042$ (right) are relevant due to the critical phase frequency. The right-hand figure indicates that vortons which are close to chiral are the most likely to be stable to all modes. The bottom left corner displays a zoomed in image of the region $\num{-5e-3} \leq \chi \leq \num{5e-3}$ for modes $15 \leq m \leq 30$ and demonstrates that there is a region (approximately $\num{-4e-4} < \chi < \num{8e-4}$) which is predicted to be completely stable.}
    \label{fig:instability_intervals}
\end{figure}

The radial (but not axially symmetric) transverse perturbations of an infinitely thin loop of string can be decomposed into the real part of the Fourier modes
\begin{equation}
    \delta r(t, \theta) = \sum_{m,j} A_{mj} e^{i(\Omega_{mj} t - m\theta)},
\end{equation}
where $m$ is the Fourier mode of the perturbation, $\Omega_{mj}$ is its frequency and $j$ labels the three possible frequencies and amplitudes for each $m$. The radial perturbations are coupled to two additional longitudinal perturbations that can be similarly decomposed. The system of equations satisfied by these three variables is an eigenvalue equation,
\begin{equation} \label{eq: stability matrix}
    \begin{pmatrix}
        2 & c_T^2+c_L^2 & (1+c_L^2)\nu_m - 2m \\
        (1+c_T^2)\nu_m - 2m & c_T^2(c_L^2+1)\nu_m - (c_T^2+c_L^2)m & 2 \\
        (1-c_T^2)\nu_m & c_T^2(c_L^2-1)\nu_m + (c_T^2 - c_L^2)m & 0
    \end{pmatrix}
    \begin{pmatrix}
        c_T\epsilon \\ \alpha \\ i\beta R
    \end{pmatrix}
    = 0 ,
\end{equation}
where $\epsilon$ and $\alpha$ are the two longitudinal perturbation variables previously mentioned and $\beta$ corresponds to the transverse perturbation. Vanishing of the determinant results in the cubic equation (for more detail see \cite{Carter1993})
\begin{equation} \label{eq: freq cubic}
    a_3\nu_m^3 + a_2\nu_m^2 + a_1\nu_m + a_0 = 0,
\end{equation} 
with $\nu_m = \Omega_m R/c_T$ and
\begin{align}
    &a_0 = 2(c_L^2 - c_T^2)(m^2 - 1)m, \\
    &a_1 = 4c_T^2(1 - c_L^2)(m^2 - 1) - (1 + c_T^2)(c_L^2 - c_T^2)(m^2 + 1), \\
    &a_2 = 2c_T^2[c_L^2 - c_T^2 - 2(1-c_L^2c_T^2)]m, \\
    &a_3 = c_T^2(1 + c_T^2)(1-c_L^2c_T^2).
\end{align}
Due to the definition of the radial perturbations, real roots to this cubic equation describe stable oscillations, complex roots with a positive imaginary component describe exponentially decaying oscillations while complex roots with a negative imaginary component describe exponentially growing oscillations. Complex roots to cubic polynomials always come in complex conjugate pairs so one of the complex roots will always describe an unstable oscillation.
Therefore, all of the roots to the cubic must be real and distinct for the string to be stable to perturbations of that mode. This can be assessed by either directly calculating the roots (either numerically or using the Cardano formula) or by computing the discriminant
\begin{equation}
    \Delta = a_1^2a_2^2 - 4a_1^3a_3 - 4a_0a_2^3 - 27a_0^2a_3^2 + 18a_0a_1a_2a_3,
\end{equation}
which has the property that when $\Delta>0$ all of the roots are real and distinct, when $\Delta=0$ all of the roots are real, but there is a repeated root, and when $\Delta<0$ there are two complex roots and one real root. Therefore, a vorton will have an instability to a mode if the discriminant is less than or equal to zero. The $m=0$ and $m=1$ modes are axially symmetric oscillations and translations respectively, which are stable if $0<c_L^2,c_T^2\leq 1$. The $m=2$ mode is a quadrupolar stretching and squeezing, elliptical oscillation, while the higher modes correspond to rotating regular $m$ sided shapes. Modes with $m\geq2$ require the calculation of the discriminant to determine whether they are stable or not. We can use our computation of the sound speeds as function of $\chi$ to predict the range of $\chi$ for which there will be an instability to each mode. The intervals of instability for modes between $m=2$ and $m=40$ are shown in Figure \ref{fig:instability_intervals} for parameter sets A and B.

Although the parameter sets used are not drastically different, the intervals of instability are clearly very different. Prima facie these plots encourage hope for a completely stable (stable to each individual mode) loop in both cases when the vorton is close to chiral. However, this is not the case for the first parameter set because only strings with $\chi > 1.02$ satisfy the additional vorton formation constraint given in equation (\ref{eq: vorton formation constraint}). Chiral vortons do appear to be accessible in the second parameter set as the additional constraint is satisfied for $\chi > -0.042$ and there is in fact a fully stable region in that case. The bottom left corner of Figure \ref{fig: CB full instability intervals} zooms into the vicinity of the chiral limit. This makes it clear that that there is a very small region (approximately $\num{-4e-4}<\chi<\num{8e-4}$) which is expected to be completely stable to all modes of perturbation. This would imply the existence of completely stable (at least classically) vorton solutions. Since the region is so narrow, there could be significant corrections to the thin string approximation. See the conclusion section for further discussion of this point.

We will briefly comment on the required condition to create a string loop that is stable to all modes of perturbation. It is easy to confirm that the discriminant is a polynomial of degree six in $m$, but only even powers of $m$ appear. Therefore it can be viewed more simply as a cubic polynomial in $m^2$. The important features of the discriminant are that $\Delta \to \infty$ as $m \to \pm \infty$ and that it is positive when $m=0$ (for physical values of the sound speeds). This clearly means that the curve must cross $\Delta = 0$ to become negative at any point so the loop will be stable to all modes if the sextic equation has no real roots or - equivalently - if the cubic equation has no real and positive roots. However, the discrete nature of $m$ will allow this rule to be mildly broken if the discriminant changes sign twice without crossing an integer value of $m$. In practice we have found this method to be of limited use and the brute force method used in \cite{Martin1994} (the results of which are presented in Figure \ref{fig:cl_ct instability}) is much more convenient for illustrating the stable regions. Nonetheless, it is a useful picture to have in mind.

It appears that complete stability with $c_L^2<c_T^2$ is only possible when the longitudinal speed is not significantly less than the transverse speed, or in the narrow regions which approach the limit $c_L^2=c_T^2=1$. The narrow zones of stability can be eliminated by allowing $m$ to take non-integer values, suggesting that they are the regions of the parameter space that evade our analysis in the previous paragraph. The largest narrow zone corresponds to the region where the discriminant dips below zero and goes back above inbetween $m=2$ and $m=3$, the next largest zone does the same between $m=3$ and $m=4$ and so on. It is these areas which enable the small region of stability shown in figure \ref{fig: CB full instability intervals}.

For any set of sound speeds equation (\ref{eq: freq cubic}) can be solved for the frequency, $\Omega_m$ of perturbations of Fourier mode $m$. The real part of this frequency determines the oscillation frequency, while the imaginary part determines the growth rate of the amplitude of the perturbations. In the next section we will construct vorton solutions and compare their radii to our predicted values. We will then compare our predictions for the stability and frequency of each mode of oscillation to the simulated dynamics of vortons.

\section{Vorton Construction and Dynamics} \label{sec: construction and dynamics}

In this section we will construct vortons, simulate their dynamical evolution and test the predictions of the thin string approximation. Initially, we take advantage of the cylindrical symmetry of vortons by either using the cartoon method \cite{Battye2009a,Alcubierre1999}, or the cut-off method (see appendices \ref{sec: cartoon} and \ref{sec: cut off} for details). Both of these methods allow the numerical relaxation to be performed in only $2$ dimensions and we also only need to consider $z>0$ due to the additional reflection symmetry. This is particularly useful for vorton construction and can also be used to test radial stability and the zero mode frequencies, but it will provide no insight into the stability or frequencies of higher order modes. This requires the simulation of the full three dimensional dynamics which we discuss in section \ref{sec: 3d}.

\subsection{Construction}

We will be constructing vortons using a gradient flow algorithm with an initial field configuration that comes from either an extension of the analytical solution of kinky vortons found in \cite{Battye2008} or, more commonly, the straight string solutions discussed in the previous section. The equations of motion for the fields are
\begin{equation}
    \mathcal{D}_\mu\mathcal{D}^\mu \phi + \frac{\lambda_\phi}{2}(|\phi|^2 - \eta_\phi^2)\phi + \beta|\sigma|^2\phi = 0,
\end{equation}
\begin{equation}
    \partial_\mu\partial^\mu \sigma + \frac{\lambda_\sigma}{2}(|\sigma|^2 - \eta_\sigma^2)\sigma + \beta|\phi|^2\sigma = 0,
\end{equation}
\begin{equation} \label{eq: gauge field EoM}
    \partial_\nu F^{\mu\nu} = ig[\phi^*\mathcal{D}^\mu\phi - \phi(\mathcal{D}^\mu\phi)^*].
\end{equation}
The gradient flow algorithm replaces the second order time derivatives in these equations with first order time derivatives, which results in the energy of the system being driven towards a nearby minima rather than oscillating around it. We make the ansatz $\sigma = e^{i(\omega t +N\theta)}\psi$, but leave the winding of $\phi$ to be enforced by the initial field configuration. The time dependence of the magnitude and phase of $\psi$ are separated so that the gradient flow algorithm reaches the pseudo-stationary state in which $|\psi|$ does not change with time. The resulting $\omega^2$ term is replaced with the conserved Noether charge using $Q = \omega\int|\sigma|^2 d^3x$. Within the cut-off method the winding of the condensate is treated exactly, but within the cartoon method it is approximated by interpolation. Techniques from lattice gauge theory (see appendix \ref{sec: lattice gauge theory}) must be used to discretise this system for non-zero gauge couplings. Without implementing this approach, the condition set by the time component of equation (\ref{eq: gauge field EoM}) (Gauss's law) is violated when the system is evolved under the equations of motion and the numerical evolution quickly diverges from the continuum equations.

\subsubsection{Solutions for parameter sets A and B}

\begin{figure}
    \centering
    \subfloat[$Q = 749$ and $N=50$ vorton with a radius of $R = 28.8$ in the parameter set $\eta_\sigma=0.35$, $\lambda_\sigma=36$, $\beta=6.6$ and $G=0.2$ (parameter set A). Associated with the straight string shown in Figure \ref{fig: LS straight string profile}.]{
        \centering
        \includegraphics[trim={0.2cm 0 1.5cm 0},clip,width=0.46\linewidth]{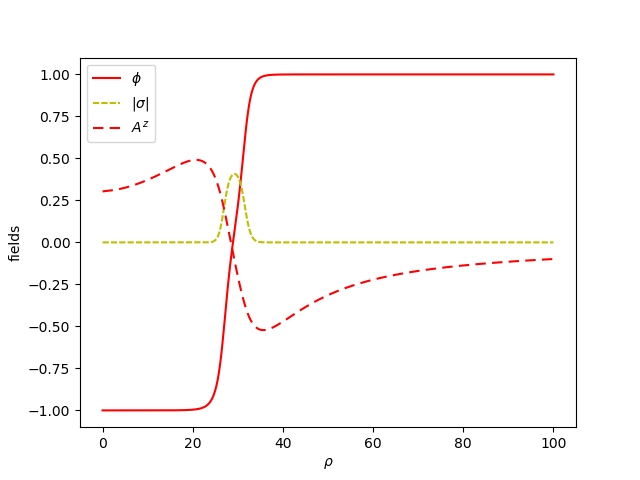}
        \label{fig: LS vorton profile}
    }\hspace{1em}
    \subfloat[$Q = 1594$ and $N=50$ vorton with a radius of $R = 56.6$ in the parameter set $\eta_\sigma=0.61$, $\lambda_\sigma=10$, $\beta=3$ and $G=0.5$ (parameter set B). Associated with a straight string that has $\chi=\num{2e-4}$ (similar to the one shown in Figure \ref{fig: CB straight string profile}).]{
        \centering
        \includegraphics[trim={0.2cm 0 1.5cm 0},clip,width=0.46\linewidth]{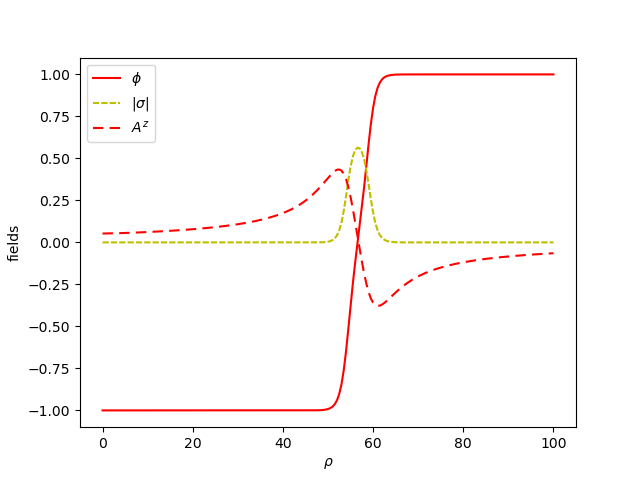}
        \label{fig: CB vorton profile}
    }
    \\
    \subfloat[Differences between the solution (shown above) and the field profiles generated from placing the straight string solution at $R=28.8$.]{
        \centering
        \includegraphics[trim={0.2cm 0 1.5cm 0},clip,width=0.46\linewidth]{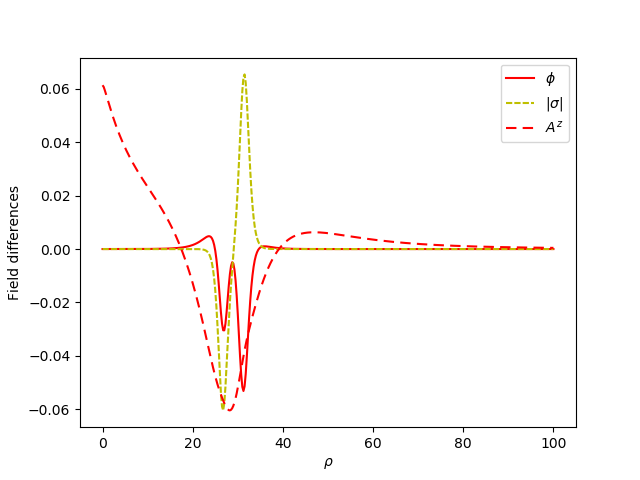}
        \label{fig: difs A}
    }\hspace{1em}
    \subfloat[Differences between the solution (shown above) and the field profiles generated from placing the straight string solution at $R=56.6$.]{
        \centering
        \includegraphics[trim={0.2cm 0 1.5cm 0},clip,width=0.46\linewidth]{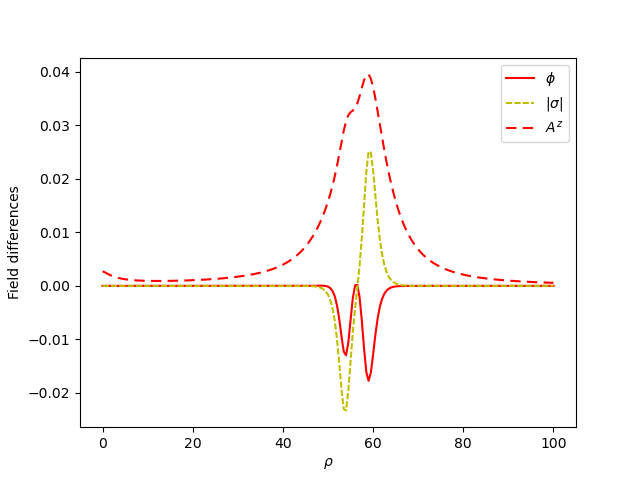}
        \label{fig: difs B}
    }
    \caption{The field profiles of two energy minimising vorton solutions (top) with the difference between them and the associated straight string profiles shown (bottom). It is important to note that the straight string profiles are placed at the correct vorton radius for comparison purposes and not at the predicted radius from the semi-analytic approach, which is slightly different. Only the $z$ component of the gauge field is non-zero in the $z=0$ plane. Notice that, although there are differences between the straight string profiles and the vorton solutions, the differences are small - a few percent. These differences can be broadly explained by three effects. The gauge field is most notably modified due to the axial symmetry forcing its first derivative to be zero at the centre, which is not the case in the straight string approximation. There is also a splitting between the radius as measured by the core of the string and the radius defined by the peak of the condensate, although, as illustrated in figure \ref{fig: radius splitting}, this effect is small. Finally, there is a slight kink at the core of the string which is enhanced in the vorton solution. The small differences between the straight string approximation and the solutions suggests that the prediction of the radius should be reasonably accurate.}
    \label{fig: vorton profiles}
\end{figure}

Figure \ref{fig: vorton profiles} displays vorton solutions with $N=50$ for parameter sets A and B in the $z=0$ plane. We also plot the differences between these solutions and the field profiles produced by placing the straight string solutions (Figure \ref{fig:straight string profiles}) at the vorton radius. In both cases the grid spacing is the same in the $x$ and $z$ directions and we advance with timesteps of $\Delta t = 0.1(\Delta x)^2$ until the system reaches the stationary state. Care must be taken to choose a timestep that satisfies the Courant-Friedrichs-Lewy (CFL) condition, $\Delta t \sum_{i=1}^d \Delta x_i^{-2} \lesssim \frac{1}{2} $, in $d$ dimensions (the exact condition depends upon the numerical scheme, but this is a useful guide), so that the algorithm is numerically stable. For parameter set B we use $\Delta x = 0.5$ and the size of the grid is $0\leq x \leq 200$ (likewise for the $z$ direction), while for parameter set A we use $\Delta x = 0.25$, so that the winding is properly resolved, and $0\leq x \leq 100$ - although we have increased this for some of the larger vortons where this would clearly not be appropriate.

\begin{figure}[t]
    \centering
    \includegraphics[trim={0.5cm 0 1.5cm 1.4cm},clip,scale=0.7]{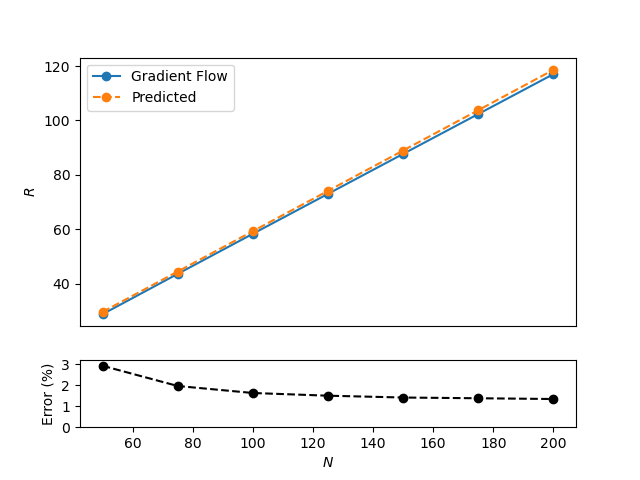}
    \caption{A comparison of the predicted radii from the TSA to the radii of solutions found via gradient flow with model parameters $\eta_\sigma=0.35$, $\lambda_\sigma=36$, $\beta=6.6$ and $G=0.2$ (parameter set A) with $Q/N=14.98$ kept constant.}
    \label{fig: rLS RvN}
\end{figure}

\begin{figure}[t]
    \centering
    \includegraphics[trim={0.5cm 0 1.5cm 1.4cm},clip,scale=0.7]{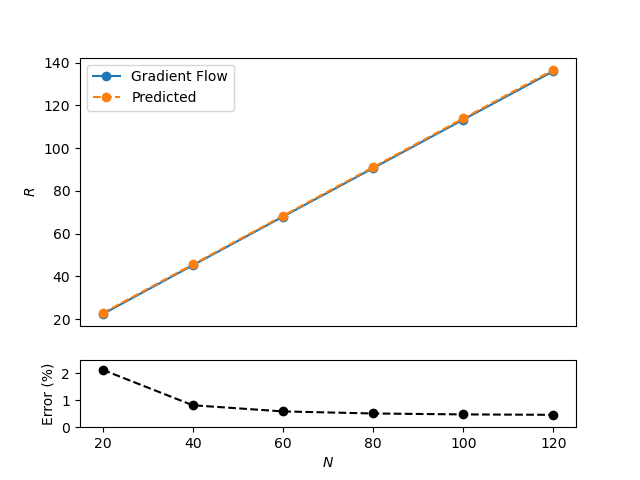}
    \caption{A comparison of the predicted radii from the TSA to the radii of solutions found via gradient flow with model parameters $\eta_\sigma=0.61$, $\lambda_\sigma=10$, $\beta=3$ and $G=0.5$ (parameter set B) with $Q/N=31.89$ kept constant.}
    \label{fig: pB RvN}
\end{figure}

The predicted radii are $R=29.7$ and $R=56.9$ respectively which corresponds to less than a $3\%$ difference in set A and much lower for set B. Figures \ref{fig: rLS RvN} and \ref{fig: pB RvN} compare the radii of vortons constructed via gradient flow to the predicted radii from the associated straight string solution, for a range of $N$. We keep the ratio of $N$ to $Q$ as a constant in this plot so that the initial conditions for each vorton can be produced by wrapping the same straight string solution into loops of different sizes. All results are obtained using the cartoon method since these vortons are small enough for this to be numerically feasible and there is no need to introduce an additional boundary condition that will make the solutions less accurate. There is clearly a very good agreement between the predictions of the TSA and the vortons that we have constructed. The relationship between the radius and the winding number is evidently linear as predicted by the theory and the percentage error decreases as the size of the vorton increases, due to the effects of curvature becoming less important.

% \begin{figure}[t]
%     \centering
%     \subfloat[$\eta_\sigma=0.35$, $\lambda_\sigma=36$, $\beta=6.6$ and $G=0.2$ (parameter set A) with $Q/N=14.98$ kept constant.]{
%         \centering
%         \includegraphics[width=0.46\linewidth]{Plots/rLS_RvN.png}\label{fig: rLS RvN}
%     }\hspace{1em}
%     \subfloat[$\eta_\sigma=0.61$, $\lambda_\sigma=10$, $\beta=3$ and $G=0.5$ (parameter set B) with $Q/N=31.89$ kept constant.]{
%         \centering
%         \includegraphics[width=0.46\linewidth]{Plots/pB_radius_with_preds.png}
%     }
%     \caption{A comparison of the predicted radii from the semi-analytic method to the radii of solutions found via gradient flow.}
%     \label{fig: radius comparisons}
% \end{figure}

Additionally, we should expect more localised strings to have more accurate predictions as they rely on the approximation that all components of the energy are confined to an infinitesimally thin string, despite the reality being that the energy is spread over some region. In particular, the mass per unit length of the string (without considering the effects of the condensate), $\mu$, is logarithmically divergent in the zero gauge coupling (global) limit of the theory. This divergence introduces infinities into the semi-analytic calculations and the straight string analysis may no longer produce useful predictions, something which was commented on in \cite{Battye2009a}. In reality, there will be a cut off scale set by the radius of the vorton and straight strings can still make predictions if we know this scale. We should, therefore, expect that the predictions improve as the gauge coupling increases - at least up to the BPS limit at which point the limiting factor switches from the gauge field mass to the mass of the vortex field.

\begin{figure}[t]
    \centering
    \includegraphics[trim={0.5cm 0 1.5cm 1.4cm},clip,scale=0.7]{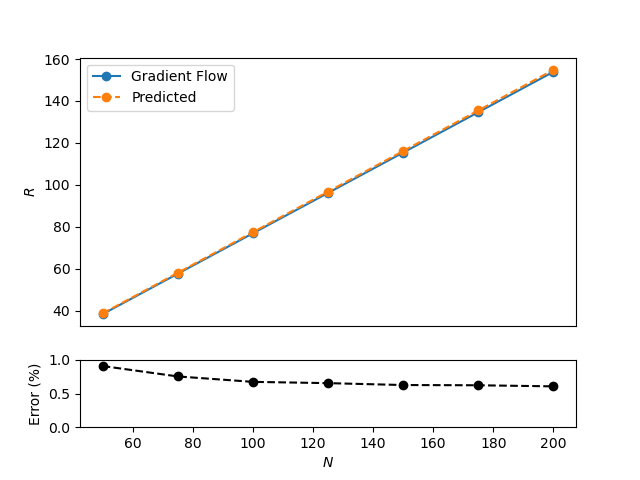}
    \caption{A comparison of the predicted radii to the radii of vorton solutions in parameter set C, to be compared with Figure \ref{fig: rLS RvN}, where the parameter set is the same, except set C has a stronger gauge coupling of $G=1$. The ratio $Q/N = 31.89$ is kept fixed. This has improved the accuracy of the predictions and increased the energy minimising radii.}
    \label{fig: rLS G1 RvN}
\end{figure}

To illustrate this improvement we have constructed vortons in parameter set C, which is very similar to set A, except that it uses $G=1$ rather than $G=0.2$. The change to the gauge coupling modifies the range of $\chi$ for which straight string solutions exist, so unfortunately we are unable to make the same choice of $\chi$ for the sake of comparison. Instead, we have chosen to use $\chi = 1.514$ which sets the ratio $Q/N$ to the value used previously. We compare the radius predictions to the radii of constructed vortons for this parameter set in Figure \ref{fig: rLS G1 RvN}. This shows that the effect of a stronger gauge coupling is to increase the radii of vortons and, by comparing the percentage error of the predictions, it is clear that it has improved their accuracy, due to the improved localisation of the energy.

There are some subtle differences (which are to be expected) between the predicted field profiles and the true solutions. We have noticed two main areas in which they differ. The first, and most obvious from Figure \ref{fig: vorton profiles}, is that the axial symmetry of the vorton forces the derivative of $A^z$ to zero at the centre. In the profiles produced from the straight string, $A^z \propto 1/\rho$ at large (gauge coupling dependent) distances from the string core and obviously this does not change at the centre of the loop.

\begin{figure}[t]
    \centering
    \includegraphics[trim={0.5cm 0 1.5cm 1.4cm},clip,scale=0.75]{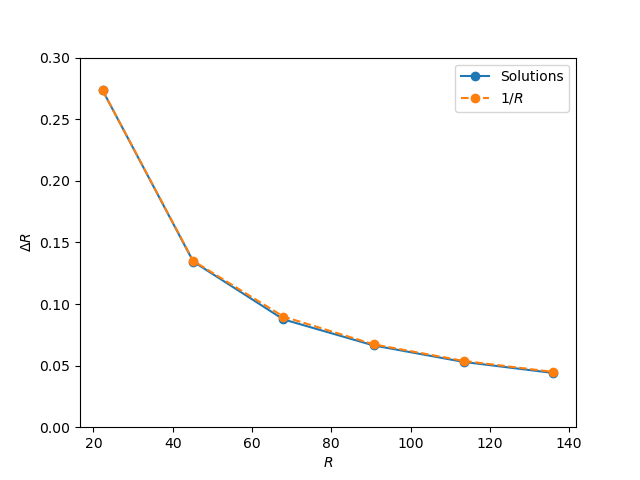}
    \caption{The radius splitting as a function of the radius (as measured by the core of the string) using parameter set B, with $Q/N=31.89$ kept constant. The effect is reduced for larger vortons (higher $N$) and appears to be inversely proportional to the radius. This is consistent with there being a curvature correction to the thin string approximation.}
    \label{fig: radius splitting}
\end{figure}

The second effect that we have noticed is a splitting between the radius defined by the core of the string and the radius defined by the peak of the condensate. Qualitatively, this is caused by the competition between the angular momentum of the condensate, which wants to cause expansion, and the tension of the string loop, which wants to cause contraction. There is a force between the condensate and the string which grows as the splitting between them increases. At some level of splitting, this force balances the competition between the angular momentum and tension. Figure \ref{fig: radius splitting} shows how this splitting is reduced for larger vortons as the curvature effects become less important and that the splitting is consistent with a $1/R$ curve. The shape of the string core also tends toward the straight string prediction for larger vortons, suggesting that this effect is a curvature correction to the thin string approximation.

\begin{figure}
    \centering
    \includegraphics[trim={0.4cm 0 1.5cm 1.4cm},clip,scale=0.75]{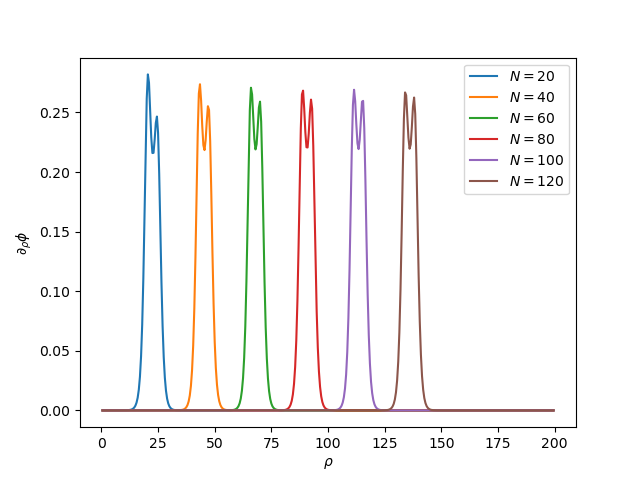}
    \caption{The radial derivative of the vortex field for vortons with various winding numbers (but fixed Q/N) using parameter set B. There is an asymmetry around the vorton radius (the local minima inside each peak) which is reduced for larger vortons. Again, this is consistent with curvature corrections to the thin string approximation.}
    \label{fig: phi deriv}
\end{figure}

Finally, there is a slight enhancement of the kink in $\phi$ at the string core. To make this effect more apparent, in Figure \ref{fig: phi deriv} we plot the radial derivative of $\phi$ for a few vortons with different winding numbers (fixed Q/N ratio). The radius of each vorton corresponds to the local minima inside the peak. There is an asymmetry around this minima that enhances the kink and is not present in the straight string profiles. Note that the double peak structure is expected from the straight string analysis, but both peaks should be the same size and shape. The effect is clearly reduced for the larger vortons, suggesting that this is also a curvature correction and we believe this may be caused by the splitting effect already alluded to.

\subsubsection{Comparison with Battye \& Sutcliffe [\protect\hyperlink{page.46}{20}]}

\begin{figure}[t]
    \centering
    \subfloat[Global vorton field profiles with $R=25.7$.]{
        \centering
        \includegraphics[trim={0.4cm 0 1.5cm 1.4cm},clip,width=0.46\linewidth]{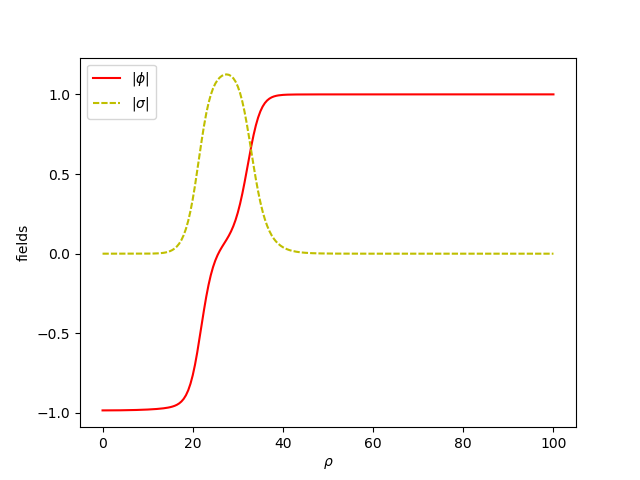}
        \label{fig: BS global}
    }\hspace{1em}
    \subfloat[Gauged vorton ($G=0.1$) field profiles with $R=32.5$.]{
        \centering
        \includegraphics[trim={0.4cm 0 1.5cm 1.4cm},clip,width=0.46\linewidth]{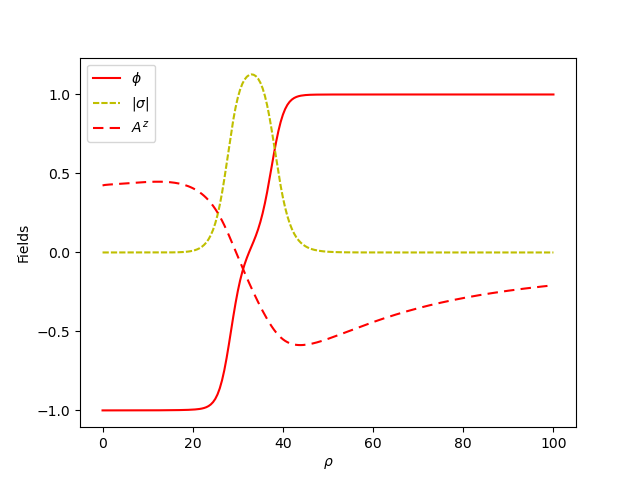}
        \label{fig: BS gauged}
    }
    \caption{$Q=9000$, $N=10$ vortons with $\eta_\sigma = 1$ and $\lambda_\sigma=\beta=2/3$ (parameter sets D and E respectively). There is a significant kink in the vortex field at the string core which is reduced in the gauged model and the inclusion of the gauge fields also act to increase the vorton radius.}
    \label{fig: pD vs pE vortons}
\end{figure}

Figure \ref{fig: BS global} displays the global vorton solution with $Q=9000$, $N=10$ and $R=25.7$ in parameter set D that was constructed in \cite{Battye2009a} (although using rescaled parameters). This is in good agreement with the previously constructed vorton with $R=15.5$, after rescaling lengths by the required factor of $\sqrt{3}$ due to the rescaling of $\lambda_\phi=3$ to $\lambda_\phi=1$. We also present a vorton in a gauged extension of this model with $G=0.1$ (parameter set E) in figure \ref{fig: BS gauged} - it has the same charge and winding number but a larger radius of $R=32.5$.

Both of these vortons correspond to strings on the higher charge branch. In the global case, a string with $q_p = 38$ predicts the existence of a vorton with $N=10$, $Q=8998.5$ and $R=25.1$ while in the gauged case, a string with $q_p = 33.38$ predicts the existence of a vorton with $N=10$, $Q=9000.5$ and $R = 33.19$. There are no strings that satisfy $\omega<\omega_c$ - see equation (\ref{eq: phase freq cond}) - on the lower charge branch which indicates that there are no vortons that can be constructed with strings on the lower charge branch in this parameter set. Therefore, we should not expect any fully stable vortons due to the inevitable longitudinal instability experienced by the higher charge strings.

In fact, the global vorton was found to be unstable to both square and triangular modes in \cite{Battye2009a} which is not the pinching instability expected when $c_L^2<0$. We expect that this mode simply had a larger growth rate of instability than the pinching mode and that larger vortons will be destroyed by the pinching instability instead, as the growth rate for the square and triangular modes is inversely proportional to $R$ and the growth rate for the pinching instability is independent of $R$. We will discuss this in more detail and test our prediction in a subsequent paper on the pinching instability.

\subsubsection{Comparisons to Lemperiere \& Shellard [\protect\hyperlink{page.46}{18}]}

\begin{figure}[t]
    \centering
    \includegraphics[trim={0.5cm 0 1.5cm 1.4cm},clip,scale=0.7]{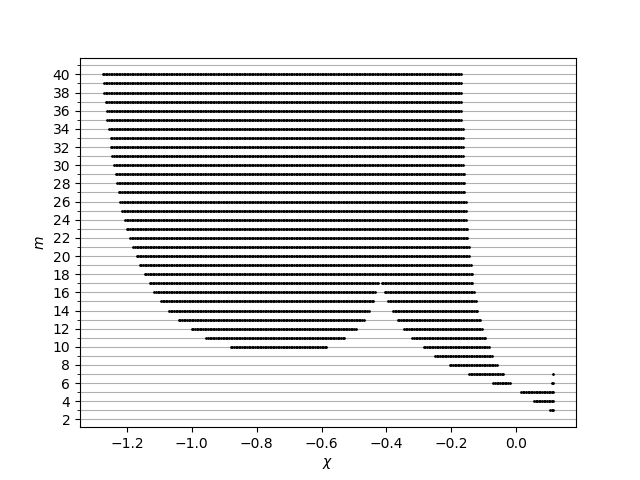}
    \caption{The predicted regions of instability for a model with $\eta_\phi=1$, $\eta_\sigma=0.35$, $\lambda_\phi=1$, $\lambda_\sigma=36$, $\beta'=6.6$, $G=0.2$ (parameter set F) and a modified interaction term $V_\text{int} = \beta'|\sigma|^2|\phi|^6$. There isn't a clear improvement in the stability of the predicted vorton, but this is a difficult comparison to make because the modified interaction term has drastically changed the range of $\chi$ for which there are superconducting solutions.}
    \label{fig: mod instabilities}
\end{figure}

In \cite{Lemperiere2003b} a global vorton was constructed in a model with a modified interaction term, $\beta|\phi|^2|\sigma|^2 \to \beta'|\phi|^6|\sigma|^2$. This was done to increase the strength of the potential seen by the condensate and make it more difficult for the condensate to split off from the string. They claim to have used the thin string approximation to construct a global vorton and additionally found that it was stable to the $n = 2$ (elliptical) mode. We have been unable to reproduce their global vorton, but we have found vortons, using the TSA, in a gauged version of their parameter set - which we call parameter set F. This is very similar to parameter set A except that it uses the modified potential. It is perhaps not surprising that there are difficulties in the global case as the strings are less localised than in the gauged case. The string mass per unit length is logarithmically divergent so any predictions will depend upon the cut-off applied during integration. Nevertheless, we have managed to use the TSA to construct global vortons in parameter set D so it is not clear why we could not manage it here. The modified potential does have the effect of widening the condensate, thereby making the string less localised (the core width and the width of the condensate have approximately doubled when compared to parameter set A), so perhaps the effects of the global limit and the modified potential combine to decrease the accuracy of the TSA enough to cause problems with our energy minimisation algorithm. Unfortunately, there does not appear to be sufficient information (e.g. the charge, $Q$ and the winding number, $N$, of their vorton) in \cite{Lemperiere2003b} to be completely sure.

In Figure \ref{fig: mod instabilities} we show the intervals of instability in parameter set F. There does not seem to be an obvious improvement in the predicted stability of the vorton, compared to parameter set A. However, this may be largely due to the drastic shift in the allowed range of $\chi$. It should be noted that there is a region of stability near the chiral state, but this is a generic feature of near-chiral strings and not necessarily a result of the modified interaction.

\subsection{Radial Dynamics}

Vortons constructed via energy minimisation in this way should be approximately stationary solutions of the equations of motion. This can be tested with dynamical simulations that evolve the system under the full equations of motion. As an initial test, in this section we will continue to impose axial symmetry and, therefore, test the stability to radial perturbations. In section \ref{sec: 3d} we will discuss non-radial perturbations whose stability is a more stringent test. We use initial field configurations that are the result of the gradient flow algorithm discussed in section \ref{sec: construction and dynamics} for parameter sets A and B. At the initial time step, the phase of the condensate field is rotated by $\omega \Delta t$, and all other fields are left the same. We use the value of $\omega$ calculated during gradient flow here, not the predicted value from the straight string analysis which is slightly different. After this, the system is evolved under the equations of motion, but with radial symmetry imposed. We use the same grid spacing and grid size that was used for gradient flow, but we change the time step to $\Delta t = 0.1\Delta x$ since the CFL condition is significantly weaker, $\Delta t \sum_{i=1}^d \Delta x_i^{-1} \lesssim 1$.

\begin{figure}[t]
    \centering
    \subfloat[Evolution of the radius for a static vorton.]{
        \centering
        \includegraphics[trim={0.4cm 0cm 1cm 1cm},clip,width=0.46\linewidth]{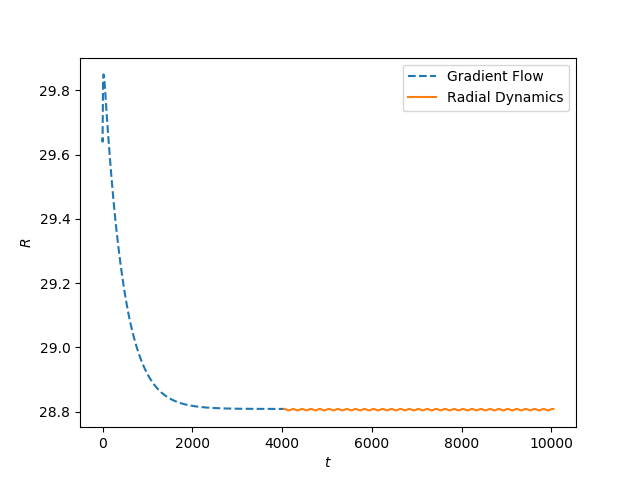}
        \label{fig: radius over time a}
    }\hspace{1em}
    \subfloat[Evolution of the radius for an oscillating vorton.]{
        \centering
        \includegraphics[trim={0.4cm 0 1cm 1cm},clip,width=0.46\linewidth]{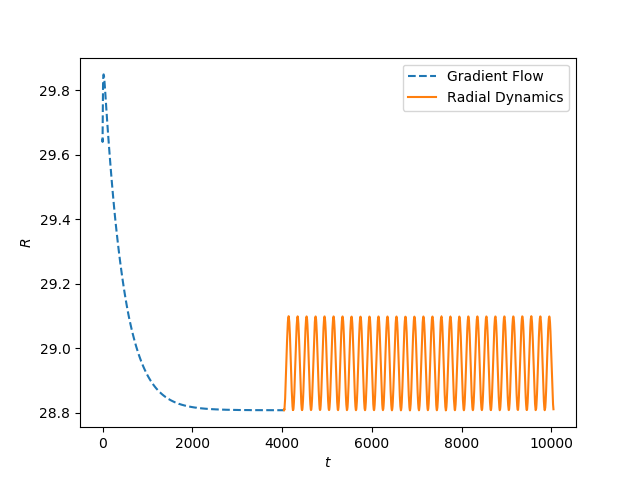}
        \label{fig: radius over time b}
    }
    \\
    \subfloat[Evolution of the energy for a static vorton.]{
        \centering
        \includegraphics[trim={0.4cm 0 1cm 1cm},clip,width=0.46\linewidth]{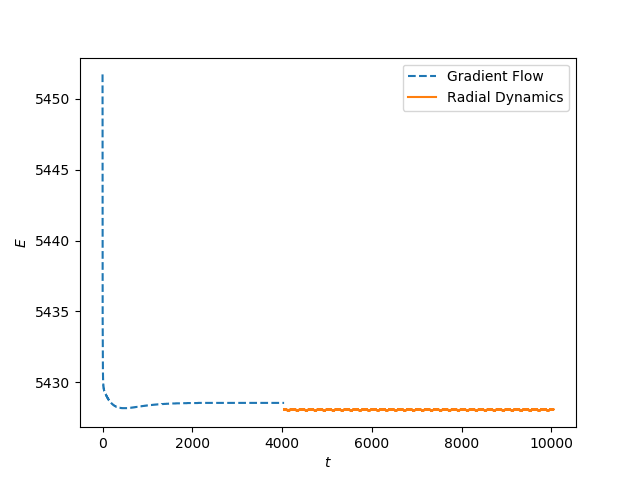}
        \label{fig: energy over time a}
    }\hspace{1em}
    \subfloat[Evolution of the energy for an oscillating vorton.]{
        \centering
        \includegraphics[trim={0.4cm 0 1cm 1cm},clip,width=0.46\linewidth]{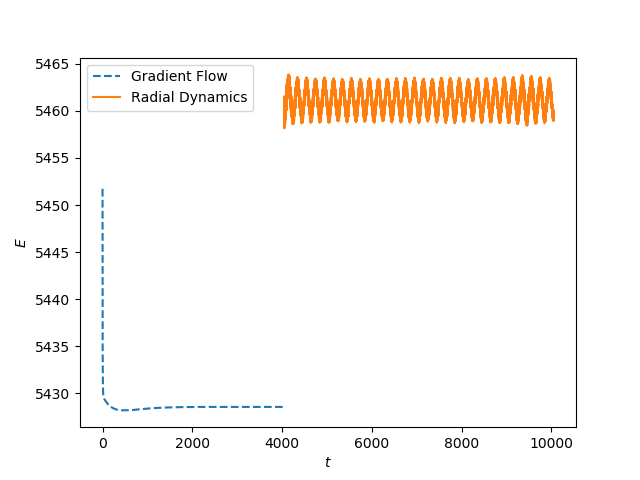}
        \label{fig: energy over time b}
    }
    \caption{The evolution of the radius is shown in Figures \ref{fig: radius over time a} and \ref{fig: radius over time b} while the evolution of the energy is shown in Figures \ref{fig: energy over time a} and \ref{fig: energy over time b} during both the gradient flow process and full radial dynamics simulations. This vorton is the one displayed in Figure \ref{fig: LS vorton profile} (parameter set A). The static vorton oscillates slightly around its equilibrium radius, while the perturbed vorton oscillates more dramatically around a larger equilibrium radius. There is a very small reduction in the energy of the static vorton due to the numerical approximations. The perturbed vorton has a significantly increased energy and it slightly oscillates although at far less than the percent level.}
    \label{fig: E and R over time}
\end{figure}

Figure \ref{fig: E and R over time} shows the evolution of the radius and energy for the static vorton solution presented in Figure \ref{fig: LS vorton profile} and also one in which we have artificially increased the initial phase frequency by $1\%$ (from $\omega = 1.99$ to $\omega = 2.01$) to create a larger oscillation about a larger energy minimising radius, similar to what was presented for global vortons in \cite{Battye2009a} with parameter set D. We calculate the radius by finding where $|\phi| = 0$ along the $y=z=0$ slice. The energy in Figure \ref{fig: energy over time b} is increased due to the smaller phase frequency and does not remain exactly constant during radial dynamics due to numerical effects, but it is only a variation of less than $\pm0.1\%$ about the average energy. The static vorton also has a slight oscillation which is due to the phase frequency being treated exactly in gradient flow, while it is approximated with finite difference operators in the dynamical code. This is visible in Figure \ref{fig: energy over time a} as a reduction in the energy of less than $0.01\%$. The violation of the constraint equation remains very small during the radial time evolution - see appendix \ref{sec: lattice gauge theory} for more details.

\begin{figure}[t]
    \centering
    \subfloat[$\eta_\sigma=0.35$, $\lambda_\sigma=36$, $\beta=6.6$ and $G=0.2$ (parameter set A) with $Q/N=14.98$ kept constant.]{
        \centering
        \includegraphics[trim={0 0 1.5cm 1cm},clip,width=0.46\linewidth]{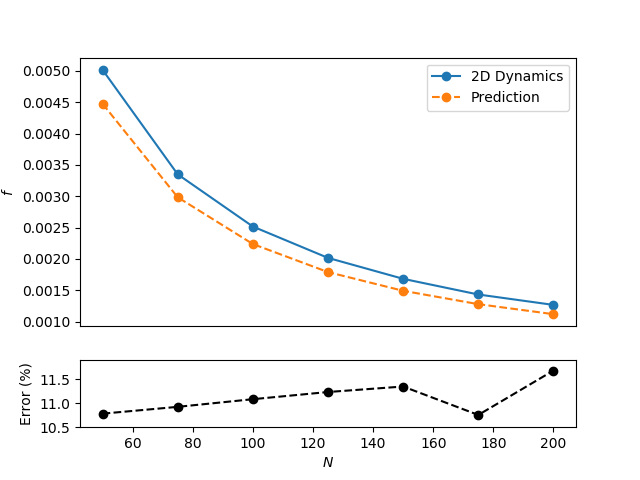}
    }\hspace{1em}
    \subfloat[$\eta_\sigma=0.61$, $\lambda_\sigma=10$, $\beta=3$ and $G=0.5$ (parameter set B) with $Q/N=31.89$ kept constant.]{
        \centering
        \includegraphics[trim={0 0 1.5cm 1cm},clip,width=0.46\linewidth]{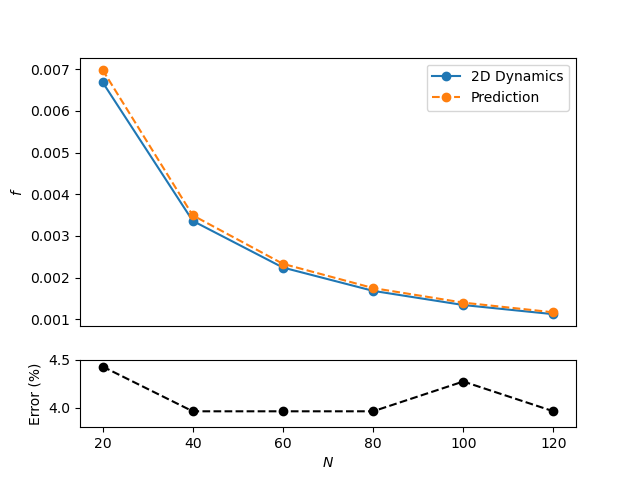}
        \label{fig: CB zero modes}
    }
    \caption{A comparison of the predicted zero mode frequencies to the zero mode frequencies calculated from the radial oscillations produced when the dynamics of vortons is simulated with axial symmetry imposed. This provides a good quantitative test of the thin string approximation. We have used the frequency defined by $f_m = \Omega_m/2\pi$.}
    \label{fig: zero_mode}
\end{figure}

All vortons that we have tested are stable to axially symmetric dynamics which suggests that the $n=0$ mode is always stable, as predicted by the thin string analysis. In figure \ref{fig: zero_mode} we compare the predicted frequency of the zero mode to the frequency of the radial oscillations during our simulations. This shows a good agreement between the predictions and simulated dynamics, although with a larger percentage error than the predicted radius and no improvement for larger vortons. This appears to be a good quantitative test of the thin string approximation. It should be noted that for the calculation of these frequencies we simulated the dynamics up to $t=60000$ and $t=100000$ respectively and, therefore, the frequency resolutions are $\Delta f = 1.67\times 10^{-5}$ and $\Delta f = 10^{-5}$, respectively.

\subsection{3D Dynamics} \label{sec: 3d}

\begin{figure}[t]
    \centering
    \subfloat[$t=0$]{
        \centering
        \includegraphics[trim={5cm 3cm 4cm 7cm},clip,width=0.32\linewidth]{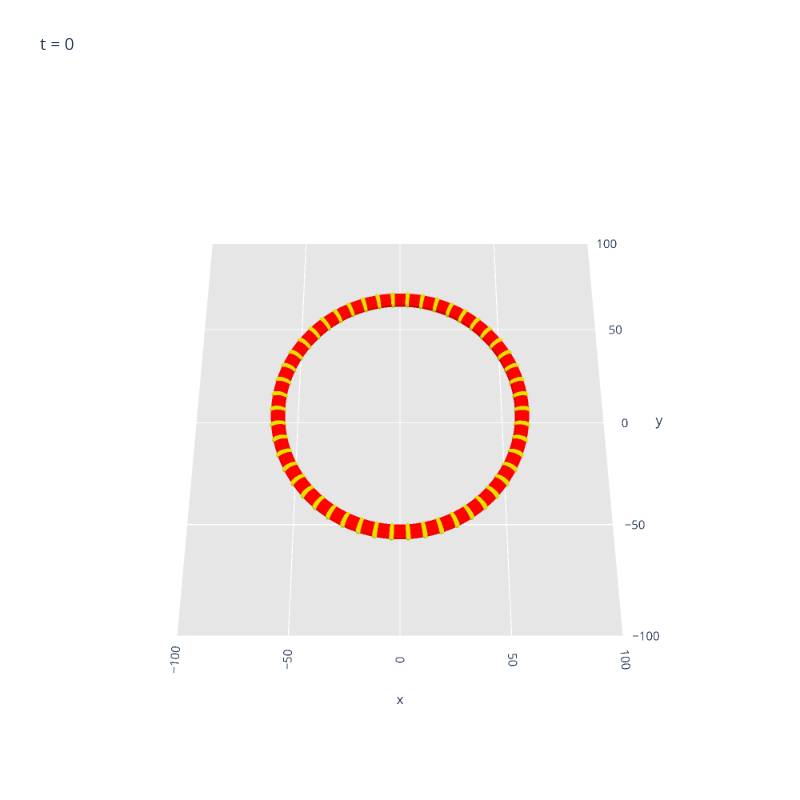}
    }
    \subfloat[$t=1500$]{
        \centering
        \includegraphics[trim={5cm 3cm 4cm 7cm},clip,width=0.32\linewidth]{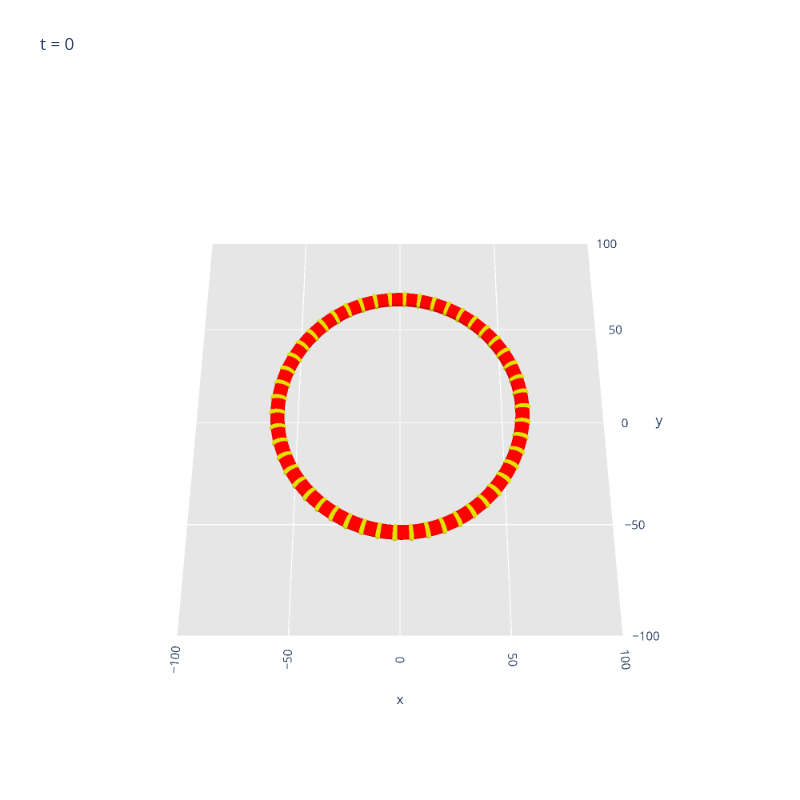}
    }
    \subfloat[$t=3000$]{
        \centering
        \includegraphics[trim={5cm 3cm 4cm 7cm},clip,width=0.32\linewidth]{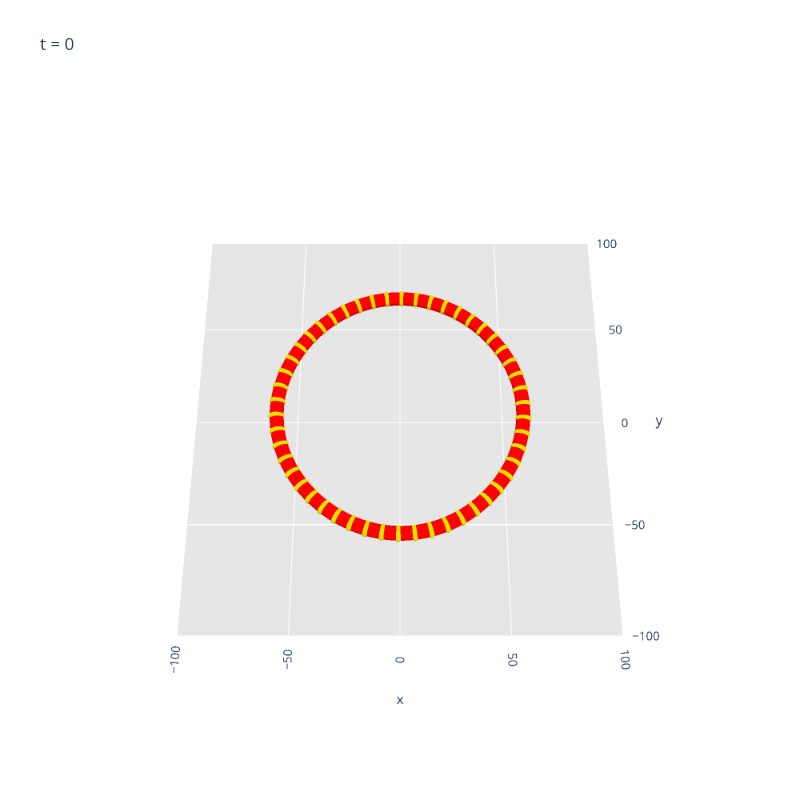}
    }
    \caption{Isosurfaces of a vorton with $Q = 1594$ and $N = 50$ in the parameter set $\eta_\sigma = 0.61$, $\lambda_\sigma = 10$, $\beta = 3$ and $G = 0.5$ (parameter set B). $|\phi|=\frac{3}{5}$ is shown in red and Re$(\sigma)=\frac{1}{5}\eta_\sigma$ is shown in yellow. There are no signs of any instabilities, but this does not yet indicate that the vorton is completely stable - only that the modes it could be unstable to are not excited in this simulation, or that the growth rate is small enough so that the instability has not significantly grown by the end of the simulation.\vspace{2em}}
    \label{fig: CB isosurfaces}
\end{figure}

\begin{figure}[t]
    \centering
    \subfloat[$t=900$]{
        \centering
        \includegraphics[trim={5cm 3cm 4cm 7cm},clip,width=0.32\linewidth]{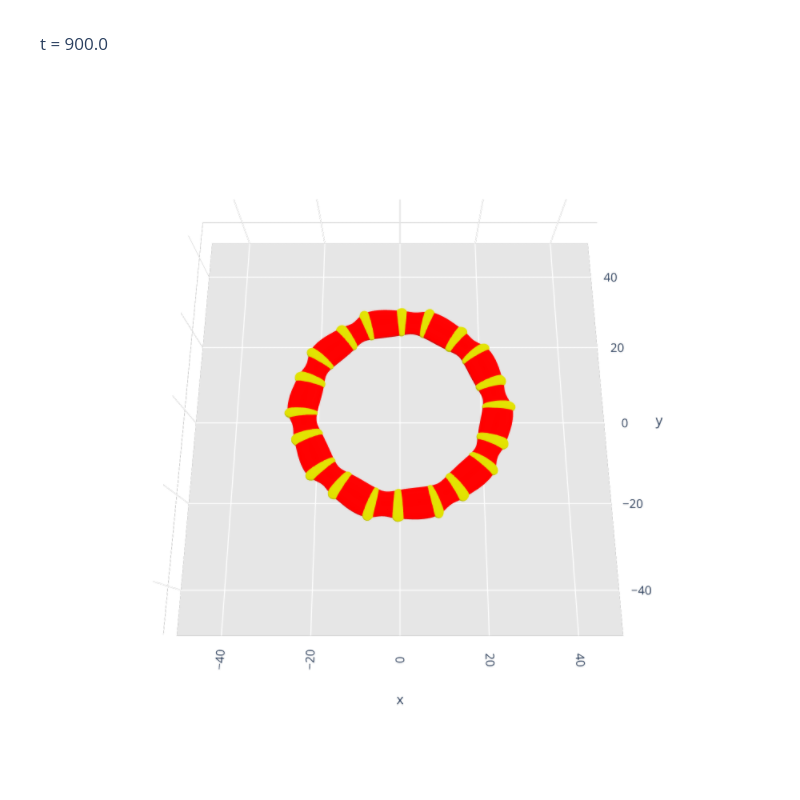}
    }
    \subfloat[$t=950$]{
        \centering
        \includegraphics[trim={5cm 3cm 4cm 7cm},clip,width=0.32\linewidth]{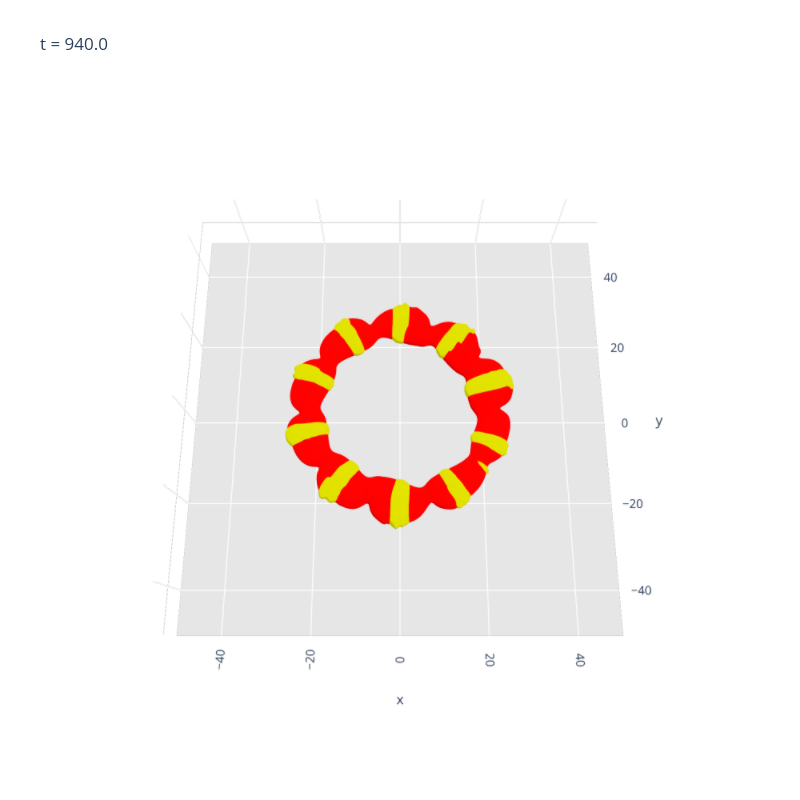}
    }
    \subfloat[$t=1000$]{ \label{fig: bubble}
        \centering
        \includegraphics[trim={5cm 3cm 4cm 7cm},clip,width=0.32\linewidth]{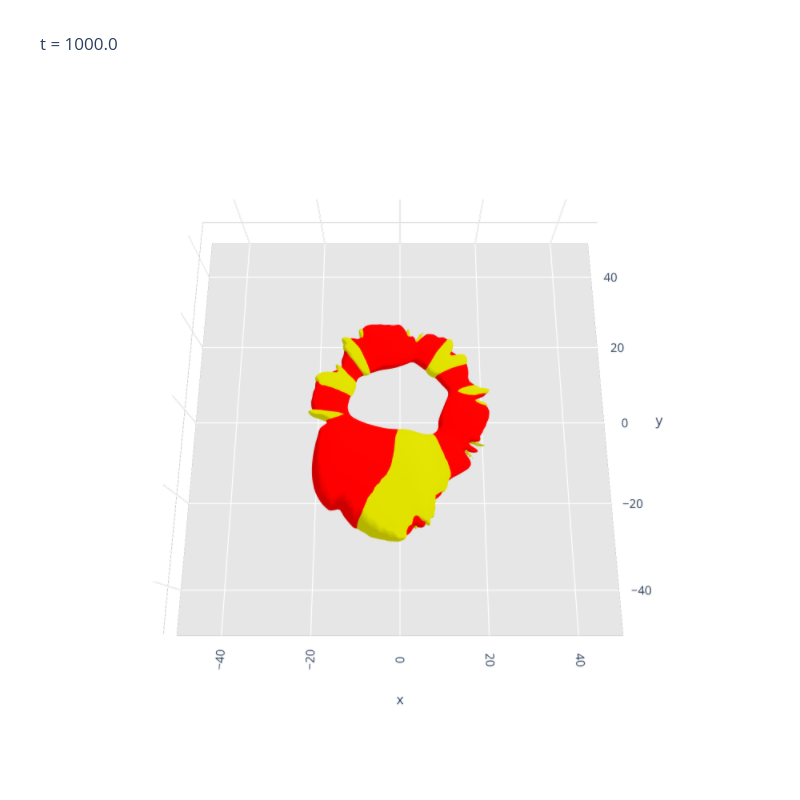}
    }
    \caption{Isosurfaces of a vorton with $Q = 637.7$ and $N = 20$ in the parameter set $\eta_\phi = 1$, $\eta_\sigma = 0.61$, $\lambda_\phi = 1$, $\lambda_\sigma = 10$, $\beta = 3$ and $G = 0.5$ (parameter set B). $|\phi|=\frac{3}{5}\eta_\phi$ is shown in red and Re$(\sigma)=\frac{1}{5}\eta_\sigma$ is shown in yellow. We have excited all of the modes between $m=2$ and $m=30$ with an amplitude of 0.001. At $t \sim 950$ the pinching instability to the $m=10$ mode becomes clear. There may be instabilities due to additional modes which will not appear because the rate of growth is smaller. The vorton is ultimately destroyed by the pinching instability - with a ten-fold symmetry - developing into a bubble of the true vacuum, as seen in Figure \ref{fig: bubble}, which quickly expands. Note that it is visually very different to the vorton destruction mechanism seen later in figure \ref{fig: LS isosurfaces}.}
    \label{fig: pert CB isosurfaces}
    \vspace{1em}
\end{figure}

\begin{figure}[t]
    \centering
    \subfloat[$t=0$.]{
        \centering
        \includegraphics[trim={5cm 3cm 4cm 7cm},clip,width=0.32\linewidth]{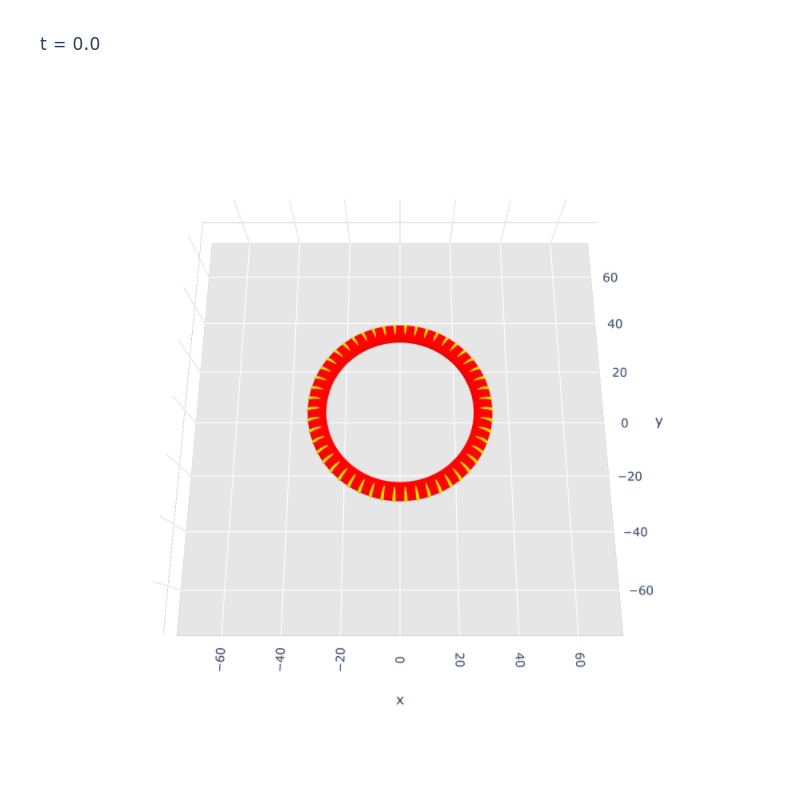}
    }
    \subfloat[$t=300$.]{
        \centering
        \includegraphics[trim={5cm 3cm 4cm 7cm},clip,width=0.32\linewidth]{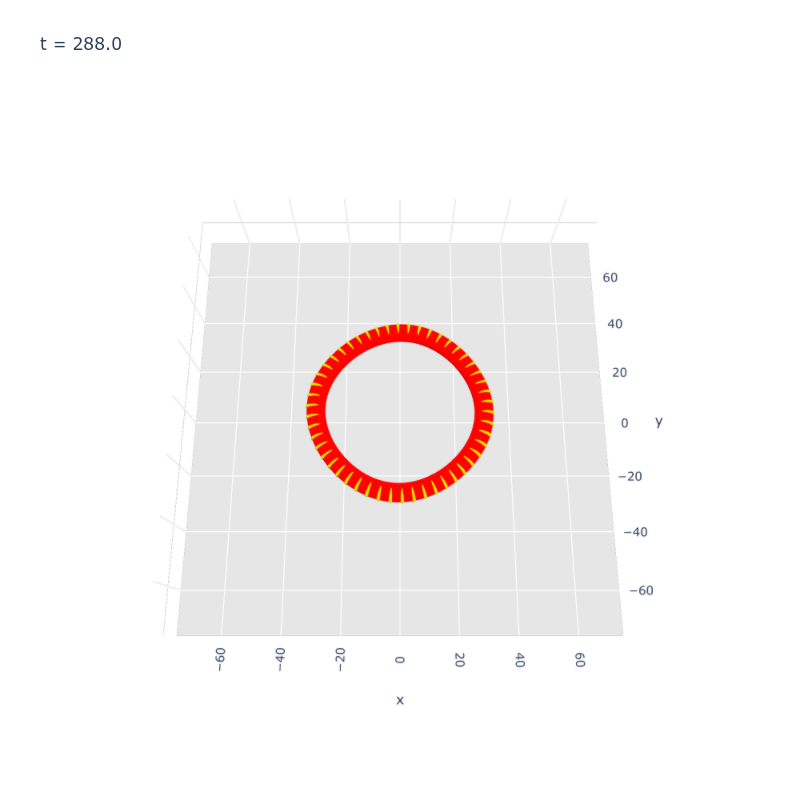}
    }
    \subfloat[$t=600$.]{
        \centering
        \includegraphics[trim={5cm 3cm 4cm 7cm},clip,width=0.32\linewidth]{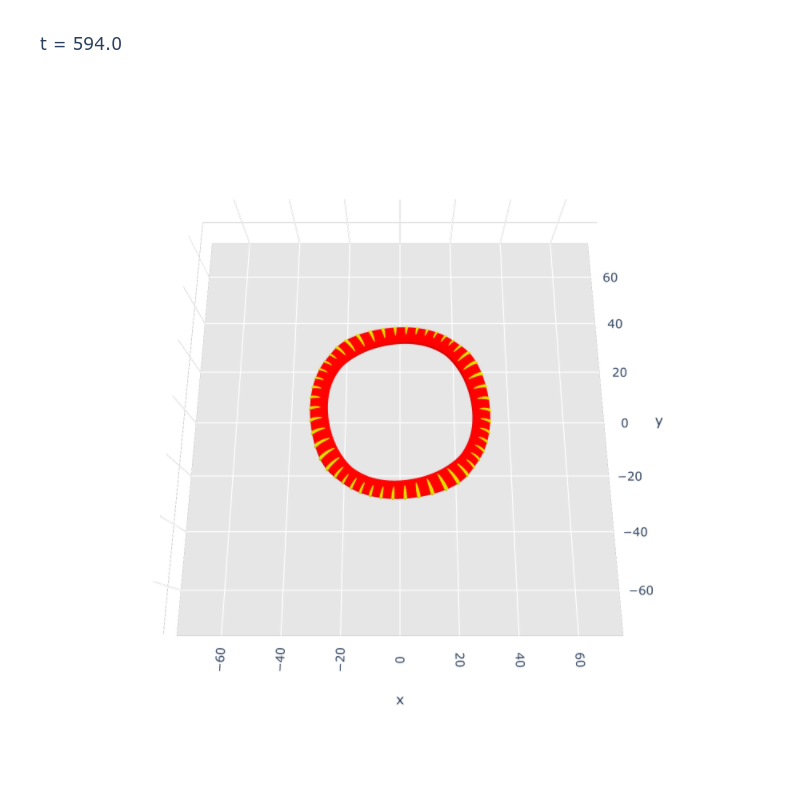}
    }
    \\
    \subfloat[$t=780$.]{
        \centering
        \includegraphics[trim={5cm 3cm 4cm 7cm},clip,width=0.32\linewidth]{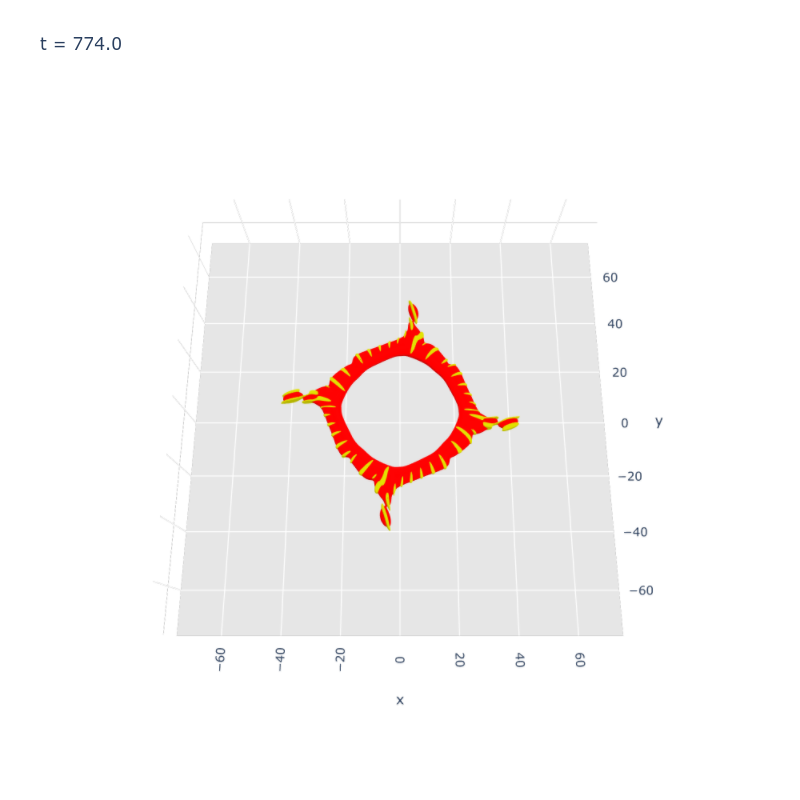}
    }
    \subfloat[$t=810$.]{
        \centering
        \includegraphics[trim={5cm 3cm 4cm 7cm},clip,width=0.32\linewidth]{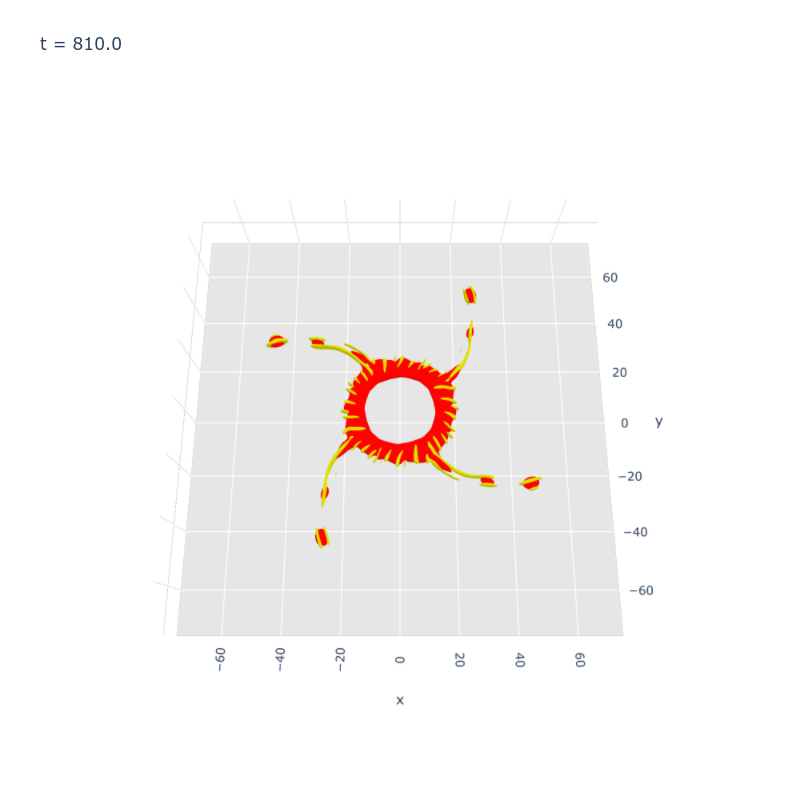}
    }
    \subfloat[$t=840$.]{
        \centering
        \includegraphics[trim={5cm 3cm 4cm 7cm},clip,width=0.32\linewidth]{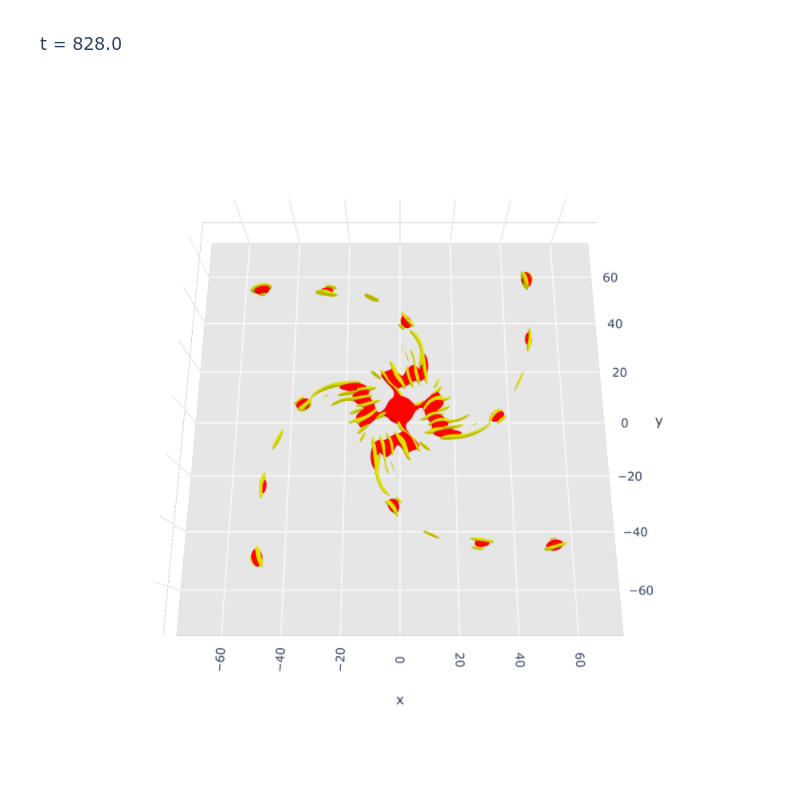}
    }
    \caption{Isosurfaces of a vorton with $Q = 749$ and $N = 50$ in the paramater set $\eta_\sigma = 0.35$, $\lambda_\sigma = 36$, $\beta = 6.6$ and $G = 0.2$ (parameter set A). $|\phi|=\frac{3}{5}\eta_\phi$ is shown in red and Re$(\sigma)=\frac{1}{5}\eta_\sigma$ is shown in yellow. By the end of the simulation, it is clear that the vorton is unstable to square perturbations, probably sourced by the discretized grid, and that this instability eventually destroys the vorton. The development of an unstable square mode is similar to what is seen in parameter set D in \cite{Battye2009a}. \vspace{-1em}}
    \label{fig: LS isosurfaces}
\end{figure}

In order to test our predictions for the stability to non-axial perturbations and the oscillation frequencies of higher order modes, a fully three-dimensional simulation is necessary. To achieve this we can either interpolate the solution found using the cartoon method to create a three dimensional Cartesian grid solution, or use the cut-off method for which extending the solution to 3D is trivial. Care must be taken with the lattice spacing used so that the variation due to the winding of the condensate is accurately resolved. The dynamics in 3D are significantly more numerically demanding and, as such, our simulations are run over a shorter time period than the radial dynamics. Some videos of the vortons shown in Figures \ref{fig: CB isosurfaces}, \ref{fig: pert CB isosurfaces} and \ref{fig: LS isosurfaces} can be viewed in the ancillary files.

\begin{figure}
    \centering
    \includegraphics[scale=0.7]{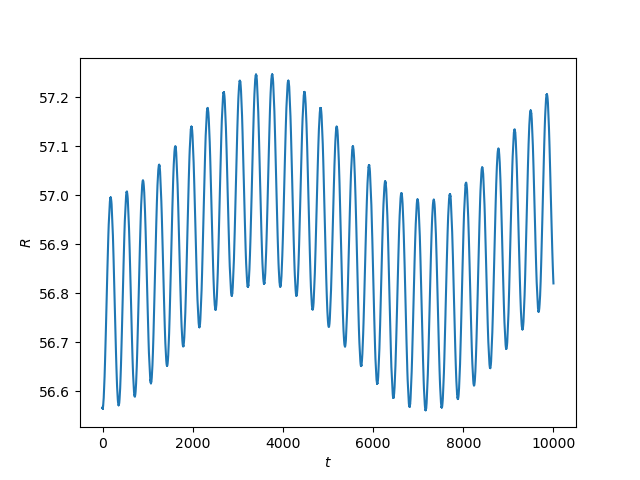}
    \caption{The evolution of the position of the core of the string along $y=z=0$ during 3D dynamics. There is no evidence of any growing frequencies of oscillation. There are two main frequency components, a high frequency component which agrees well with the prediction for the zero mode frequency and another unexplained, low frequency oscillation, which we believe to be caused by curvature effects that have induced a small departure from the expected third frequency of $f_0=0$.}
\label{fig: CB 3d radius}
\end{figure}

Initially, we will consider parameter set B and discuss the fully stable, chiral vorton that we presented in \cite{PhysRevLett.127.241601}. In Figure \ref{fig: CB isosurfaces} we show the isosurfaces of the vorton shown in figure \ref{fig: CB vorton profile} ($Q = 1594$, $N = 50$ and $R = 56.57$) at a few snapshots. During the simulation, energy was conserved to within less than $0.1\%$ and the average violation of the gauge condition reaches a maximum of  $\sim 10^{-3}$. No obvious instability is apparent in these plots and since we expect that only the square mode (due to the grid) and the zero mode (due to the initial numerical solution not being exactly perfect) are excited in this simulation it supports the prediction of Figure \ref{fig: CB full instability intervals} that the square mode is stable - although of course it only places an upper limit on the growth rate. We have run the simulation until $t=10000$ which is longer than any other presently published in the literature. The current completes around $28$ full rotations during this period. Figure \ref{fig: CB 3d radius} shows how the radius of the vorton changes over time and demonstrates that it is stable for a long period, with no evidence for any instability. The Fourier transform shows that this is predominantly the superposition of a radial oscillation, $f_0 = \num{2.8e-3}$, (with $f_0 = \num{2.79e-3}$ predicted by the thin string approximation) and another, unexplained, low frequency component of $f \approx \num{1.4e-4}$ (note that this is calculated by eye rather than by the Fourier transform as the value is very close to the resolution limit). We think that this unexplained component is a perturbation away from $f_0=0$ that is caused by curvature effects. This is supported by the fact that this frequency decreases much faster than $1/R$ as the radius is increased. For example the $N=20$ version of this vorton (with a smaller radius by a factor of $\sim2/5$) has frequency components $f_0 = \num{7.3e-3}$ and $\num{2.7e-3}$ -  the frequency that is not predicted by the TSA. The former has increased by $\sim5/2$ as expected from the $1/R$ scaling, while the latter is an order of magnitude larger than $1/R$ scaling would predict.

We can test the stability to other (non-axial) modes by applying a perturbation to the initial field configuration and then dynamically evolving, as before. We do this by making the modification, $\sigma \to \sigma(1 + \epsilon\sin(m\theta))$, where $\epsilon$ is the amplitude of the perturbation and $m$ is the mode we wish to excite \cite{Battye2009a}. We have used this to test the prediction that this vorton solution is stable, as predicted in Figure \ref{fig: CB full instability intervals}. We artificially excite, individually, the modes between $m = 2$ and $m = 10$, each with an amplitude of $\epsilon = 0.1$, and evolve the system up to $t=3000$ using our 3D dynamics code.

\begin{figure}[t]
    \centering
    \includegraphics[width=\linewidth, height = 9.5cm]{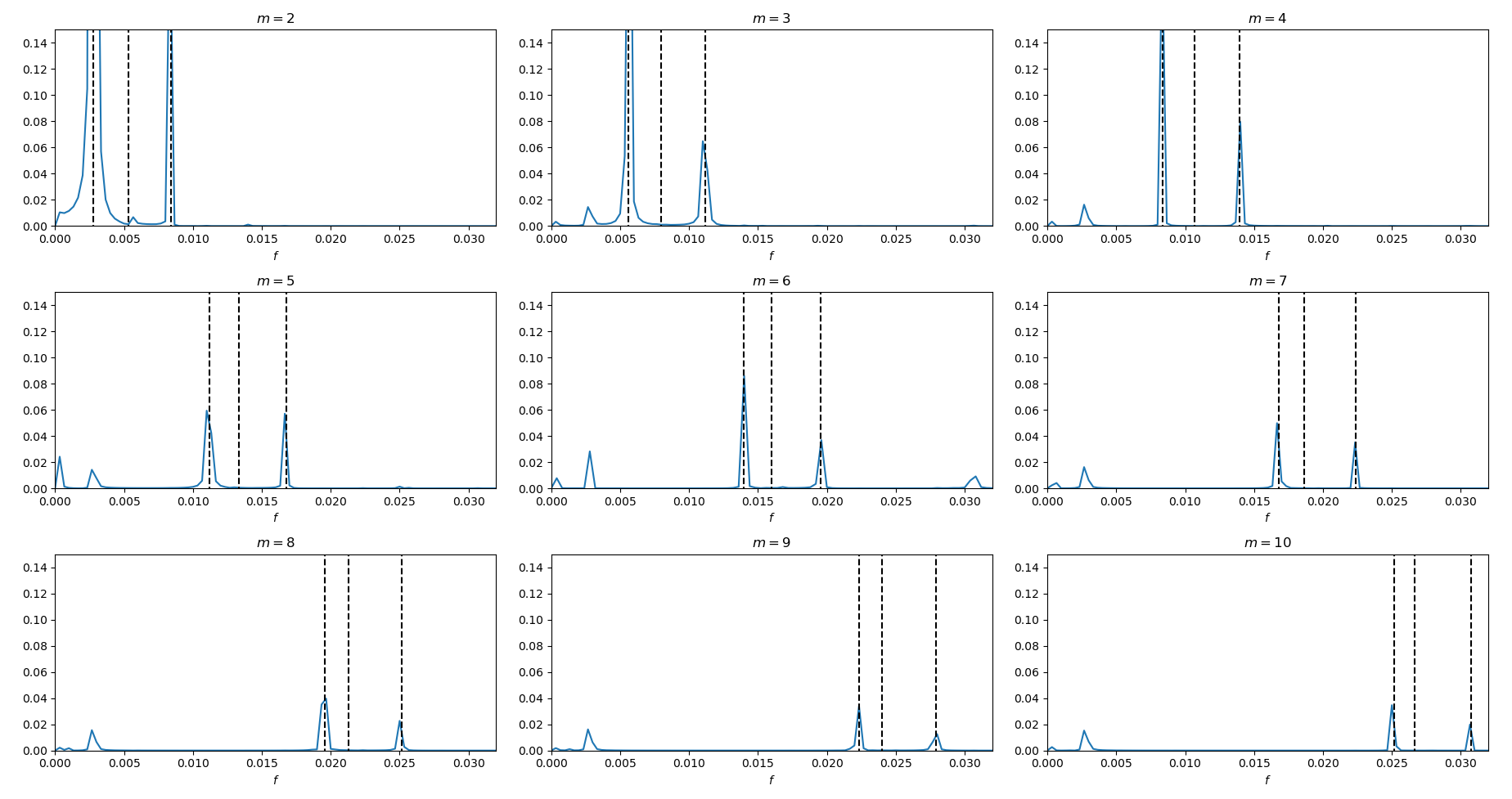}
    \caption{A comparison of the dominant frequency components to the predicted frequencies of oscillation for modes $2$ to $10$, all excited with $\epsilon = 0.1$, in parameter set B. They show a very good level of agreement for the higher and lower frequencies, with the intermediate frequency not appearing because it is very close to a purely longitudinal oscillation for vortons near the chiral limit.}
\label{fig: fourier transforms}
\end{figure}

We use these simulations to test our predictions of the frequencies of oscillation, as presented in Figure \ref{fig: fourier transforms}. Clearly, the predictions of the smallest and largest frequencies of oscillation are very good, but it is interesting that the intermediate frequency either has a much smaller peak, or doesn't appear to be there at all. We believe that this is because the frequency is associated with almost purely longitudinal oscillations, which are not picked up by the position of the string, and there is only a very small coupling to transverse oscillations. This effect is caused by the vorton being close to the chiral limit and we expect that it wouldn't occur, in general, for the rest of the parameter space. Near the chiral limit, there is an eigenvector solution with a transverse component that is very small (and two much larger, approximately equal, longitudinal components) when the associated frequency is $\nu_m \approx m(c_T^2 - c_L^2)/(1-c_T^2c_L^2)$ - see equation (\ref{eq: stability matrix}). Note also that we have performed similar simulations for the $N=20$ vorton and we get a similar level of agreement.

All of the modes that we excited on the $N=50$ vorton were stable during this time period, except for $m=6$, which develops a pinching instability that destroys the vorton at $t\sim2700$. This pinching instability does not exist in corresponding simulations of straight strings with periodic boundary conditions, indicating that it is caused by curvature effects. As such, we might expect that larger vortons, with the same ratio of $Q/N$, will experience a weaker pinching instability, or possibly none at all, while smaller vortons will be less stable. Our simulations suggest that this is exactly what happens, and actually the $N=20$ vorton is unstable to many modes, and it is destroyed by these instabilities much earlier.

% \begin{figure}[t]
%     \centering
%     \includegraphics[scale=0.7]{Plots/CB_m2to30_3D_radius.png}
%     \caption{The evolution of the position of the core of the string along $y=z=0$ during 3D dynamics in parameter set B. We have cut the plot off at $t=945$ after which our method for determining the radius becomes unreliable. A growing mode is clearly visible and shortly after $t\approx950$ the vorton is destroyed.}
% \label{fig: pert CB 3d radius}
% \end{figure}

To illustrate, we have performed a simulation of the $N=20$ vorton and perturbed all the modes between $m=2$ and $m=30$ at once, all with the amplitude $\epsilon = 10^{-3}$. From the isosurfaces in Figure \ref{fig: pert CB isosurfaces} it is very clear that there is a pinching instability to the $m=10$ mode, which didn't appear in the case of the $N=50$ vorton, and the vorton is destroyed by $t\sim1000$ which is much earlier than for $N=50$. The perturbations grow in a way that is visually very different from extrinsic instabilities, which we will present later in Figure \ref{fig: LS isosurfaces}. The instability clearly manifests itself as oscillations in the width of the string as opposed to distortions in the shape of the vorton. Notice that the instability causes the condensate to unwind (from $N=20$ initially, to $N=10$ by $t=950$) which then results in the collapse of the vorton. By individually exciting the modes, we can also see that $m=10$ is not the only one that has a pinching instability - in fact all of the modes from $m=2$ up to at least $m=20$ are unstable and the vorton is destroyed before $t=3000$, often much earlier. The growth rate is simply largest for $m=10$ which causes it to dominate.

\begin{figure}[t]
    \centering
    \includegraphics[scale=0.7]{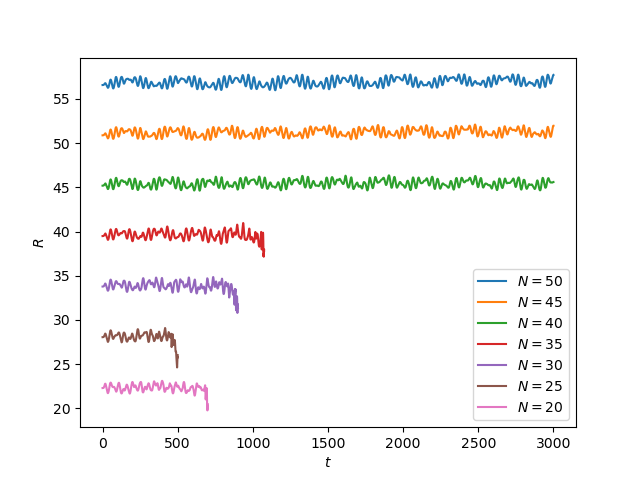}
    \caption{The evolution of the position of the core of the string along $y=z=0$ for vortons with different values of the winding number, $N$, in parameter set B, with the ratio of $N$ to $Q$ kept fixed, and using a fourth order stencil for the spatial derivatives - higher than the previous simulations. The $N=20$ vorton is the same one shown in Figure \ref{fig: pert CB isosurfaces}, although here it is excited with only the $m=4$ mode with an amplitude of $0.1$. All the vortons in this plot are excited with the $m=N/5$ mode (so that the wavenumber $m/R$ remains roughly constant) with an initial amplitude of $0.1$. Note that the bottom four plots have been cut off at roughly the point where the vortons are destroyed, while the top three survive until the end of the simulation at $t=3000$.}
\label{fig: radius evol same p}
\end{figure}

Since the pinching instability is an internal instability (not dependant upon the positional oscillations of the string for small curvatures), we should expect that, if the pinching instability is not caused by curvature effects, then vortons of different radii will be destroyed at roughly the same time. Instead, we have shown in Figure \ref{fig: radius evol same p} that the pinching instability either no longer exists, or at least takes significantly longer to develop for the vortons with $N\geq40$. This indicates that the pinching instability disappears when curvature effects become negligible. Note that it is the wavelength of the perturbation that is important for the sake of comparison, which means that we need to increase $m$ proportionally with $N$ if we are to examine the same instability.

In simulations of straight strings with periodic boundary conditions, we find that it is necessary to be very careful with the resolution because spurious pinching instabilities can appear in simulations with larger grid spacings, and the same problem applies to vortons as well. We have performed thorough convergence tests that confirm that the instabilities in our simulations are real effects and have also upgraded to a fourth order stencil for the simulations in Figure \ref{fig: radius evol same p}. Pinching instabilities can therefore be categorised as being caused by insufficient resolution (not a real instability), curvature effects which stop being relevant for larger vortons or instabilities in the underlying straight string solutions - which will be examined in more detail in an upcoming paper.

The evidence in Figure \ref{fig: radius evol same p} suggests that the slight $m=6$ pinching instability that the $N=50$ vorton suffers from is a left-over effect of curvature and will disappear, like the instabilities to the other modes did, at larger radii. Our simulations of a vorton with $N=60$ confirm these expectations as the modes $m=6$,$7$ and $8$ with $\epsilon=0.1$ are all stable up to $t=3000$. Similarly, the vorton shows no signs of instability when all the modes between $m=2$ and $m=30$ are excited with $\epsilon = 10^{-3}$, up to $t=3000$, whereas the $N=20$ vorton was destroyed by these perturbations at $t\sim 1000$. In addition, we have performed simulations that excite the $m=21$ and $m=22$ modes, which are the most likely to be unstable to extrinsic oscillations rather than the pinching instability (according to the TSA - see Figure \ref{fig: CB full instability intervals}), and there are no signs of an instability by $t=3000$. It is, therefore, very likely that larger vortons of this type will be fully stable.

Next, we will consider parameter set A which was predicted to be unstable to modes with $3\leq m \leq 6$ in Figure \ref{fig: LS instability intervals}. In Figure \ref{fig: LS isosurfaces} we show isosurfaces at six different snapshots in time when the vorton (shown in Figure \ref{fig: LS vorton profile} with $Q = 749$, $N = 50$ and $R = 28.8$) is evolved under 3D dynamics, with no perturbations applied, on a grid with $\Delta x = 0.3$, $-75\leq x \leq 75$, in all directions, and $\Delta t = 0.03$. The first set of three were chosen to display the instability to square modes ($m=4$) that was predicted, while the second set were chosen to show how the vorton is destroyed. We believe that a small excitation of this mode is produced either by small scale effects of discretisation onto a Cartesian grid or the boundary conditions imposed at the edges. A similar outcome was seen in \cite{Battye2009a} when evolving global vortons in parameter set D. Notice that this is clearly a different type of instability to that shown in Figure \ref{fig: pert CB isosurfaces} as there are only slight changes to the width of the string.

\begin{figure}[t]
    \centering
    \includegraphics[scale=0.65]{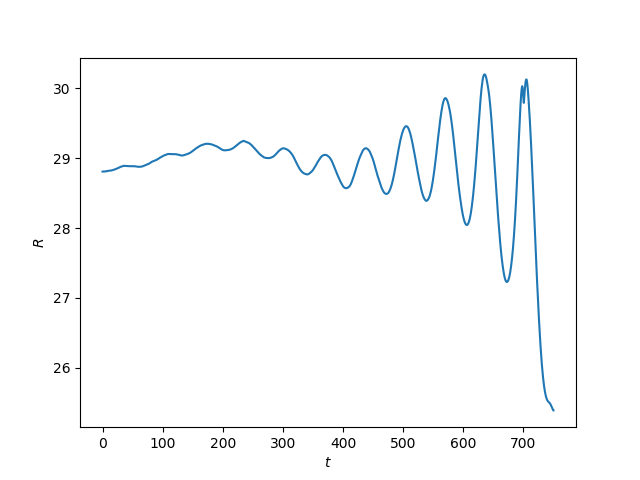}
    \caption{The evolution of the position of the core of the string along $y=z=0$ during 3D dynamics. We have stopped the simulation at $t=750$, at which point the unstable vorton begins to break apart and collapse. This plot is predominantly described by a superposition of the zero mode and an exponentially growing square mode.}
    \label{fig: LS 3d radius}
\end{figure}

 In Figure \ref{fig: LS 3d radius} we plot the radius of the vorton (the position of the core of the string along the $y=z=0$ slice) as a function of time which appears to be a superposition of a stable radial oscillation and the oscillation caused by the growing square mode. A spectral analysis of the radius evolution up to $t=750$ shows that the evolution of the radius is primarily composed of two frequencies, $f_0 = 0.003$ and $f_4 = 0.015$, which we assume to be the zero mode and (the growing) square mode frequencies respectively. For comparison, the predicted frequencies are $f_0 = 0.0045$ and $f_4 = 0.013$. It is interesting that the zero mode frequency is not the same in the 2D and 3D simulations. The frequency resolution here is $\Delta f = 1.3\times 10^{-3}$, comparable to the discrepancy between the simulation and predictions, and worse than in the 2D dynamics as we are only analysing the signal up to $t=750$, rather than to $t=60000$. This could be improved upon, in principle, by reducing the size of the initial excitation (caused by numerical approximations) so that the vorton survives for a longer period of time

\begin{figure}
    \centering
    \includegraphics[scale=0.65]{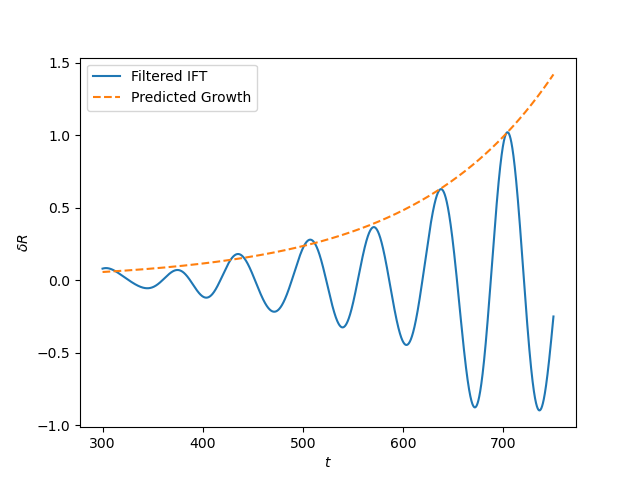}
    \caption{Comparison of the approximate oscillations caused by the square mode and the rate of growth predicted by the semi-analytic method, in parameter set A. The square mode was isolated by using a top hat function to select a small frequency range (that it lies within) and then performing an inverse Fourier transform, which has been cut off at early times when the square mode was not the dominant oscillation. The predicted growth of $Ae^{\text{Im}(\Omega_4)t}$ with $\text{Im}(\Omega_4) = 0.0072$ fits the envelope of this signal well, where $A$ is chosen so that the envelope intersects the largest peak.}
    \label{fig: growth of mode}
\end{figure}

 The imaginary part of the $m=4$ frequency provides an estimate for the rate of growth of the unstable mode. By filtering the Fourier transform to isolate only the square mode and then performing an inverse Fourier transform, the effect of this mode on the position of the string can be seen more clearly. We choose to filter with a top-hat function that is only non-zero when $0.01\leq f \leq 0.022$. The inverse Fourier transform is displayed in Figure \ref{fig: growth of mode} with the predicted growth - $Ae^{\text{Im}(\Omega_4)t}$ with $\text{Im}(\Omega_4) = 0.0072$ - overlaid. The initial amplitude of the mode, $A$, is chosen such that it intersects the largest peak. There is clearly a very good agreement between the predicted growth rate of the mode and the dynamical growth rate. We consider this to be a good quantitative test of the thin string model in the unstable regime.

\FloatBarrier
\section{Conclusions}

We have shown that it is possible to predict the existence and properties of vortons from straight, superconducting string solutions to a good degree of accuracy, albeit with a new constraint on the phase frequency that has not been previously recognised. We have used a gradient flow algorithm and techniques from lattice gauge theory to construct gauged vortons and simulate their dynamics. Using this, we have shown that the thin string approximation reliably predicts the frequencies and stability of each mode of oscillation. We have also found regions of the parameter space that admit completely stable vorton solutions, and have provided strong numerical evidence that they are indeed stable, once the vortons are large enough to sufficiently reduce the effects of curvature - as would be the case for vortons that are relevant in cosmology.

The thin string approximation and straight string analysis is a powerful tool for future studies on vortons because the parameter space can be explored with much more ease than by constructing individual vortons - which is both more numerically challenging and less general. For a given set of parameters the allowed vorton solutions can quickly be assessed though the following process:

\begin{itemize}
    \item Calculating the $\chi$ range for which there are superconducting string solutions. The lower limit will always be set by equation (\ref{eq: schrodinger}) which can be solved numerically after finding a non-superconducting string solution. The upper limit will either be set by condition (\ref{eq: mass max cond}), or by another limit, related to the existence of a lower energy vacuum state where the $U(1)_\sigma$ symmetry is broken and the $U(1)_\phi$ symmetry is unbroken, that must be determined by trial and error. In the electric regime, this process will only find one of two possible solutions - the one that has lower charge and energy - however this shouldn't be important when searching for stable vortons because we expect that the higher charge solutions will always be unstable to pinching instabilities.
    \item By sweeping through the $\chi$ range and solving equations (\ref{eq: 1D EoMs start} - \ref{eq: 1D EoMs end}), which is relatively quick and easy to do as they are just a system of 1D ODEs, a few useful integrated quantities can be calculated which allow for the properties of a vorton formed with each string solution to be determined. In particular, the required ratio of the winding number, $N$, to the Noether charge, $Q$, and the radius (for a specific choice of $N$ and $Q$) can be calculated.
    \item The solutions must also satisfy condition (\ref{eq: vorton formation constraint}) on the phase frequency, otherwise the condensate can delocalise from the string. This can also be assessed using the integrated quantities from the straight string solutions. This is everything that is required to construct vorton solutions.
    \item The stability properties of each vorton solution to oscillations in the position of the string can be investigated by calculating the transverse and longitudinal sound speeds of the straight string solutions. All tested parameter sets have supersonic sound speeds - meaning that the transverse speed is greater than the longitudinal speed. All loops with subsonic sound speeds are fully stable, while those with supersonic sound speeds, which are relevant for the case of vortons, have a complex structure of stability to different modes of oscillation. There are, however, still regions of complete stability - see Figure \ref{fig:cl_ct instability}. In particular, vortons that are very close to the chiral limit are typically predicted to be stable. The growth rate of unstable oscillations can also be predicted which allows for the typical lifetime of unstable vortons to be estimated.
    \item We will discuss the stability of vortons to pinching instabilities in more detail in a follow up paper, but we believe that all of the pinching instabilities presented here are caused by curvature effects. It should be noted that insufficient resolution can also cause spurrious pinching instabilities.
\end{itemize}

We have verified that the predictions made by the thin string approximation are in good quantitative agreement by constructing and simulating the dynamics of vortons. The errors in our predictions can be explained by curvature corrections and we have provided evidence that (with the exception of the frequencies of oscillation) this effect is diminished for larger vortons.

Whether vortons are a cosmologically relevant phenomenon has been an unanswered question since they were proposed. The answer crucially depends on how they form, their stability and how ubiquitous they are within the parameter space. It is important to stress that, although fully stable vortons are perhaps the most interesting solution, unstable vortons with a small rate of growth may last for long enough to have an impact on cosmology. Additionally, the growth rate of extrinsic oscillations is inversely proportional to the vorton radius which means that large vortons (as are relevent to cosmology) will decay more slowly than the ones that we have presented, unless they suffer from a pinching instability that is not caused by curvature corrections. We believe that the methods presented in this paper, and the confirmation that fully stable vortons exist, represents a significant step forward in answering this question.

\appendix

\section{Numerical methods}

Here, we will discuss in more detail the techniques that we have employed to simulate vortons. In order to discretise a system with a local symmetry group we used methods from lattice gauge theory (see \cite{Creutz1983} for a review and \cite{Moriarty1988} for an example of implementation for the Abelian-Higgs model). Furthermore, a vorton has a cylindrical symmetry which can be taken advantage of. The natural coordinate system to choose is clearly cylindrical polar coordinates, however these are numerically unstable due to the coordinate singularity at the origin. We can avoid this problem by either using the cartoon method \cite{Alcubierre1999} which uses a single 2D plane in Cartesian coordinates and calculates the perpendicular derivative by symmetry arguments, or by cutting out a cylinder which includes the origin and then using cylindrical coordinates. The latter method relies on the fields being well localised because it introduces a new boundary upon which we must set a boundary condition. Numerical errors will be introduced in the cartoon method by the interpolations necessary for calculated the normal derivative and by the additional boundary in the cut-off method.

\subsection{Lattice gauge theory} \label{sec: lattice gauge theory}

In order to retain the local $U(1)_\phi$ symmetry during simulations on a discretised grid, we cannot rely on the continuum form of the covariant derivative. Attempting to run dynamical simulations with regular gauge fields results in the violation of the gauge condition. Comparisons between neighbouring points can only be performed by mapping between neighbouring fibres, using an element of the symmetry group. This is analogous to the role of Christoffel symbols in General Relativity. We will assume from now on that the lattice spacing, $a$, is the same in all directions as it simpler to write and easy to generalise. We also work in the temporal gauge ($A_t = 0$) so that the time derivatives can be treated as regular finite difference operators and then $A_t$ provides no contribution to the Yang-Mills term. We define our covariant finite difference operators as
\begin{equation}
    \Delta_i\phi(\boldsymbol{x}) = \frac{\phi(\boldsymbol{x}+a\hat{\boldsymbol{i}}) - U_i(\boldsymbol{x})\phi(\boldsymbol{x})}{a},
\end{equation}
where we define $U_i = e^{igA_i a}$ such that the usual covariant derivative is recovered in the continuum limit. It is convenient to define the lattice link variables as $\theta_i=gA_i a$ and use these as the dynamical variables. The only part of the Lagrangian that cannot be replaced with finite difference operators is the Yang-Mills term because it would not result in a gauge-invariant quantity. Furthermore, we do not want to be restricted to the continuum limit so it is better to express it in terms of group elements rather than gauge fields. The only gauge invariant quantity that can be constructed on a lattice, purely from group elements, is the trace of a closed loop - known as a Wilson loop. This "plaquette action" is defined as

\begin{equation}
    P_{ij}(\boldsymbol{x}) = U_i(\boldsymbol{x})U_j(\boldsymbol{x} + a\hat{\boldsymbol{i}})U_i^{-1}(\boldsymbol{x} + a\hat{\boldsymbol{j}}) U_j^{-1}(\boldsymbol{x}),
\end{equation}

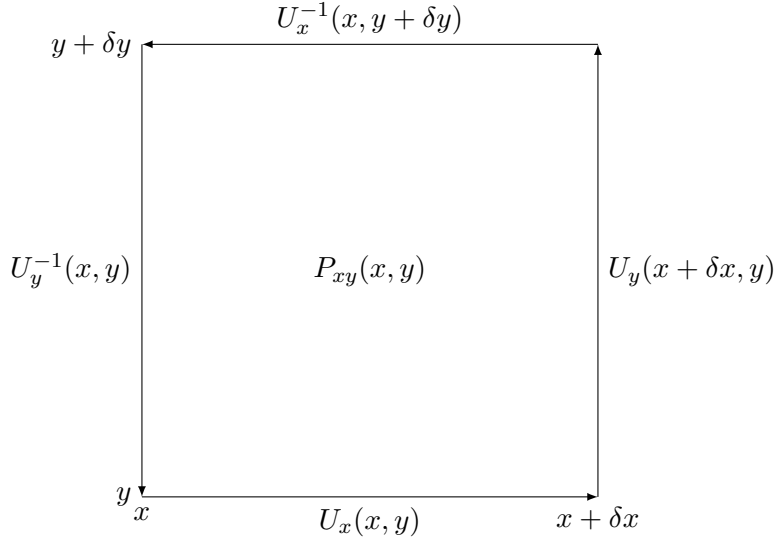
\begin{figure}[t]
\centering
\begin{tikzpicture}[scale=3]
    \draw [->] (-1,-1) -- (1,-1);
    \draw [->] (1,-1) -- (1,1);
    \draw [<-] (-1,1) -- (1,1);
    \draw [<-] (-1,-1) -- (-1,1);
    \node [below] at (-1,-1) {$x$};
    \node [below] at (1,-1) {$x+\delta x$};
    \node [left] at (-1,-1) {$y$};
    \node [left] at (-1,1) {$y+\delta y$};
    \node [below] at (0,-1) {$U_x(x,y)$};
    \node [right] at (1,0) {$U_y(x + \delta x,y)$};
    \node [above] at (0,1) {$U_x^{-1}(x,y + \delta y)$};
    \node [left] at (-1,0) {$U_y^{-1}(x,y)$};
    \node at (0,0) {$P_{xy}(x,y)$};
\end{tikzpicture}
\caption{A diagram of a Wilson loop.}
\end{figure}

\noindent which is related to the discrete version of the field strength tensor by

\begin{equation}
    P_{ij} = \exp(ig F_{ij} a^2).
\end{equation}

\noindent Taking the taylor expansion of this gives

\begin{equation}
    P_{ij} = 1 + igF_{ij}a^2 - \frac{1}{2}(gF_{ij}a^2)^2 + \mathcal{O}(g^3),
\end{equation}

\noindent which allows the Yang-Mills term to be be expressed in a gauge invariant way as

\begin{equation}
    \frac{1}{4}F_{ij}^2 \approx \frac{1}{2(g a^2)^2}[1-\text{Re}(P_{ij})].
\end{equation}

\noindent Now the Lagrangian can be completely expressed using discrete, lattice quantities that do not break the symmetry of the model, \cite{Moriarty1988}

\begin{equation}
\begin{split}
    L = a^3\sum_{\boldsymbol{x}} \bigg\{& |\dot{\phi}(\boldsymbol{x})|^2 - \sum_i \bigg(\frac{\phi(\boldsymbol{x} + a\hat{\boldsymbol{i}}) - e^{i\theta_i(\boldsymbol{x})}\phi(\boldsymbol{x})}{a}\bigg)\bigg(\frac{\phi^*(\boldsymbol{x} + a\hat{\boldsymbol{i}}) - e^{-i\theta_i(\boldsymbol{x})}\phi^*(\boldsymbol{x})}{a}\bigg) \\
    + & |\dot{\sigma}(\boldsymbol{x})|^2 - \sum_i \bigg(\frac{\sigma(\boldsymbol{x} + a\hat{\boldsymbol{i}}) - 
    \sigma(\boldsymbol{x})}{a}\bigg)\bigg(\frac{\sigma^*(\boldsymbol{x} + a\hat{\boldsymbol{i}}) - \sigma^*(\boldsymbol{x})}{a}\bigg) \\
    + & \frac{1}{2}\sum_i\bigg(\frac{\dot{\theta}_i(\boldsymbol{x})}{ag}\bigg)^2 - \frac{1}{2g^2}\sum_{i,j}\frac{1-cos[\theta_i(\boldsymbol{x}) + \theta_j(\boldsymbol{x} + a\hat{\boldsymbol{i}}) - \theta_i(\boldsymbol{x} + a\hat{\boldsymbol{j}}) - \theta_j(\boldsymbol{x})]}{a^4} \\
    - & \frac{\lambda_\phi}{4}(|\phi(\boldsymbol{x})|^2 - \eta_\phi^2)^2 - \frac{\lambda_\sigma}{4}(|\sigma(\boldsymbol{x})|^2 - \eta_\sigma^2)^2 - \beta|\phi(\boldsymbol{x})|^2|\sigma(\boldsymbol{x})|^2 + \frac{\lambda_\sigma}{4}\eta_\sigma^4 \bigg\}.
\end{split}
\end{equation}

\noindent The equations of motions are then derived by varying the fields at each lattice site and requiring the action to be minimised, as usual. This must be done with some care because the contributions from neighbouring lattice sites are easy to miss since they are implicit in the summations.

\begin{equation}
    \Ddot{\phi}(\boldsymbol{x}) = \sum_i\frac{e^{-i\theta_i(\boldsymbol{x})}\phi(\boldsymbol{x} + a\hat{\boldsymbol{i}}) - 2\phi(\boldsymbol{x}) + e^{i\theta_i(\boldsymbol{x} - a\hat{\boldsymbol{i}})}\phi(\boldsymbol{x} - a\hat{\boldsymbol{i}})}{a^2} - \bigg[\frac{\lambda_\phi}{2}(|\phi(\boldsymbol{x})|^2 - \eta_\phi^2) + \beta|\sigma(\boldsymbol{x})|^2\bigg]\phi(\boldsymbol{x}),
\end{equation}

\begin{equation}
    \Ddot{\sigma}(\boldsymbol{x}) = \sum_i\frac{\sigma(\boldsymbol{x} + a\hat{\boldsymbol{i}}) - 2\sigma(\boldsymbol{x}) + \sigma(\boldsymbol{x} - a\hat{\boldsymbol{i}})}{a^2} - \bigg[\frac{\lambda_\sigma}{2}(|\sigma(\boldsymbol{x})|^2 - \eta_\sigma^2) + \beta|\phi(\boldsymbol{x})|^2\bigg]\sigma(\boldsymbol{x}),
\end{equation}

\begin{equation}
\begin{split}
    \Ddot{\theta_i}(\boldsymbol{x}) = &-ig^2\Big[e^{-i\theta_i(\boldsymbol{x})}\phi^*(\boldsymbol{x})\phi(\boldsymbol{x} + a\hat{\boldsymbol{i}}) - e^{i\theta_i(\boldsymbol{x})}\phi(\boldsymbol{x})\phi^*(\boldsymbol{x} + a\hat{\boldsymbol{i}})\Big] \\
     &-\sum_j \frac{1}{a^2}
     \bigg(\sin\Big[\theta_i(\boldsymbol{x}) + \theta_j(\boldsymbol{x} + a\hat{\boldsymbol{i}}) - \theta_i(\boldsymbol{x} + a\hat{\boldsymbol{j}}) - \theta_j(\boldsymbol{x})\Big] \\
      &-\sin\Big[\theta_i(\boldsymbol{x} - a\hat{\boldsymbol{j}}) + \theta_j(\boldsymbol{x} + a\hat{\boldsymbol{i}} - a\hat{\boldsymbol{j}}) - \theta_i(\boldsymbol{x}) - \theta_j(\boldsymbol{x} - a\hat{\boldsymbol{j}})\Big]\bigg).
\end{split}
\end{equation}

\noindent The additional gauge condition can be derived by reintroducing $A_t$ back into the Lagrangian by taking

\begin{equation}
    |\dot{\phi}(\boldsymbol{x})|^2 \to |\dot{\phi}(\boldsymbol{x}) - igA_t(\boldsymbol{x})\phi(\boldsymbol{x})|^2 \quad \quad \bigg(\frac{\dot{\theta}_i(\boldsymbol{x})}{ag}\bigg)^2 \to \bigg(\frac{\dot{\theta}_i(\boldsymbol{x})}{ag} - \frac{A_t(\boldsymbol{x} + a\hat{\boldsymbol{i}}) - A_t(\boldsymbol{x})}{a}\bigg)^2,
\end{equation}

\noindent and then minimising the action with respect to variations in $A_t(\boldsymbol{x})$.

\begin{equation}
    \sum_i \frac{\dot{\theta}_i(\boldsymbol{x}) - \dot{\theta}_i(\boldsymbol{x} - a\hat{\boldsymbol{i}})}{a^2g} + ig\Big[\phi^*(\boldsymbol{x})\dot{\phi}(\boldsymbol{x}) - \phi(\boldsymbol{x})\dot{\phi}^*(\boldsymbol{x})\Big] = 0.
\end{equation}

\begin{figure}[t]
    \centering
    \subfloat[The evolution of the deviation parameter during radial dynamics.]{
        \centering
        \includegraphics[trim={0 0 1.5cm 1cm},clip,width=0.46\linewidth]{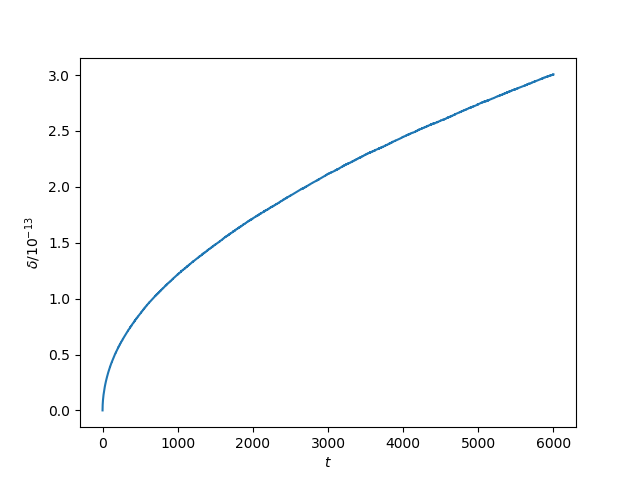}
    }\hspace{1em}
    \subfloat[The evolution of the deviation parameter during 3D dynamics.]{
        \centering
        \includegraphics[trim={0 0 1.5cm 1cm},clip,width=0.46\linewidth]{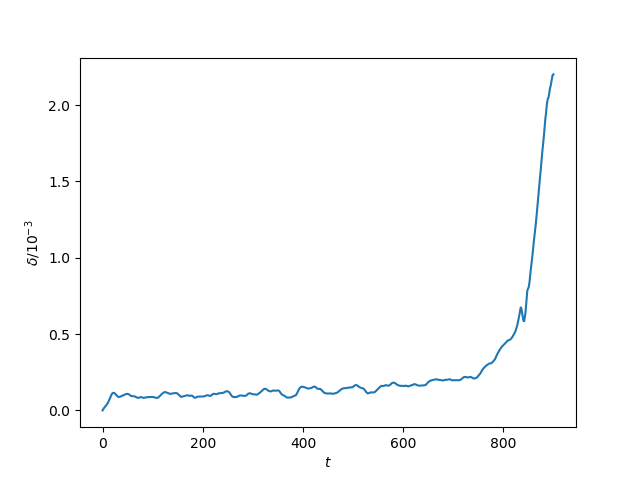}
    }
    \caption{Examples of how the deviation parameter evolves during both radial and full 3D dynamics. Both of these show the evolution of vortons constructed in parameter set A with $Q=749$ and $N=50$. It remains incredibly small during radial dynamics and grows smoothly. This is in contrast to the 3D dynamics where it is nearly $10$ orders of magnitude larger (although still within acceptable limits) and varies more wildly, growing particularly quickly after the vorton starts to break apart at $t\approx 750$, but remaining sufficiently small before this point.}
    \label{fig:dev_param}
\end{figure}

\noindent This condition is never directly enforced by our code but should remain approximately satisfied if the system is evolved according to the equations of motion above. We define the deviation parameter, $\delta$, to be the absolute value of the left hand side of this equation, averaged over all grid points. Figure \ref{fig:dev_param} shows how this deviation parameter evolves during some of our simulations. It is exactly zero initially because, except for the condensate, all of the fields are set to be equal for the first two time steps.

\subsection{Cartoon method} \label{sec: cartoon}

This is a technique developed in \cite{Alcubierre1999} for simulating axisymmetric systems using cartesian coordinates in only one plane (which we usually choose to be $y=0$). In order to get derivatives in the $y$ direction, the neighbouring planes are considered but not evolved. By assuming cylindrical symmetry, the fields on each point in the neighbouring planes will be equal to some point in the $y=0$ plane. Typically this requires interpolation because the corresponding position in the $y=0$ plane is not included in the simulation grid. For the case of a vorton, all fields are cylindrically symmetric except for the condensate field due to the winding of the phase around the loop. However, the magnitude is symmetric and can be interpolated. Then the value of the field can be deduced using the phase factor $\exp(iN\theta)$. Furthermore, only one quadrant of the $x,z$ plane needs to be used as the $x<0$ and $z<0$ sectors are related to the $x\geq0,z\geq0$ region by reflection symmetries.

However, this method is not obviously compatible with lattice gauge theory. The fields are easy to interpolate because they are defined at a single point. The link variables on the other hand, are mappings between two points. Rotating these mappings will only provide information about radial mappings - not the Cartesian mappings required in the cartoon method.
A coordinate transformation is required to relate the polar link variables to the Cartesian ones. Since $A_x = A_\rho\cos(\phi)$ and $\Delta x = \Delta \rho \cos(\phi)$ then $\theta_x = \theta_\rho \cos^2(\phi).$ The value of $\theta_\rho$ can be calculated by interpolation although care must be taken to be mapping between the correct two points.

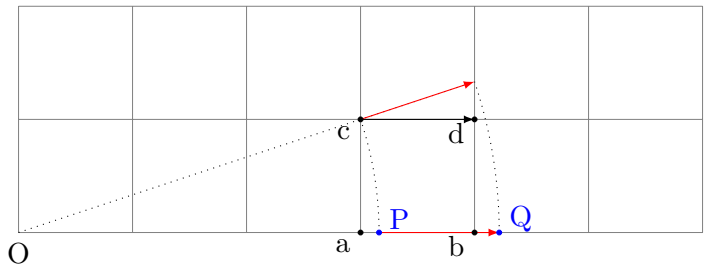
\begin{figure}[t]
\centering
\begin{tikzpicture}[scale=3]
    \tikzmath{\x = 1.5; \a = 0.5; \d = 0.03;}
    \draw[step=0.5cm,gray,very thin] (0,0) grid (3,1);
    \draw [->] (\x,\a) -- (\x + \a,\a);
    \draw [dotted] (0,0) -- (\x + \a,{\a*(\x + \a)/\x});
    \draw [red,->] (\x,\a) -- (\x + \a,{\a*(\x + \a)/\x});
    \draw [red,->] ({sqrt(\x^2 + \a^2)},0) -- ({sqrt((\x + \a)^2 + (\a*(\x + \a)/\x)^2)},0);
    \draw [dotted] ({sqrt(\x^2 + \a^2)},0) arc (0:{atan(\a/\x)}:{sqrt(\x^2 + \a^2});
    \draw [dotted] ({sqrt((\x + \a)^2 + (\a*(\x + \a)/\x)^2)},0) arc (0:{atan(\a/\x)}:{sqrt((\x + \a)^2 + (\a*(\x + \a)/\x)^2)});
    % \draw [blue,->] (\x, \d) -- ({sqrt(\x^2 + \a^2)}, \d);
    % \draw [blue,->] (\x + \a, \d) -- ({sqrt((\x + \a)^2 + (\a*(\x + \a)/\x)^2)}, \d);
    
    \node [left] at (\x, -2*\d) {a};
    \node [left] at (\x + \a, -2*\d) {b};
    \node [left] at (\x + \a, \a - 2*\d) {d};
    \node [left] at (\x, \a - 2*\d) {c};
    \node [right, blue] at ({sqrt(\x^2 + \a^2)}, 2*\d) {P};
    \node [right, blue] at ({sqrt((\x + \a)^2 + (\a*(\x + \a)/\x)^2)}, 2*\d) {Q};
    \node [below] at (0,0) {O};
    
    \filldraw (\x,0) circle (0.3pt);
    \filldraw (\x + \a, 0) circle (0.3pt);
    \filldraw (\x, \a) circle (0.3pt);
    \filldraw (\x + \a, \a) circle (0.3pt);
    \filldraw [blue] ({sqrt(\x^2 + \a^2)},0) circle (0.3pt);
    \filldraw [blue] ({sqrt((\x + \a)^2 + (\a*(\x + \a)/\x)^2)},0) circle (0.3pt);
\end{tikzpicture}
\caption{The link variable required is the one that maps $c \to d$. This can be calculated from the upper red link, which is equal to the lower one by cylindrical symmetry. The lower red link is given by $\theta_{PQ} = -\theta_{aP} + \theta_{ab} + \theta_{bQ}$ where interpolation is used to approximate the mapping from $a \to P$ and $b \to Q$.}
\end{figure}

There are severe limitations placed on the parameter space due to the numerical feasibility of this method. If the ratio of the vorton radius to string width is too large, the number of grid points required to run accurate simulations will become too large for the available computing resources. In particular, this is an issue for predicting vortons using the semi-analytic method because the cartoon method operates in precisely the regime in which the straight string approximation breaks down.

\subsection{Cut off method} \label{sec: cut off}

This method uses cylindrical coordinates, but with a minimum $\rho$ at which we impose boundary conditions to avoid the coordinate singularity at the origin. We usually either use fixed boundary conditions or set the radial derivative (covariant for gauged fields) to be zero. Clearly, for the latter to be a valid approximation, $\rho_{\text{min}}$ must be far from the string core and the fields must be well localised to it. Both the vortex field and the condensate are likely to be well localised to the string as this is the nature of a soliton, however the gauge field is long range. The extent of this issue can be assessed by comparison with the straight string solutions. At some distance from the core, $A_\theta \to n/g$, so we can expect that (so long as the straight string profiles are a good approximation to a vorton) that $A_z \propto \rho^{-1}$ beyond this distance. Of course, this does not guarantee that this is the case outside the plane of the vorton where other field components are non-zero and the phase of $\phi$ varies with $\rho$.

The benefits of this technique are that the winding of the condensate can be treated exactly (in 2D simulations) and it allows vortons with a much larger radius to be simulated. There is no need for any interpolation which means that the code can also run faster.
The method is quite complementary with the cartoon method as the small vortons that are not accessible to this method are perfect for the cartoon method and vice versa. Additionally, fully 3D simulations will be limited by the winding number, $N$, (since larger values will require more sampling of the angular variations) rather than the size of the vorton. We therefore expect that this method will scale better than the cartoon method for larger vortons (with the ratio of $N$ to $Q$ fixed) as the number of grid points only increases linearly with the size of the vorton, rather than quadratically.

\begin{figure}[t]
    \centering
    \includegraphics[scale=0.7]{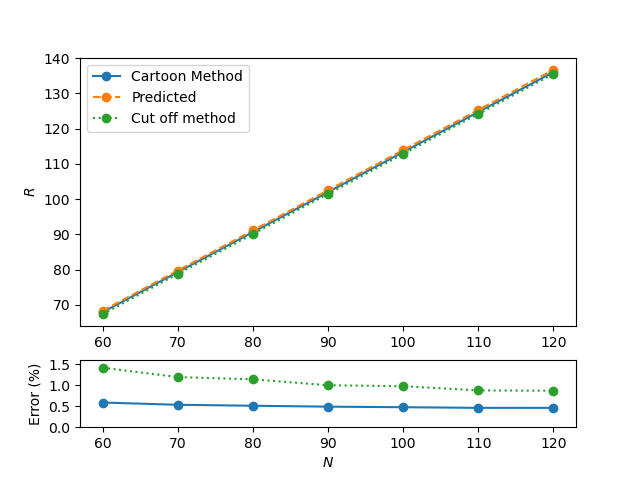}
    \caption{A comparison of the radii of vortons produced by gradient flow using the cut off method and the cartoon method in parameter set B with $Q/N = 31.89$ kept constant. The predicted radii from the thin string analysis has also been displayed. The cartoon method is in better agreement with the predicted radii than the cut off method. Both of the percentage error plots decrease with $N$, with the cut off method approaching the accuracy of the cartoon method at larger radii.}
\label{fig: method compare}
\end{figure}

Which method is more accurate depends on the trade off between errors introduced by the boundary at $\rho_\text{min}$, and errors introduced by interpolation. From Figure \ref{fig: method compare} it appears that the results agree with the TSA better when the cartoon method is used. Nevertheless, the cut off method may still prove to be a valuable technique for investigating very large vortons, particularly if improvements can be made to the boundary conditions.

\bibliographystyle{JHEP}
\bibliography{refs.bib}

\end{document}